\documentclass[3p,final,11pt]{elsarticle}
\usepackage[svgnames,dvipsnames]{xcolor}
\pdfoutput=1
\usepackage{multirow}
\usepackage{comment}
\usepackage{graphicx}
\usepackage{epstopdf}
\usepackage{amssymb,amsmath}
\usepackage{relsize}
\usepackage{placeins}
\usepackage{mathtools}
\usepackage{setspace}
\usepackage{enumitem}
\usepackage{lineno}
\usepackage{tikz}
\usepackage{pgffor}
\usepackage{booktabs}   
\usepackage{lineno,hyperref}

\biboptions{sort&compress}
\def \ie{{\it\frenchspacing i.e.\ }}
\def \eg{{\it\frenchspacing e.g.\ }}
\def \cf{{\it\frenchspacing cf.\ }}

\def \viz{{\it\frenchspacing viz.\ }}
\def \etal{{\it\frenchspacing et al.\ }}
\def \xv{\boldsymbol{x}}
\def \rv{\boldsymbol{r}}

\providecommand{\e}[1]{\ensuremath{\times 10^{#1}}}

  \newlist{steps}{enumerate}{1}
  \setlist[steps]{label=\textit{Step~\arabic*}}

  \graphicspath{{figures/}}

\newcommand{\grad}{\nabla}

\newcommand{\tv}{\mathbf{ t}}
\newcommand{\uv}{\mathbf{ u}}

\newcommand{\Jc}{\mathcal{J}}

\newcommand{\omegav}{\boldsymbol{\omega}}
\newcommand{\tauv}{\boldsymbol{\tau}}

\newcommand{\bogus}[1]{}

\renewcommand{\url}[1]{}
\newcommand{\citeCount}[1]{}

\includecomment{mycomment}

\newlength{\tfwidth}
\newlength{\tfheight}
\newlength{\tfxa}
\newlength{\tfxb}
\newlength{\tfya}
\newlength{\tfyb}
\newcommand{\trimFigWithBox}[6]{%
\setlength\fboxsep{0pt}%
\setlength\fboxrule{1.0pt}
\fbox{\includegraphics[width=#2, clip, trim=#3 #4 #5 #6]{#1}}%
}
\newcommand{\trimFigNoBox}[6]{%
\setlength\fboxsep{1pt}
\setlength\fboxrule{0.0pt}
\fbox{\includegraphics[width=#2, clip, trim=#3 #4 #5 #6]{#1}}%
}
\newcommand{\trimFigHeightWithBox}[6]{%
\setlength\fboxsep{0pt}%
\setlength\fboxrule{1.0pt}
\fbox{\includegraphics[height=#2, clip, trim=#3 #4 #5 #6]{#1}}%
}
\newcommand{\trimFigHeightNoBox}[6]{%
\setlength\fboxsep{1pt}
\setlength\fboxrule{0.0pt}
\fbox{\includegraphics[height=#2, clip, trim=#3 #4 #5 #6]{#1}}%
}

\newsavebox\figBox

\newcommand{\trimw}[6]{%
\sbox\figBox{\includegraphics{#1}}
\setlength{\tfwidth}{\the\wd\figBox}
\setlength{\tfheight}{\the\ht\figBox}
\setlength{\tfxa}{\tfwidth*\real{#3}}%
\setlength{\tfxb}{\tfwidth*\real{#4}}%
\setlength{\tfya}{\tfheight*\real{#5}}%
\setlength{\tfyb}{\tfheight*\real{#6}}%
\trimFigNoBox{#1}{#2}{\tfxa}{\tfya}{\tfxb}{\tfyb}%
}

\newcommand{\trimwb}[6]{%

\sbox\figBox{\includegraphics{#1}}
\setlength{\tfwidth}{\the\wd\figBox}
\setlength{\tfheight}{\the\ht\figBox}
\setlength{\tfxa}{\tfwidth*\real{#3}}%
\setlength{\tfxb}{\tfwidth*\real{#4}}%
\setlength{\tfya}{\tfheight*\real{#5}}%
\setlength{\tfyb}{\tfheight*\real{#6}}%
\trimFigWithBox{#1}{#2}{\tfxa}{\tfya}{\tfxb}{\tfyb}%
}

\newcommand{\trimh}[6]{%
\sbox\figBox{\includegraphics{#1}}
\setlength{\tfwidth}{\the\wd\figBox}
\setlength{\tfheight}{\the\ht\figBox}
\setlength{\tfxa}{\tfwidth*\real{#3}}%
\setlength{\tfxb}{\tfwidth*\real{#4}}%
\setlength{\tfya}{\tfheight*\real{#5}}%
\setlength{\tfyb}{\tfheight*\real{#6}}%
\trimFigHeightNoBox{#1}{#2}{\tfxa}{\tfya}{\tfxb}{\tfyb}%
}

\newcommand{\trimhb}[6]{%

\sbox\figBox{\includegraphics{#1}}
\setlength{\tfwidth}{\the\wd\figBox}
\setlength{\tfheight}{\the\ht\figBox}
\setlength{\tfxa}{\tfwidth*\real{#3}}%
\setlength{\tfxb}{\tfwidth*\real{#4}}%
\setlength{\tfya}{\tfheight*\real{#5}}%
\setlength{\tfyb}{\tfheight*\real{#6}}%
\trimFigHeightWithBox{#1}{#2}{\tfxa}{\tfya}{\tfxb}{\tfyb}%
}
\newcommand{\figWidth}{6cm}
\newcommand{\tableFont}{\footnotesize}
\newlength{\ycbTop}
\newlength{\ycbMid}%

\journal{Journal of Computational Physics}

\begin{document}

\begin{frontmatter}

\title{Direct numerical simulation of particulate flows with an overset grid method}

\author[cam]{A.R.~Koblitz\corref{cor1}\fnref{SGRThanks,CDTThanks}}
\ead{ark44@cam.ac.uk}

\author[slb]{S.~Lovett}
\ead{SLovett@slb.com}

\author[cam]{N.~Nikiforakis}
\ead{nn10005@cam.ac.uk}

\author[rpi]{W.D.~Henshaw\fnref{DOEThanks,NSFgrantNew}}
\ead{henshw@rpi.edu}

\cortext[cor1]{Corresponding author}

\address[cam]{Department of Physics, Cavendish Laboratory, J J Thomson Avenue, Cambridge, CB3 0HE, UK}
\address[slb]{Schlumberger Gould Research Centre, High Cross, Madingley Road, Cambridge, CB3 0EL, UK}
\address[rpi]{Department of Mathematical Sciences, Rensselaer Polytechnic Institute, Troy, NY 12180, USA}

\fntext[DOEThanks]{This work was supported by contracts from the U.S. Department of Energy ASCR Applied Math Program.}
\fntext[NSFgrantNew]{Research supported by the National Science Foundation under grant DMS-1519934.}
\fntext[SGRThanks]{Research supported by the Schlumberger Gould Research Centre.}
\fntext[CDTThanks]{This work was supported by the EPSRC Centre for Doctoral Training in Computational Methods for Materials Science under grant EP/L015552/1}

\begin{abstract}
    We evaluate an efficient overset grid method for two-dimensional and three-dimensional particulate 
    flows for small numbers of particles at finite Reynolds number. The rigid particles are 
    discretised using moving overset grids overlaid on a Cartesian background grid. This allows 
    for strongly-enforced boundary conditions and local grid refinement at particle surfaces, thereby 
    accurately capturing the viscous boundary layer at modest computational cost. The incompressible 
    Navier--Stokes equations are solved with a fractional-step scheme which is second-order-accurate in 
    space and time, while the fluid--solid coupling is achieved with a partitioned approach including 
    multiple sub-iterations to increase stability for light, rigid bodies. Through a series of 
    benchmark studies we demonstrate the accuracy and efficiency of this approach compared 
    to other boundary conformal and static grid methods in the literature. In particular, we find 
    that fully resolving boundary layers at particle surfaces is crucial to obtain accurate solutions 
    to many common test cases. With our approach we are able to compute accurate solutions using as 
    little as one third the number of grid points as uniform grid computations in the literature. A 
    detailed convergence study shows a 13-fold decrease in CPU time over a uniform grid test case 
    whilst maintaining comparable solution accuracy.
\end{abstract}


\end{frontmatter}

\section{Introduction}

Flows of finite-sized particles in viscous fluids are common to many industrial as well as natural processes, 
such as primary cementing in the oil and gas industry \cite{Nelson2006} and blood flow \cite{Bagchi2007}. Being so 
ubiquitous, particulate-flow problems 
span a large range of material and flow properties. Of interest to this work are flows of an incompressible 
Newtonian fluid with rigid, spherical (circular) particles of finite Reynolds number, where the Reynolds number describes 
the relative strength of inertial to viscous forces.

Approximate solution methods have been applied to both high and low particle Reynolds number flow regimes, 
where by neglecting viscous or inertial contributions, respectively, the equations of motion can be 
linearised and solved with powerful mathematical tools; 
see~\cite{Sangani1993,Kushch2002,Brady1988} for examples in both flow regimes. It is the intermediate flow 
regime, where such approximations are not valid, that the full 
incompressible Navier--Stokes equations must be solved. 
Direct numerical 
simulation\footnote{The definition of what constitutes a direct numerical simulation is taken to be that of a simulation within a finite Reynolds number regime where no sub-grid closure models, such as turbulence or wall models, are used and the governing equations can be directly solved such that all continuum time- and length-scales can be resolved through grid refinement alone.} methods designed for such intermediate Reynolds number flow regimes broadly fit into two categories: 
those methods that use a static grid that is unaligned
with the particle boundaries, and those that use boundary-conforming grids. Below we give a brief overview 
of these two classes of methods and their application to particulate flow problems.

Arbitrary Lagrangian Eulerian (ALE) methods were developed for deforming boundary problems as a 
response to the difficulties encountered when using fully Lagrangian schemes in problems 
with large amounts of circulation, see for example, \cite{Behr1994}. 
The ALE method belongs to the class of boundary conformal methods, 
to which the boundary integral method (only applicable to inviscid, irrotational flow) and 
the overset grids method belong. 
The semi-discrete, \viz finite element in space and finite difference in time, method of 
Hughes \etal\cite{Hughes1981} has proved popular with general fluid structure interaction (FSI) 
problems and has been continually developed 
over the last few decades. In this ALE approach, nodal velocities explicitly enter the 
convective terms of the momentum equations allowing the underlying grid to distort with the 
moving boundary, while an Eulerian description can be used in regions of large circulation. 
Hu \etal\cite{Hu2001} used an ALE method with a combined fluid--solid formulation to simulate 
pairs of spheres interacting in Newtonian and visco-elastic liquids, and 2D sedimentation 
simulations of 90 circular particles.

This ALE method can be considered a special case of the deforming spatial domain or stabilised 
space-time (DSD/SST) method developed by Tezduyar \etal\cite{Tezduyar1992}, where a finite element 
formulation is used both in space and time. 
By extending the finite element formulation 
in time any grid deformation is automatically accounted for \cite{Tezduyar1992}. 
To avoid turning a three dimensional 
problem into a four dimensional problem, courtesy of the time dimension, a discontinuous-in-time 
space-time finite element formulation of the problem is solved for one space-time ``slab'' at 
a time.

Both the conventional ALE and DSD/SST methods use unstructured grids to explicitly represent 
the particle surface. Unstructured grids allow for complex 
surfaces to be represented but are generally not  
amenable to fast elliptic solvers, such as geometric multigrid.
The distortion of the underlying grid inevitably causes elements to become increasingly 
skewed. Highly skewed elements can often cause problems relating to accuracy and stability, 
requiring a new grid to be computed 
on to which the solution is then projected. Both the regridding and projection steps are costly 
procedures and are the main 
drawbacks of ALE and DSD/SST type methods \cite{Haeri2012}. Particulate flow simulations 
generally encounter large deformations, 
requiring frequent regridding. 

Various approaches have been developed to improve the quality of the deformed grid. 
Johnson and Tezduyar~\cite{Johnson1996}, for example, used the equations of linear elasticity to govern the deformation of 
the grid for their particulate flow simulations. The grid distortion was controlled 
through a stiffness parameter, effectively weighting distortion towards larger elements. This 
preserved smaller elements in the surface regions that are important to the accuracy of the 
overall solution. By promoting less important elements, \ie those far away from the particle 
surfaces, to distort more easily, regridding is reduced since the grid quality is, 
potentially, preserved for longer. However, this comes at the cost of solving an additional 
partial differential equation for the grid deformation. 

Due to the complexities and costs of the unstructured regridding associated with 
boundary-conforming grid methods, the static grid methods have gained favour with many 
researchers for approaching FSI problems.

Static grid methods can be split into two categories based on the treatment of the solid--fluid 
interface: diffuse interface and sharp interface methods. 
An efficient class of diffuse 
interface methods is the immersed boundary (IB) method. The original IB method 
of Peskin~\cite{Peskin1972} was developed to study flow patterns around heart valves. The entire 
computational domain is represented by a Cartesian grid and the interface is 
represented by a collection of massless Lagrangian points. The velocity of these Lagrangian 
points is used to compute the stress on the elastic interface through a constitutive relation, 
such as Hooke's law. This singular force distribution is transmitted from the Lagrangian 
interface points to the Eulerian fluid through a forcing term in the momentum equation. The 
information transfer is facilitated by a discrete Dirac delta function, smoothly spreading 
the force from the Lagrangian to the Eulerian grid. Aside from inherent stability benefits, this 
approach is attractive because only the right-hand side of the momentum equation is 
affected, so no changes to the underlying solver are required. Assuming constant 
density across the entire domain also allows fast Poisson solvers to be used without modification 
\cite{Kempe2012a}.

Directly applying Peskin's original IB method to rigid boundary problems is not 
straightforward because the constitutive relation for the elastic boundary is generally not 
well behaved in the rigid limit \cite{Mittal2005}. Uhlmann~\cite{Uhlmann2005} used the direct forcing 
formulation of Mohd-Yusof~\cite{Mohd1997} whilst retaining the smooth information transfer between Lagrangian 
and Eulerian grid points. Direct forcing methods use a virtual forcing term determined by the 
difference between the desired interface velocities and the velocities interpolated from 
the underlying Eulerian grid. Uhlmann proposed evaluating the forcing term at the Lagrangian 
interface points before smoothly transferring it to Eulerian grid points in order to reduce 
spatial oscillations common to direct forcing methods. With this direct forcing IB method, 
Uhlmann and Doychev~\cite{Uhlmann2014} performed sedimentation simulations with $\mathcal{O}(10^4)$ 
spherical particles in tri-periodic domains.

Kempe and Fr\"ohlich~\cite{Kempe2012a} modified the interpolation and spreading procedure of \cite{Uhlmann2005}, 
greatly improving on the Courant--Friedrichs--Lewy (CFL) time step restriction and the stability for 
light bodies, achieving particle--fluid density ratios as low as 
$\rho_r=0.3$ for spherical particles. With this IB method Vowinckel \etal\cite{Vowinckel2014} 
investigated turbulent channel flow with mobile beds consisting of $\mathcal{O}(10^{4})$ spherical 
particles.

Glowinski \etal\cite{Glowinski2001} developed a distributed Lagrange multiplier/fictitious domain method 
(DLM/FD) reminiscent 
of Peskin's IB method 
using the principle of variational inequality in a finite element context. A combined 
variational formulation of the fluid-solid coupling was used---similar to that of 
\cite{Hu1996}---but the rigidity was enforced in the particle subdomains through 
Lagrange multipliers, thereby allowing the governing equations to be solved on stationary grids. 
The original DLM/FD method imposed velocity on the particle subdomain. Patankar \etal\cite{Patankar2000} 
imposed the deformation-rate tensor instead in order to simplify treatment for 
irregularly shaped particles, while Wachs~\cite{Wachs2009} used this formulation in conjunction with 
a discrete element method to develop an improved collision mechanism for irregularly shaped 
bodies. 

The smoothing intrinsic to diffuse interface methods reduces the solution accuracy in the 
immediate vicinity of the interface. This can be addressed through increased grid resolution, 
but at the cost of 
greatly increased calculation size given the grid uniformity. This problem is particularly 
severe in high Reynolds number flows, where the viscous boundary layer is thin. 
One popular way to improve accuracy at the interface is through the use of sharp interface 
methods, which generally reconstruct the solution near the interface, thereby 
enforcing the boundary conditions strongly.
Fadlun \etal\cite{Fadlun2000} developed a hybrid approach that borrowed on the momentum forcing 
approach of \cite{Peskin1972} but with the goal of enforcing boundary conditions exactly. 
Fadlun \etal\cite{Fadlun2000} used the direct forcing formulation of \cite{Mohd1997}, but imposed boundary 
conditions by directly 
modifying the coefficients of the linear system rather than explicitly calculating the 
forcing. This was done in order to avoid solving an implicit system for the forcing, which came about 
because of the implicit treatment of the viscous terms in their fractional step 
time advancement scheme \cite{Fadlun2000}. In effect, this is equivalent to applying 
momentum forcing inside the fluid domain, leading to problems with moving grids when 
previously ``solid'' grid points are converted to ``fluid'' points. 
Yang and Balaras~\cite{Yang2006} demonstrated that unphysical information carried by a previously solid point 
into the flow field makes the evaluation of the right-hand side of the momentum equation 
in the fractional step algorithm problematic. They developed a ``field extension'' 
strategy, whereby the velocity and pressure fields are extended into the solid phase at the 
end or beginning of each sub-step. 
Borazjani \etal\cite{Borazjani2008}, developed a sharp interface method based on front-tracking
for general curvilinear background grids that can handle an arbitrary number of bodies in three dimensions. Added-mass effects
were treated with sub-iterations and an Aitken acceleration technique.
Recently, Yang and Stern~\cite{Yang2015} improved on the field extension strategy through the introduction 
of temporary non-inertial reference frames attached to the solid phases. This is based on 
the fluid--solid coupling strategy of Kim and Choi~\cite{Kim2006} and allows the velocity coupling to be 
done in a non-iterative fashion. Crucially, \cite{Yang2015} only use the non-inertial 
reference frame for the forcing, solving the flow field on the inertial reference frame and 
thereby allowing for an arbitrary number of solid phases. By allowing non-iterative 
methods to be used for the forcing and rigid-body motion calculations, \cite{Yang2015} 
improved the performance by up to an order of magnitude. However, such field extension 
strategies reduce the sharpness of the immersed boundary, diminishing the advantage over 
diffuse methods \cite{Seo2011}.

Flow reconstruction based sharp interface methods suffer from force oscillations 
in moving boundary problems \cite{Luo2012}. This is caused by the instantaneous role reversal 
of a grid point from an interpolation to a discrete Navier--Stokes point, and vice versa, 
when crossed by the boundary. This role reversal involves an immediate change in computational 
stencil, causing a temporal discontinuity that even the field extension of 
\cite{Yang2006} does not address~\cite{Luo2012}.
Seo and Mittal~\cite{Seo2011} explored the source of the temporal discontinuity in their study using 
    a cut-cell sharp interface method and were able to reduce the oscillations by roughly 
an order of magnitude through improved geometric mass conservation near the interface.

Much recent effort has been focused on static grid methods for particulate flow, owing to 
the high efficiency of such methods and the desire to simulate engineering flows containing huge numbers of 
particles. Broadly speaking, boundary conformal methods can offer superior accuracy near the solid/fluid interface 
but may be inefficient for problems  with large spatial deformations due to the
costs associated with regridding at each time step.
On the other hand, static grid methods---which can be highly efficient---
have difficulty resolving the boundary layers near particles and may require
very fine grids or adaptive grid refinement to obtain accurate results~\cite{Wachs2015}.

In this work we will evaluate the overset grid, or 
Chimera grid, method for viscous particulate flow. 
Overset grid methods have been widely used for problems with moving geometries.
They were recognised early on to be a useful technique for treating
rigid moving bodies, such as aircraft store separation~\cite{Dougherty}, and have
subsequently been applied to many other moving-grid aerodynamic
applications, see for example~\cite{Meakin93,mog2006,Zahle2007,Chan2009,ChandarDamodaran2010,LaniSjogreenYeeHenshaw2012,}. 
English \etal\cite{English2013} present a novel overset grid approach using a 
Voronoi grid to link Cartesian overset grids. This differs to
the method used here, where interpolation stencils are directly substituted into the 
coefficient matrix and solid boundaries are represented using curvilinear grids.
The basic approach of moving overset grids used in this article was developed 
for high-speed compressible and reactive flows by~\cite{mog2006} and included the support for adaptive grid refinement.
The deforming composite grid (DCG) approach was developed in~\cite{fsi2012} for treating deforming bodies 
with overset grids, and a partitioned scheme was developed for light deforming bodies that was stable
without sub-iterations.
A method to overcome the added-mass instability with compressible flows and rigid bodies 
was developed in~\cite{lrb2013}.
More recently, stable partitioned schemes for incompressible flows and deforming solids
have been developed~\cite{Banks2014,fis2014,beamins2016} and extended to non-linear solids~\cite{flunsi2016}.

The method described in this paper retains much of the efficiency of static structured 
grid methods whilst still allowing for sharp representation of solid boundaries.
The overset grid method can be seen as a bridge between the static grid methods such as IBM
and boundary conformal grid methods; the 
curvilinear particle grids allow for higher than first-order accuracy and boundary conditions to be 
implemented strongly, while grid connectivity with the static Cartesian background grid is only locally 
updated. Since the grid connectivity 
is only updated locally, the regridding procedure is less costly and complex than for unstructured 
body conformal methods, such as ALE. Local grid refinement allows boundary layers to be fully resolved 
without appreciably affecting the total grid point count. This is in contrast with general static grid 
methods where the solver efficiency is offset by the unfavourable scaling associated with uniform grids, 
making large fully resolved simulations very costly~\cite{Wachs2015}. For these reasons, we evaluate the 
suitability of the method for fully resolved simulations of incompressible fluid flow with 
rigid particles. 
{Note that the scheme described here is implemented in the Cgins solver that is available 
as part of the Overture framework of codes ({\tt overtureFramework.org}). Although past papers have described the
use of Cgins for bodies undergoing specified motions (\eg\cite{BroeringLianHenshaw2012}), 
the discussion here of the algorithm involving freely-moving rigid-bodies is new.

In section 2 the overset grid method is summarised, before stating the mathematical formulation in 
section 3. These equations are discretised in space and time in section 4 and 5, respectively, while 
section 6 presents the fluid--solid coupling methodology. A grid convergence study is performed 
in section 7.1 using a representative test case to determine appropriate spatial resolutions 
for the wake structures captured by the background grid, and boundary layer captured by the 
particle grid. Following this, validation cases in both two and three space dimensions
are presented, comparing against published experimental and numerical results. The results of 
our evaluation are summarised in section 8, along with an outlook towards future work.

\section{Overset grids}
The method of overset grids (also called overlapping, overlaid or Chimera grids) bridges 
the stationary and boundary conformal grid methods. A complex domain is represented by 
multiple body-fitted curvilinear grids that are allowed to overlap, as shown in Fig.~\ref{fig:overlappingGridCartoon}. 
Overset grids bring flexibility to 
grid generation since component grids are not required to align along block boundaries. 
This flexibility allows component grids to be added 
in a relatively independent manner, requiring only local changes to grid connectivity. 

A composite grid $\boldsymbol{\mathcal{G}}$ consists of logically rectangular component grids 
$\boldsymbol{\mathcal{G}}_k$, with $k=1,2,\dots N_g$.
As illustrated in Fig.~\ref{fig:overlappingGridCartoon}
 the grid points of $\boldsymbol{\mathcal{G}}$ are classified as 
interior points, boundary points, interpolation points and exterior or un-used points.
The algorithm for generating a composite grid from 
a collection of component grids is intricate; a detailed 
description is beyond the remit of this work and so it will be only briefly discussed, with 
the intention of showing why the method of overset grids is appropriate for 
particulate flow problems. The interested reader is referred to~\cite{CGNS} and~\cite{OGEN} for 
a full description of the algorithm and implementation. 

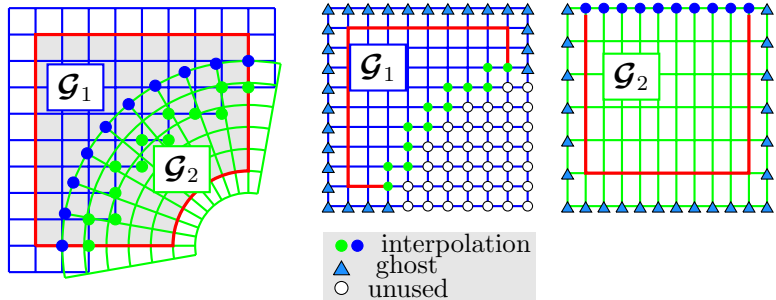
\begin{figure}[hbt]
\begin{center}
%
\begin{tikzpicture}[scale=.7]
\useasboundingbox (.75,1.25) rectangle (15.5,6);  
%
\begin{scope}[xshift=1cm,yshift=1cm]
\fill[black!10!white,xshift=.5cm,yshift=.5cm] (0,0) -- (2.583333,0) arc (180:90:1.416667) -- (4.,4.) -- (0,4.) -- (0,0);
\draw[-,thick,blue,yshift=.0 cm] 
   \foreach \x/\y in {1.5/0,1.5/.5,2/1,2/1.5,2.5/2,3/2.5,4/3,5/3.5,5/4,5/4.5,5/5}{ (0,\y) -- (\x,\y) }
   \foreach \x/\y in {0/0,.5/0,1/0,1.5/0,2/1,2.5/2,3/2.5,3.5/3,4/3,4.5/3.5,5/3.5}{ (\x,\y) -- (\x,5) };
  \begin{scope}[xshift=4.5cm,yshift=0.5cm]
    \draw[thick,green] \foreach \r in {1.000000,1.416667,1.833333,2.250000,2.666667,3.083333,3.500000}{ (0,\r) arc (90:190:\r)  (0,\r) arc (90:80:\r) };
    \draw[thick,green]
     (0.173648,0.984808)  -- (0.607769,3.446827)
     (0.000000,1.000000)  -- (0.000000,3.500000)
     (-0.173648,0.984808) -- (-0.607769,3.446827)
     (-0.342020,0.939693) -- (-1.197071,3.288924)
     (-0.500000,0.866025) -- (-1.750000,3.031089)
     (-0.642788,0.766044) -- (-2.249757,2.681156)
     (-0.766044,0.642788) -- (-2.681156,2.249757)
     (-0.866025,0.500000) -- (-3.031089,1.750000)
     (-0.939693,0.342020) -- (-3.288924,1.197071)
     (-0.984808,0.173648) -- (-3.446827,0.607769)
     (-1.000000,0.000000) -- (-3.500000,0.000000)
     (-0.984808,-0.173648) -- (-3.446827,-0.607769);
  \end{scope}
  \draw[very thick,red,xshift=.5cm,yshift=.5cm] (0,0) -- (2.583333,0) arc (180:90:1.416667) -- (4.,4.) -- (0,4.) -- (0,0);
%
   \filldraw[green] (1.5,.5)  circle (3pt)
                 (1.5,1 )  circle (3pt)
                 (2  ,1 )  circle (3pt)
                 (2 ,1.5)  circle (3pt)
                 (2 , 2 )  circle (3pt)
                 (2.5,2 )  circle (3pt)
                 (2.5,2.5) circle (3pt)
                 (3 , 2.5) circle (3pt)
                 (3  ,3 )  circle (3pt)
                 (3.5,3 )  circle (3pt)
                 (4  ,3. ) circle (3pt)
                 (4  ,3.5) circle (3pt)
                 (4.5,3.5) circle (3pt);
%
  \begin{scope}[xshift=4.5cm,yshift=0.5cm]
      \filldraw[blue]
       (0.000000,3.500000)    circle (3pt)
       (-0.607769,3.446827)   circle (3pt)
       (-1.197071,3.288924)  circle (3pt) 
       (-1.750000,3.031089)  circle (3pt) 
       (-2.249757,2.681156)  circle (3pt) 
       (-2.681156,2.249757)  circle (3pt) 
       (-3.031089,1.750000)  circle (3pt) 
       (-3.288924,1.197071)  circle (3pt) 
       (-3.446827,0.607769)  circle (3pt) 
       (-3.500000,0.000000)  circle (3pt);
  \end{scope}
  \draw (1.25,3.5) node[thick,draw=blue,fill=white] {\large$\boldsymbol{\mathcal{G}}_1$};
  \draw (3.25,1.95) node[thick,draw=green,fill=white] {\large$\boldsymbol{\mathcal{G}}_2$};
\end{scope}
%
\definecolor{ghostColour}{named}{DodgerBlue}
\newcommand{\mytrix}{(\x,-.15) -- ++(.3,0) -- ++(-.15,.26) -- (\x,-.15)}
\newcommand{\mytriy}{(-.15,\y) -- ++(.3,0) -- ++(-.15,.26) -- (-.15,\y)}
\begin{scope}[xshift=7cm,yshift=2.25cm,scale=.75]
\draw[-,thick,blue,yshift=.0 cm] 
   \foreach \x in {0,.5,...,5}{ (\x,0) -- (\x,5) }
   \foreach \y in {0,.5,...,5}{ (0,\y) -- (5,\y) };
  \draw[very thick,red,xshift=.5cm,yshift=.5cm] (1.,0) -- (.0,0) -- (.0,4.) -- (4.,4.) -- (4.,3.);
   \filldraw[green] (1.5,.5)  circle (3pt)
                 (1.5,1 )  circle (3pt)
                 (2  ,1 )  circle (3pt)
                 (2 ,1.5)  circle (3pt)
                 (2 , 2 )  circle (3pt)
                 (2.5,2 )  circle (3pt)
                 (2.5,2.5) circle (3pt)
                 (3 , 2.5) circle (3pt)
                 (3  ,3 )  circle (3pt)
                 (3.5,3 )  circle (3pt)
                 (4  ,3. ) circle (3pt)
                 (4  ,3.5) circle (3pt)
                 (4.5,3.5) circle (3pt);
  \filldraw[fill=white,draw=black]  \foreach \x in {2,2.5,...,5}{ (\x,.0) circle (3.5pt) };
  \filldraw[fill=white,draw=black]  \foreach \x in {2,2.5,...,5}{ (\x,.5) circle (3.5pt) };
  \filldraw[fill=white,draw=black]  \foreach \x in {2.5,3,...,5}{ (\x,1.) circle (3.5pt) };
  \filldraw[fill=white,draw=black]  \foreach \x in {2.5,3,...,5}{ (\x,1.5) circle (3.5pt) };
  \filldraw[fill=white,draw=black]  \foreach \x in {3,3.5,...,5}{ (\x,2.0) circle (3.5pt) };
  \filldraw[fill=white,draw=black]  \foreach \x in {3.5,4,...,5}{ (\x,2.5) circle (3.5pt) };
  \filldraw[fill=white,draw=black]  \foreach \x in {4.5,5}      { (\x,3.0) circle (3.5pt) };
  \draw[fill=ghostColour,xshift=-.15cm,yshift=0cm]  \foreach \x in {.5,1.,1.5}{ \mytrix };  
  \draw[fill=ghostColour,xshift=-.15cm,yshift=5cm]  \foreach \x in {.5,1.,...,5}{ \mytrix };  
  \draw[fill=ghostColour,xshift=0cm,yshift=-.15cm]  \foreach \y in {0,.5,...,5}{ \mytriy };
  \draw[fill=ghostColour,xshift=5cm,yshift=-.15cm]  \foreach \y in {3.5,4,4.5}{ \mytriy };
  \draw (1.25,3.5) node[thick,draw=blue,fill=white] {\large$\boldsymbol{\mathcal{G}}_1$};
\end{scope}
\begin{scope}[xshift=11.5cm,yshift=2.25cm,scale=.75]
\draw[-,thick,green,yshift=.0 cm] 
   \foreach \x in {0,.454545,...,5}{ (\x,0) -- (\x,5) }
   \foreach \y in {0,.833333,...,5}{ (0,\y) -- (5,\y) };
 \draw[very thick,red,xshift=.454545cm,yshift=.833333cm] (0.,4) -- (.0,0) -- (4.0909,0.) -- (4.0909,4);
 \filldraw[blue]  \foreach \x in {.454545,.909090,...,4.545454}{ (\x,5) circle (3.5pt) };
 \draw[fill=ghostColour,xshift=-.15cm]  \foreach \x in {.454545,.909090,...,4.545454}{ \mytrix };
 \draw[fill=ghostColour,yshift=-.15cm]  \foreach \y in {0,.833333,...,5}{ \mytriy };
 \draw[fill=ghostColour,xshift=5cm,yshift=-.15cm]  \foreach \y in {0,.833333,...,5}{ \mytriy };
\end{scope}
\begin{scope}[xshift=7cm,yshift=.7cm]
  \fill[black!10!white,xshift=-.1cm,yshift=-.25cm] (0,0) -- (4,0) -- (4.,1.3) -- (0,1.3) -- (0,0);
  \filldraw[green,xshift=.0cm,yshift=.8cm] (.25,.0)  circle (3pt);
  \filldraw[blue,xshift=.3cm,yshift=.8cm] (.25,.0)  circle (3pt);
  \draw[xshift=.0cm,yshift=.8cm] (.5,0) node[anchor=west,xshift=6] {\small interpolation};
  \draw[fill=ghostColour,xshift=.0cm,yshift=.4cm] (.35,0) \foreach \x in {.1}{ \mytrix } node[anchor=west,xshift=12,yshift=3] {\small ghost};
  \draw[fill=white,draw=black,xshift=.0cm,yshift=.0cm] (.25,0) circle (3.5pt) node[anchor=west,xshift=6] {\small unused};
\end{scope}
\begin{scope}[xshift=11.5cm,yshift=2.25cm,scale=.75]
    \draw (1.6,3.27) node[thick,draw=green,fill=white] {\large$\boldsymbol{\mathcal{G}}_2$};
\end{scope}
\end{tikzpicture}
\end{center}
\caption{Left: an overlapping grid consisting of two
    structured curvilinear component grids, $\xv=\boldsymbol{\mathcal{G}}_1(\rv)$ and $\xv=\boldsymbol{\mathcal{G}}_2(\rv)$. Middle and right: 
component grids for the square and annular grids in the unit square parameter space $\rv$. Grid
 points are classified as discretisation points, interpolation points or unused points. Ghost points
 are used to apply boundary conditions. 
 }
  \label{fig:overlappingGridCartoon}
\end{figure}

When moving grids are used, as is the case in particulate flow problems where each 
particle is represented by a separate component grid, the relative position of 
overset grids changes continuously. As a result, overlapping connectivity information, 
\ie Chimera holes (regions of exterior points in the overset component grids) and 
interpolation points, must be recomputed at every time-step. Crucially, this is cheaper 
than complete grid regeneration and the required connectivity information recomputation 
can be locally confined to those grids affected by the moving grid.

Values of the solution at interpolation points are determined by standard
tensor-product Lagrange-interpolation. We use quadratic interpolation
(three-point stencil in each direction) for the results in this paper, as required for second-order
accuracy~\citep{CGNS}. This interpolation is not locally conservative. Locally conservative interpolation
on overset grids is possible~\citep{CGCN94} but has not been found necessary in
our experience. Corrections to ensure global conservation are also possible and have
been shown to have advantages, see for example~\citep{TangJonesSotiropoulos2003}.

\section{Governing equations}
Incompressible flow is governed by the Navier--Stokes equations,
\begin{linenomath}
\begin{align*}
    \frac{\partial\boldsymbol{u}}{\partial t} + 
    (\boldsymbol{u}\cdot\nabla)\boldsymbol{u} &=
    -\frac{1}{\rho}\nabla p +
    \nu\nabla^2\boldsymbol{u}+\boldsymbol{f},    \\ 
    \grad\cdot\boldsymbol{u}&=0, 
\end{align*}
\end{linenomath}
where 
$\boldsymbol{u}$ is the vector of Cartesian components of the velocity $u_i$, $p$ the pressure field, 
$\rho$ the fluid density, and $\nu=\mu/\rho$ the kinematic viscosity. 
For discretising the equations on a moving grid (on an overset grid some grids are static while others are
attached to, and move with the body), 
we make a change of dependent variables $\xv$ and $t$  to a frame that moves with 
the grid.
As a result, on moving domains, the governing equations transform to 
\begin{linenomath}
\begin{align}
    \frac{\partial\boldsymbol{u}}{\partial t} + 
    ((\boldsymbol{u}-\boldsymbol{w}))\cdot\nabla\boldsymbol{u} &=
    -\frac{1}{\rho}\nabla p +
    \nu\nabla^2\boldsymbol{u}+\boldsymbol{f},   \label{EQ: NSa} \\ 
    \grad\cdot\boldsymbol{u}&=0, \label{EQ: NSb}
\end{align}
\end{linenomath}

where $\boldsymbol{w}$ is the velocity of a point attached to the moving domain. The partial derivative in time 
in the moving frame, as appearing in~\eqref{EQ: NSa}, is therefore 
the derivative in time when keeping the spatial location fixed to a point that is attached to the moving domain.
We solve an alternative formulation of system~\eqref{EQ: NSa}-\eqref{EQ: NSb}. A pressure Poisson equation is derived by 
taking the divergence of the momentum equation~\eqref{EQ: NSa} and using~\eqref{EQ: NSb}. The 
resulting velocity--pressure formulation of the initial-boundary value problem is

\begin{linenomath}
\begin{align}
		\frac{\partial \boldsymbol{u}}{\partial t}
        +({(\boldsymbol{u}-\boldsymbol{w})}\cdot{\nabla})\boldsymbol{u}
		+\frac{1}{\rho}{\nabla}p
		-\nu{\nabla}^2\boldsymbol{u} - \boldsymbol{f} &= 0,
		\qquad\qquad\forall\boldsymbol{x}\in\Omega,  \\
		\mathcal{J}({\nabla}\boldsymbol{u})
		+\frac{1}{\rho}{\nabla}^2p
		-{\nabla}\cdot\boldsymbol{f} &= 0,
		\qquad\qquad\forall\boldsymbol{x}\in\Omega , \label{EQ: PPE}\\
        {B}(\boldsymbol{u},p) &= \boldsymbol{0},  \qquad\qquad\forall\boldsymbol{x}
		\in\partial\Omega , \\
		{\nabla}\cdot\boldsymbol{u} &= 0,
        \qquad\qquad\forall\boldsymbol{x}\in\partial\Omega,  \label{EQ: velocity divergence}\\
		\boldsymbol{u}(\boldsymbol{x},0) &= \boldsymbol{u}_0(\boldsymbol{x}),  \qquad\text{at}\quad t=0,
\end{align}
\end{linenomath}
where $\Jc(\grad\boldsymbol{u})\equiv\grad\boldsymbol{u} : \grad\boldsymbol{u}$ and 
$\Omega$ denotes the fluid domain in $n_d$ space dimensions.
There are 
$n_d$ primary boundary conditions, denoted by $B(\boldsymbol{u},p)=\boldsymbol{0}$. 
The velocity--pressure formulation 
requires an additional boundary condition for the pressure. Here, the velocity divergence \eqref{EQ: velocity divergence} is applied as the boundary condition on the pressure, making the velocity--pressure 
formulation equivalent to the velocity--divergence formulation \citep{Henshaw1994}.
For the second-order accurate scheme 
used here, boundary conditions are required to determine $\boldsymbol{u}$ and $p$ at a line of 
fictitious (ghost) points outside the domain boundary. Some of the numerical boundary conditions are 
{\em compatibility conditions}, derived 
by applying the momentum and pressure equations on the boundary. 

The motion of a rigid body immersed in the fluid is governed by the Newton--Euler equations, 
\begin{linenomath}
\begin{align*}
	\frac{\mathrm{d}\boldsymbol{x}_{b}}{\mathrm{d}t} = \boldsymbol{v}_{b}, \quad
	m_b\frac{\mathrm{d}\boldsymbol{v}_{b}}{\mathrm{d}t} =
	\boldsymbol{F}, \quad
	A\frac{\mathrm{d}\boldsymbol{\omega}}{\mathrm{d}t} = 
	- \omegav\times A\boldsymbol{\omega} + \boldsymbol{T},\quad
	\frac{\mathrm{d}\boldsymbol{e}_i}{\mathrm{d}t} =
	\boldsymbol{\omega}\times\boldsymbol{e}_i. 
\end{align*}
\end{linenomath}
Here $\boldsymbol{x}_b(t)$ and $\boldsymbol{v}_b(t)$ are the position and velocity of the centre of mass, 
respectively, $m_b$ is the mass of the body, $\boldsymbol{\omega}$ is the angular velocity,
$A$ is the moment of inertia matrix, 
$\boldsymbol{e}_i$ are the principal axes of inertia,
$\boldsymbol{F}(t)$ is the applied force, 
and $\boldsymbol{T}(t)$ is the applied torque about 
the centre of mass of the body. 
The principal axes of inertia are integrated over time to find the rotation matrix 
which is used to update positions, velocities and acceleration of points attached 
to the body surface.

The force and torque on the body are determined from both body forces, such as gravity, and hydrodynamic forces arising 
from the stresses exerted by the fluid on the body surface, $\Gamma$, 
\begin{linenomath}
\begin{align*}
    \boldsymbol{F} = \int_{\Gamma} (-p\boldsymbol{n}+\tauv\cdot\boldsymbol{n})\;\mathrm{d}s +
	\boldsymbol{f}_b,\qquad
	\boldsymbol{T} = \int_{\Gamma}
	(\boldsymbol{x}-\boldsymbol{x}_{b})\times(-p\boldsymbol{n}+\tauv\cdot\boldsymbol{n})\;\mathrm{d}s
	+ \boldsymbol{t}_b,
\end{align*}
\end{linenomath}
where 
$\xv$ is a point on $\Gamma$, 
$\tauv=\mu(\grad\boldsymbol{u} + (\grad\boldsymbol{u})^T)$ is the viscous stress tensor,
$\boldsymbol{n}$ is the unit normal vector to the body surface (outward pointing from the fluid domain), 
$\boldsymbol{f}_b$ is any external body force and 
$\boldsymbol{t}_b$ is any external body torque.

\section{Spatial discretisation}

The equations of the velocity-pressure formulation are discretised to second-order accuracy in space using finite difference methods on overset grids, see~\cite{Henshaw1994} and~\cite{splitStep2003}.
An overset grid consists of logically rectangular grids that cover a grid region and overlap where 
they coincide. The solutions between adjacent grids are connected via interpolation conditions.
Each component grid (numbered $k=1,2,\dots,N_g$) is associated with a transformation 
$\boldsymbol{d}_k:\mathbb{R}^3\rightarrow\mathbb{R}^3$ from the unit square, with coordinates denoted by
$\boldsymbol{r}=(r_1,r_2,r_3)$, into physical space, $\boldsymbol{x}=(x_1,x_2,x_3)$, 
and denoted by $\boldsymbol{d}_k(\boldsymbol{r},t)=\boldsymbol{x}(\boldsymbol{r},t)$,
which allows for body fitted grids of non-rectangular shapes. 
Consider solving the equations in 
three space dimensions on a square component grid $\boldsymbol{\mathcal{G}}_k$, with grid spacing 
$h_m=1/N_m$, for a positive integer $N_m$:
\begin{linenomath}
\begin{equation}
	\boldsymbol{\mathcal{G}}_k = \{ \boldsymbol{x}_{\boldsymbol{i},k} ~|~ \boldsymbol{i}=(i_1,i_2,i_3),~
	N_{m,a,k}-1\leq i_m \leq N_{m,b,k}+1,~ m=1,2,3\}, \nonumber
\end{equation}
\end{linenomath}
where $\boldsymbol{i}=(i_1,i_2,i_3)$ is a multi-index and $a$ and $b$ denote the beginning and end grid line 
numbers, respectively. Ghost points are included at the boundaries, 
$i_m = N_{m,a,k}$ or $i_m=N_{m,b,k}$, to facilitate discretising to second order. The component grid number $k$ will be dropped in the following discussion. 
Let $\boldsymbol{U}_{\boldsymbol{i}}\approx\boldsymbol{u}(\boldsymbol{x}_{\boldsymbol{i}},t)$, 
$\boldsymbol{W}_{\boldsymbol{i}} \approx \boldsymbol{w}(\xv_{\boldsymbol{i}},t)$, and 
$P_{\boldsymbol{i}}\approx p(\boldsymbol{x}_{\boldsymbol{i}},t)$ 
 be the numerical approximations to 
$\boldsymbol{u}$, $\boldsymbol{w}$ and $p$, respectively.
The momentum and pressure equations are discretised with second-order finite difference stencils. 
The derivatives with respect to $\boldsymbol{r}$ 
are standard second-order centred finite difference approximations, for example,
\begin{linenomath}
\begin{align*}
	\frac{\partial\boldsymbol{u}}{\partial r_m}\approx D_{r_m}
	\boldsymbol{U}_{\boldsymbol{i}} &\coloneqq 
    \frac{\boldsymbol{U}_{\boldsymbol{i}+\boldsymbol{e}_m} - \boldsymbol{U}_{\boldsymbol{i}-\boldsymbol{e}_m}}{2h_m},\qquad
	\frac{\partial^2\boldsymbol{u}}{\partial r_m^2} 
	\approx D_{r_mr_m}\boldsymbol{U}_{\boldsymbol{i}} \coloneqq
    \frac{\boldsymbol{U}_{\boldsymbol{i}+\boldsymbol{e}_m} - 2\boldsymbol{U}_{\boldsymbol{i}-\boldsymbol{e}_m} + \boldsymbol{U}_{\boldsymbol{i}-\boldsymbol{e}_m}}{h_m^2},
\end{align*}
\end{linenomath}
where  $\boldsymbol{e}_m$ is the unit vector in the $m$-th coordinate direction.
Using the chain rule the derivatives with respect to $\boldsymbol{x}$ are defined as
\begin{linenomath}
\begin{align*}
	\frac{\partial \boldsymbol{u}}{\partial x_m} &= 
	\sum_n \frac{\partial r_n}{\partial x_m}\frac{\partial \boldsymbol{u}}{
	\partial r_n} \approx D_{x_m}\boldsymbol{U}_{\boldsymbol{i}} \coloneqq
	\sum_n\frac{\partial r_n}{\partial x_m}D_{r_n}\boldsymbol{U}_{\boldsymbol{i}},\\
	\frac{\partial^2\boldsymbol{u}}{\partial x_m^2} &= 
	\sum_{n,l} \frac{\partial r_n}{\partial x_m}
	\frac{\partial r_l}{\partial x_m}
	\frac{\partial^2\boldsymbol{u}}{\partial r_n\partial r_l}
	+ \sum\frac{\partial^2 r_n}{\partial x_m^2}
	\frac{\partial \boldsymbol{u}}{\partial r_n}\\
	&\approx D_{x_m}D_{x_m}\boldsymbol{U}_{\boldsymbol{i}} \coloneqq
	\sum_{n,l}\frac{\partial r_n}{\partial x_m}
	\frac{\partial r_l}{\partial x_m}D_{r_mr_l}\boldsymbol{U}_{\boldsymbol{i}}
	+\sum_{n}\left(D_{x_m}\frac{\partial r_n}{\partial x_m}\right)
	D_{r_n} \boldsymbol{U}_{\boldsymbol{i}},
\end{align*}
\end{linenomath}
where the entries in the Jacobian matrix, $\partial r_m/\partial x_n$ are 
obtained from the mapping $\xv=\boldsymbol{d}_k(\rv,t)$. 

The discretised governing equations are
\begin{linenomath}
\begin{align*}
	\frac{\mathrm{d}}{\mathrm{d}t}\boldsymbol{U}_{\boldsymbol{i}}
	+ ({(\boldsymbol{U}_{\boldsymbol{i}}-\boldsymbol{W}_{\boldsymbol{i}})}\cdot\nabla_2)\boldsymbol{U}_{\boldsymbol{i}}
	+ \frac{1}{\rho}\nabla_2 P_{\boldsymbol{i}} - 
    \nu\nabla_2^2\boldsymbol{U}_{\boldsymbol{i}}
    -\boldsymbol{f}_{\boldsymbol{i}}&= 0, \\
	\frac{1}{\rho}\nabla_2^2P_{\boldsymbol{i}} + 
	\mathcal{J}(\nabla_{2}\boldsymbol{U}_{\boldsymbol{i}}) -
	\nabla_2\cdot\boldsymbol{f}_{\boldsymbol{i}} &= 0,
\end{align*}
\end{linenomath}
where
$    \nabla_2 \boldsymbol{U}_{\boldsymbol{i}} = 
    (D_{x_1}\boldsymbol{U}_{\boldsymbol{i}},
    D_{x_2}\boldsymbol{U}_{\boldsymbol{i}},
    D_{x_3}\boldsymbol{U}_{\boldsymbol{i}})$,
$    \nabla_2^2 \boldsymbol{U}_{\boldsymbol{i}} = 
    (D_{x_1x_1}+D_{x_2x_2}+D_{x_3x_3})\boldsymbol{U}_{\boldsymbol{i}}$, and 
$	\nabla_2\cdot\boldsymbol{U}_{\boldsymbol{i}} = 
	D_{x_1}\boldsymbol{U}_{1,\boldsymbol{i}}+D_{x_2}\boldsymbol{U}_{2,\boldsymbol{i}}
    +D_{x_3}\boldsymbol{U}_{3,\boldsymbol{i}} $.

\section{Temporal discretisation}\label{sec:temporalDiscretization}
The method of lines is used for solving the equations in time. After 
discretising the governing equations in space 
they can be regarded as a system of ODEs,
\begin{linenomath}
\begin{equation*}
	\frac{\mathrm{d}}{\mathrm{d}t}\boldsymbol{U} = 
    \boldsymbol{F}(\boldsymbol{U},\boldsymbol{t}),
\end{equation*}
\end{linenomath}
where pressure is considered a function of the velocity, $P = P(\boldsymbol{U})$. 
The equations are integrated in time using either a fully explicit or semi-implicit scheme, 
depending on the stability restriction imposed by the viscous time-step characteristic to the problem. 
The explicit scheme uses a second-order accurate Adams-Bashforth predictor followed by 
a second-order accurate Adams-Moulton corrector.
For light rigid bodies, multiple correction steps are used to stabilise
the scheme, under-relaxing the computed forces on the bodies.
The semi-implicit scheme treats the viscous term of the momentum equation implicitly with a second-order 
Crank-Nicolson method, which is once again combined with Adams-Moulton corrector steps if under-relaxed 
sub-iterations are required.
To illustrate this we will use the momentum equations as 
an example. Splitting the equations into explicit and implicit components we have
\begin{linenomath}
\begin{align*}
	\frac{\mathrm{d}\boldsymbol{u}}{\mathrm{d}t} &= 
    -({(\boldsymbol{u}-\boldsymbol{w})}\cdot{\nabla})\boldsymbol{u}
    -\frac{1}{\rho}\nabla p 
    ~~ +\nu\nabla^2\boldsymbol{u} 
    ~~ \equiv
	\boldsymbol{F}_E
	+\boldsymbol{F}_I,
\end{align*}
\end{linenomath}
where $\boldsymbol{F}_E$ and $\boldsymbol{F}_I$ are the explicit and implicit components, 
respectively
\begin{linenomath}
\begin{align}
	\boldsymbol{F}_E = 
    -((\boldsymbol{u}-\boldsymbol{w})\cdot{\nabla})\boldsymbol{u} - \frac{1}{\rho}{\nabla}p, \quad
	\boldsymbol{F}_I = 
    \nu\nabla^2\boldsymbol{u}.
	\label{EQ: implicit-explicit momentum term}
\end{align}
\end{linenomath}
The equations are integrated using either fully explicit or semi-implicit schemes.
The explicit integration scheme used in the present work is the second-order in time Adams 
predictor--corrector method. It consists of an Adams--Bashforth predictor
\begin{linenomath}
\begin{equation*}
	\frac{\boldsymbol{u}^p - \boldsymbol{u}^n}{\Delta t} = 
	\beta_0 \boldsymbol{F}^n +
	\beta_1 \boldsymbol{F}^{n-1},
\end{equation*}
\end{linenomath}
with the constants $\beta_0=1 + \frac{\Delta t}{2\Delta t_1}$ and 
$\beta_1= -\frac{\Delta t}{2\Delta t_1}$ chosen for second-order accuracy even with a variable time-step
where $\Delta t_1 = t_n - t_{n-1}$,
and an Adams--Moulton corrector
\begin{linenomath}
\begin{equation*}
	\frac{\boldsymbol{u}^{n+1} - \boldsymbol{u}^n}{\Delta t} = 
	\frac{1}{2}\boldsymbol{F}^p + 
	\frac{1}{2}\boldsymbol{F}^n.
\end{equation*}
\end{linenomath}
Though only one corrector step was taken here, one may in practice correct multiple times. In fact,
this is necessary when dealing with moving, light rigid bodies and partitioned 
fluid--solid coupling, as will be discussed later.

A semi-implicit approach is taken in some diffusion dominated problems where the explicit 
diffusive time-step is overly restrictive. Generally, this is the case when the Reynolds number 
is very low or the grid is highly refined near solid boundaries. Here, the non-linear 
convective terms are treated with the explicit Adams predictor--corrector method while the 
viscous terms are treated with the implicit second-order in time Crank--Nicolson method. 
Using the notation introduced in~\eqref{EQ: implicit-explicit momentum term}
then the time-step 
consists of a predictor,
\begin{linenomath}
\begin{equation*}
	\frac{\boldsymbol{u}^p - \boldsymbol{u}^n}{\Delta t} =
	\beta_0\boldsymbol{F}_E^n - \beta_1\boldsymbol{F}_E^{n-1} +
	\alpha \boldsymbol{F}_I^p + (1-\alpha)\boldsymbol{F}_I^n,
\end{equation*}
\end{linenomath}
and a corrector
\begin{linenomath}
\begin{equation*}
	\frac{\boldsymbol{u}^c-\boldsymbol{u}^n}{\Delta t} =
	\frac{1}{2}\boldsymbol{F}_E^p + \frac{1}{2}\boldsymbol{F}_E^n +
	\alpha \boldsymbol{F}_I^c + (1-\alpha)\boldsymbol{F}_I^n,
\end{equation*}
\end{linenomath}
where the superscript $c$ denotes the corrected solution and $\alpha=\frac{1}{2}$ gives the 
second-order Crank--Nicolson method.

The basic Navier--Stokes solver uses a solution algorithm that decouples the pressure and 
velocity fields~\cite{Henshaw1994,splitStep2003} in a similar fashion to many fractional-step and projection schemes, \cf 
\cite{Almgren1998,Ferziger2002,Patankar1972} and many others.
The advantage of the current scheme over typical projection schemes is that the boundary conditions for the
pressure are well-defined and it is straightforward to obtain full second-order accuracy for all variables. 

 Assume that at time $t-\Delta t$ 
the values of $\boldsymbol{U}(t-\Delta t)$ and 
$P(t-\Delta t)$ are known at all points in the solution domain and the values of 
$\boldsymbol{F}(\boldsymbol{U}(t-\Delta t),t-\Delta t)$ are known at all interior points. To advance the 
solution in time to $t$ the fully explicit algorithm proceeds as follows:

\begin{steps}[leftmargin=5em]\small
    \item Determine an intermediate solution 
        $\boldsymbol{U}_{\boldsymbol{i}}^\ast(t)$ at all interior nodes using a 
        predictor sub-step
        \begin{linenomath}
        \begin{equation*}
            \boldsymbol{U}_{\boldsymbol{i}}^\ast(t) = 
            \boldsymbol{U}_{\boldsymbol{i}}(t-\Delta t)+\alpha\Delta t
            \boldsymbol{F}_{\boldsymbol{i}}(\boldsymbol{U}_{\boldsymbol{i}}(t-\Delta t),t-\Delta t),
            \qquad\forall \boldsymbol{i}\in\Omega
        \end{equation*}
    \end{linenomath}
    \item Determine $\boldsymbol{U}^\ast(t)$ at the boundary and ghost nodes 
        by solving the boundary conditions
        \begin{linenomath}
        \begin{equation*}
            \left.\begin{aligned}
            \boldsymbol{U}_{\boldsymbol{i}}^\ast(t)-
            \boldsymbol{u}_B(\boldsymbol{x}_{\boldsymbol{i}},t) &= 0\\
           \nabla_2\cdot\boldsymbol{U}_{\boldsymbol{i}}^\ast(t) &= 0\\
           \text{Extrapolate ghost values of $\tv_\mu\cdot \boldsymbol{U}^*_{\boldsymbol{i}}$} &
            \end{aligned}\right\rbrace
            \forall \boldsymbol{i}\in\partial\Omega
        \end{equation*}
    \end{linenomath}
        where $\mu=1,\dots,n_d-1$ and only the tangential component of the momentum equation is used. 
            \item Determine $P_{\boldsymbol{i}}(t)$ by solving the pressure Poisson equation along with the 
        remaining boundary conditions
        \begin{linenomath}
        \begin{align*}
            \nabla_2^2P_{\boldsymbol{i}}(t) =&
            -\mathcal{J}(\nabla_2\boldsymbol{U}_{\boldsymbol{i}}^\ast( t)) +
            \nabla_2\cdot\boldsymbol{f}_{\boldsymbol{i}}(t),\qquad\forall\boldsymbol{i}\in\Omega\\
            \boldsymbol{n}\cdot\nabla_2P_{\boldsymbol{i}}(t)=&
            -\boldsymbol{n}\cdot\bigg[
            \frac{\partial\boldsymbol{U}_{\boldsymbol{i}}^\ast(t)}{\partial t} +
            ({(\boldsymbol{U}_{\boldsymbol{i}}^\ast(t)-\boldsymbol{W}_{\boldsymbol{i}}(t))}\cdot\nabla_2)
            \boldsymbol{U}_{\boldsymbol{i}}^\ast(t) \\
          &\qquad\quad  + \nu\nabla_2 \times \nabla_2\times \boldsymbol{U}_{\boldsymbol{i}}^\ast(t)-\boldsymbol{f}(t)\bigg],
            \qquad\forall\boldsymbol{i}\in\partial\Omega.
        \end{align*}
    \end{linenomath}
       The normal component of the momentum equation is used here as a Neumann boundary condition for 
        the pressure Poisson equation. The $\nu\Delta\uv$ term has been replaced by $-\nu\grad\times\grad\uv$
        to avoid a viscous time-step restriction, see~\citep{Petersson2001} for more details.
        \label{ITEM: pressure correction}
    \item Given $\boldsymbol{U}^\ast(t)$ and $P(t)$ the pressure gradients can be computed 
        and $\boldsymbol{F}(\boldsymbol{U}^\ast(t),t)$ found at interior nodes.
    \item {Correction steps can now be taken to either increase the time-step, or as needed,
        to stabilise the algorithm for light rigid bodies. The correction steps consists of
        the Adams-Moulton corrector for the velocity followed by an additional pressure solve.
        For light bodies, when added mass effects are large, under-relaxed sub-iterations are used 
        during these corrector steps to stabilise the scheme. 
        Typically 3--7 corrector steps are used in the present work, depending on the significance of 
        added mass effects in the problem.}
\end{steps}
For moving grids, additional steps in the algorithm are required to evolve the rigid-body equations (as discussed
in the next section) and subsequently move the component grids. 
After the component grids have been moved the overset grid connectivity information
is regenerated.
Note that since the governing equations are solved in a reference frame moving with the
grid, no additional interpolation is needed to transfer the solution at discretisation points
from one time step to the next. As grids move, however, some unused points may become
active and values at these {\em exposed-points} are interpolated at previous times
as discussed in~\cite{mog2006}

For small problems (number of grid points $\mathcal{O}(10^4)$) the linear systems of equations for 
the velocity components and the separate system of equations for 
the pressure are effectively solved using direct solution methods. Larger problems 
necessitate iterative approaches; we use Krylov subspace methods from the PETSc library \citep{Balay2013},
algebraic multigrid solvers from the the Hypre package~\citep{Falgout2002} 
and the geometric multigrid solvers for overset grids from Overture~\citep{automg}.

\section{Fluid-solid coupling}
This system of ODEs governing the particle motion is discretised in time using a 
Leapfrog predictor and Adams--Moulton corrector scheme. The predictor consists of
\begin{linenomath}
\begin{alignat*}{3}
	\boldsymbol{v}_{b}^p &= \boldsymbol{v}_{b}^{n-1} + \frac{2\Delta t}{m_p}
	\boldsymbol{F}^n,\quad&&
	\boldsymbol{x}_{b}^p &= 2\boldsymbol{x}_{b}^n - \boldsymbol{x}^{n-1} +
	\frac{\Delta t}{m_{p}}\boldsymbol{F}^n,\\
	\boldsymbol{\omega}^p &=
	\boldsymbol{\omega}^{n-1} +
	2\Delta t(-\omegav^n\times A\boldsymbol{\omega}^n + \boldsymbol{T}^n), \quad&&
	\boldsymbol{e}_i^p &=
	\boldsymbol{e}_i^{n-1} +
	2\Delta t(\boldsymbol{\omega}^n) \times \boldsymbol{e}_i^n,
\end{alignat*}
\end{linenomath}
and is performed before Step 1 in the time-stepping algorithm of Section~\ref{sec:temporalDiscretization}.
The corrector is
\begin{linenomath}
\begin{alignat*}{3}
	\boldsymbol{v}_{b}^{n+1} &= 
	\boldsymbol{v}_{b}^n +
	\frac{\Delta t}{2m_p}(
	\boldsymbol{F}^n+\boldsymbol{F}^p), \quad&
	\boldsymbol{x}_{b}^{n+1} &=
	\boldsymbol{x}_{b}^n +
	\frac{\Delta t}{2m_p}
	(\boldsymbol{v}_{b}^n + \boldsymbol{v}_{b}^p),\\
	\boldsymbol{\omega}^{n+1} &=
	\boldsymbol{\omega}^{n} + \frac{\Delta t}{2}
	(-\omegav^n\times A\boldsymbol{\omega}^n + \boldsymbol{T}^n -
	\omegav^p\times A\boldsymbol{\omega}^p + \boldsymbol{T}^p),\quad&
 	\boldsymbol{e}_i^{n+1} &=
	\boldsymbol{e}_i^n + \frac{\Delta t}{2}
	(\boldsymbol{\omega}^n \times \boldsymbol{e}_i^n +
	 \boldsymbol{\omega}^p \times \boldsymbol{e}_i^p).
\end{alignat*}
\end{linenomath}
and is performed after Step 3 (pressure solve) in the time-stepping algorithm.
A predictor-corrector scheme is used to facilitate the 
fluid-solid coupling, and to allow for sub-time-step iterations for light bodies as discussed next.

Low solid/fluid density ratios can cause the standard time-stepping routine to become 
unstable, owing principally to the added-mass instability~\citep{Banks2014}. 
To alleviate 
this, under-relaxed sub-iterations are performed during the correction stages (\ie fluid velocity solve and pressure solve)
of the time-stepping algorithm. These sub-iterations are thus relatively expensive although the implicit systems
are not changed during these iterations. 

The approach used here is similar to that used by many previous authors, 
although we prefer to 
under-relax the force on the rigid-body as opposed to under-relaxing the entire state of the rigid-body.
Note that more sophisticated approaches exist to reduce the required number of sub-iterations 
such as those based on Aitken acceleration~\citep{KuttlerWall2008,Borazjani2008}.

We illustrate the relaxed sub-iteration through consideration of the rigid-body velocity equation, 
though this is performed for  the angular velocity equation as well. The force-relaxation sub-iteration replaces
the update for $\boldsymbol{v}_{b}^{n+1}$ in the corrector step above
by the iteration
\begin{linenomath}
\begin{equation*}
	\boldsymbol{v}_{b}^{n+1,k} = 
	\boldsymbol{v}_{b}^{n} + 
	\frac{\Delta t}{2}(
	\boldsymbol{F}^{n} + 
	\boldsymbol{F}^{n+1,k}),\qquad k=1,2,\dots
\end{equation*}
\end{linenomath}
where $k$ denotes the iteration count. The iterative forcing used to evolve the equation is 
\begin{linenomath}
\begin{equation*}
	\boldsymbol{F}^{n+1,k} = 
	(1 - \alpha)\boldsymbol{F}^{n+1,k-1}+
\alpha\widetilde{\boldsymbol{F}}^{k},\qquad \alpha\in(0,1]
\end{equation*}
\end{linenomath}
where $\alpha$ is a relaxation parameter and 
$\widetilde{\boldsymbol{F}}^{k}$ is the forcing at step $k$, which initially is 
simply the predicted force from the previous fluid solve step, \ie 
$\widetilde{\boldsymbol{F}}^1=\boldsymbol{F}^p$.
{During each sub-iteration, the fluid velocity and fluid pressure are recomputed and these updated fluid
values are used to compute the next approximation to the force on the rigid body.}
 A small $\alpha$ can ensure 
stability---at the cost of increased iterations. An optimal value of $\alpha$ is problem dependent and some 
experimentation is required to reach a good compromise between stability and computational 
cost. {For example, a value of $0.1$ was used in the pure wake interaction test case of \S\ref{SEC: pure wake interaction} where the maximum number of sub-iterations was 39 during the first few time-steps, likely 
due to the non-smooth forcing at start up, and the minimum and average number of sub iterations were 5 and 7, respectively.}
Iterations are performed until the absolute or relative change in the force fall below 
their respective convergence criteria,
$\Delta {F}^k<\tau_a$, or $\Delta F^k/(|\boldsymbol{F}^k|+\epsilon_F)<\tau_r$,
where $\Delta F^k = |\boldsymbol{F}^{n+1,k}-\boldsymbol{F}^k|.$ 

\subsection{Collision model}

A hard-sphere collision model based on the linear conservation of momentum is used to handle 
cases in which particles touch\footnote{ 
Note that in principle the particles should never actually touch, but resolving the
near contact would require a very fine grid.}.
During the predictor step of the particle advancement 
scheme the new positions are used to determine whether or not particles breach 
the minimum separation distance, as stipulated by the requirements of the interpolation 
stencils. If this minimum separation distance is breached, a collision is deemed to have 
occurred and the particle velocities are corrected. The velocity corrections are 
calculated by 
\newcommand{\er}{e_r} 
\begin{linenomath}
\begin{align*}
	\hat{\boldsymbol{v}}_{b,A}^{n+1} &=
	\boldsymbol{v}_{b,A}^{n+1} +
	\left[
		v_A^n-
		v_A^{n+1} -
		\frac{(1+\er)m_{b,B}}{m_{b,A}+m_{b,B}}
		(
		v_{A}^{n}+
		v_{B}^{n}
		)
	\right]\boldsymbol{n}_A,\\
	\hat{\boldsymbol{v}}_{b,B}^{n+1} &=
	\boldsymbol{v}_{b,B}^{n+1} +
	\left[
		v_B^n-
		v_B^{n+1} -
		\frac{(1+\er)m_{b,A}}{m_{b,A}+m_{b,B}}
		(
		v_{A}^{n}+
		v_{B}^{n}
		)
	\right]\boldsymbol{n}_B,
\end{align*}
\end{linenomath}
where $v_A=\boldsymbol{v}_{b,A}\cdot\boldsymbol{n}_A$, $v_B = \boldsymbol{v}_{b,B}
\cdot\boldsymbol{n}_B$, $\er$ is the coefficient of restitution and $\boldsymbol{n}_A=-
\boldsymbol{n}_B$ is the unit normal vector pointing from the centre of mass of particle $A$ 
to the centre of mass of particle $B$. In this work, collisions were modelled as perfectly 
elastic with a coefficient of restitution of $\er=1$. This is a frictionless model, so tangential forces 
are assumed to be zero during the collision, and angular velocities are not corrected 
by the model. This hard-sphere model is also restricted to the contact of only two particles 
at any given moment in time.

\section{Numerical results}

\subsection{Convergence study \label{SEC: convergence study}}

To accurately simulate viscous flows the grid resolution must be fine enough to fully capture 
boundary layers adjacent to solid surfaces. These can be very thin, depending on the Reynolds number of 
the problem as the boundary layer depth scales approximately
 as $1 /\sqrt{\operatorname{Re}}$, see~\citep{Batchelor1967}. 
A major advantage of boundary-conformal over static grid methods is the ability to selectively 
refine the grid near solid boundaries.
In a detailed grid independence study 
of viscous flow past a static cylinder, Nicolle~\cite{Nicolle2010} 
investigated how refinement of different areas of the grid affected the behaviour of the 
cylinder. Predictably, the surface resolution was found to most affect the cylinder behaviour, 
but the downstream wake resolution was found to affect the Strouhal number. Nonetheless, large 
ratios between surface and wake resolution were found to give very accurate results.

\begin{figure}
    \centering
    \includegraphics[width=\figWidth]{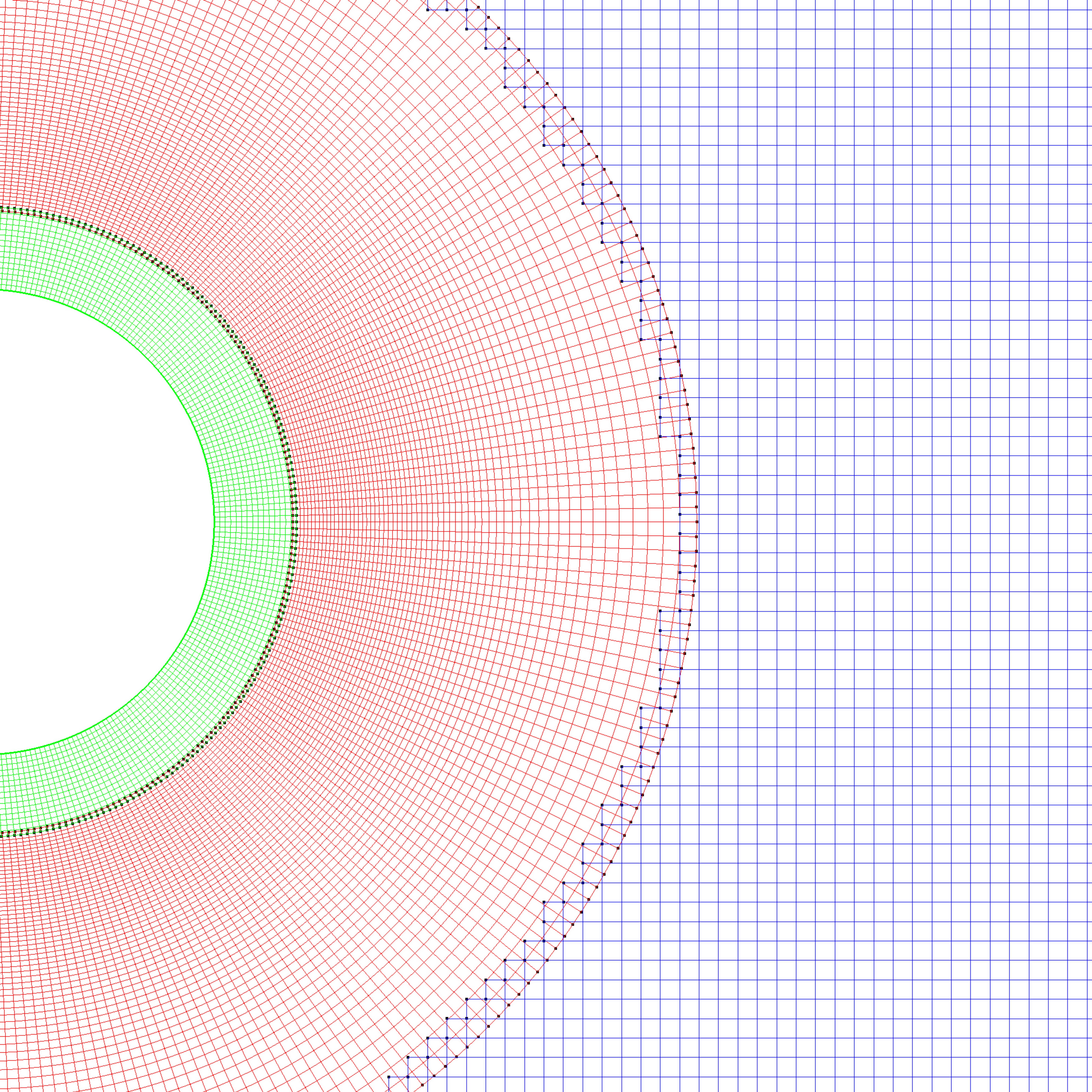}
    \def\svgwidth{5cm}
     {\small
    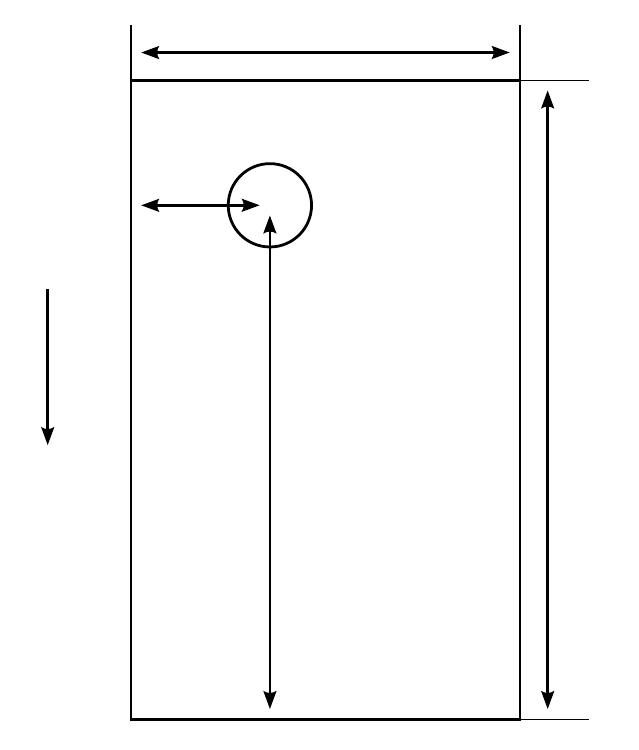
   }
    \caption{Left: Cropped view of \textit{GR4}, showing the boundary layer grid (green), 
        transition grid (red) 
    and background grid (blue) with interpolation points. Right: Geometry of the convergence study 
problem. \label{FIG: refinement grid}}
\end{figure}

In the present work, emphasis is placed on the grid characteristic length scales to optimise 
run time. Following Nicolle and Eames~\citep{Nicolle2011} we use two grid length scales to quantify the quality of the 
computational grid: the domain length scale ($\textit{DLS}$) is the {background} grid element 
size  while the surface length scale ($\textit{SLS}$) is the grid element size 
on the surface of the particle.

The grid independence study is first performed using a grid with nearly uniform 
grid spacings and is then 
repeated using grid refinement near the particle boundary. Descriptions of the grids used 
are provided in table \ref{TAB: grid description}. Because curvilinear grids are used 
to represent the particles, the cells are slightly distorted in physical space. Thus, table 
\ref{TAB: grid description} provides minimum, average and maximum cell volumes (areas). 

\begin{table}\tableFont
    \centering
    \begin{tabular}{@{}l r r r r r r@{}}
        \toprule
        Grid    &  Node count  &   \multicolumn{3}{c}{Cell volume}     &   \textit{DLS} &   \textit{SLS}\\
        \midrule
                &               & min   & ave & max       &     & \\
 \textit{GU1} & 1497931 & 0.66 & 0.678 & 0.678 & $D/96$ & $D/96$ \\
 \textit{GU2} & 670984  & 1.47 & 1.53  & 1.53  & $D/64$ & $D/64$\\
 \textit{GU3} & 380394  & 2.58 & 2.71  & 2.71  & $D/48$ & $D/48$\\
 \textit{GU4} & 171724  & 5.65 & 6.10  & 6.10  & $D/32$& $D/32$\\
 \textit{GU5} & 98102   & 9.90 & 10.8  & 10.9  & $D/24$& $D/24$\\
 \textit{GU6} & 44967   & 21.3 & 24.4  & 24.4  & $D/16$& $D/16$\\
 \textit{GR1} & 679685  & 0.67 & 1.52  & 1.53  & $D/64$& $D/96$\\
 \textit{GR2} & 390285  & 0.78 & 2.68  & 2.71  & $D/48$& $D/96$\\
 \textit{GR3} & 191265  & 0.78 & 5.68  & 6.10  & $D/32$& $D/96$\\
 \textit{GR4} & 123095  & 0.71 & 9.09  & 10.9  & $D/24$& $D/96$\\
 \textit{GR5} & 63935   & 0.78 & 18.2  & 24.4  & $D/16$& $D/96$\\
 \textit{GR6} & 31455   & 0.78 & 39.5  & 97.7  & $D/8$& $D/96$\\
        \bottomrule
    \end{tabular}
    \caption{Description of the uniform and refined grids used in the convergence 
    study. \label{TAB: grid description}}
\end{table}

The test used in both convergence studies is as follows: the domain is a rectangular channel of 
dimensions $(W^\ast,H^\ast)=(4\,D,40\,D)$ filled with an a priori quiescent fluid of density
$\rho_f=1\,\mathrm{g/cm^3}$ and kinematic viscosity of $0.05\,\mathrm{cm^2/s}$. 
The particle $(D,\rho_p)=(0.25, 1.5)$ is released from rest at $(x_0^\ast,y_0^\ast)=(D,38.4\,D)$ with 
the gravitational constant set to $\boldsymbol{g}=981\,\mathrm{cm/s^2}$ in the negative $y$-direction. The problem geometry can be seen in 
Fig.~\ref{FIG: refinement grid}. The results are non-dimensionalised as follows:
\begin{equation}
    u^\ast = \frac{u}{U_T},\quad
    v^\ast = \frac{v}{U_T},\quad
    x^\ast = \frac{x}{D},\quad
    y^\ast = \frac{y}{D},\quad
    \omega^\ast = \frac{\omega D}{U_T},\quad
    t^\ast = \frac{t U_T}{D}
    \label{EQ: non-dimensionalization}
\end{equation}
where {$U_T=8.6041\,\mathrm{cm}\,\mathrm{s}^{-1}$} is the measured terminal settling velocity.

\begin{figure}[hbt]
    \centering
    \begin{minipage}[t]{0.49\linewidth}
        \includegraphics[width=\figWidth]{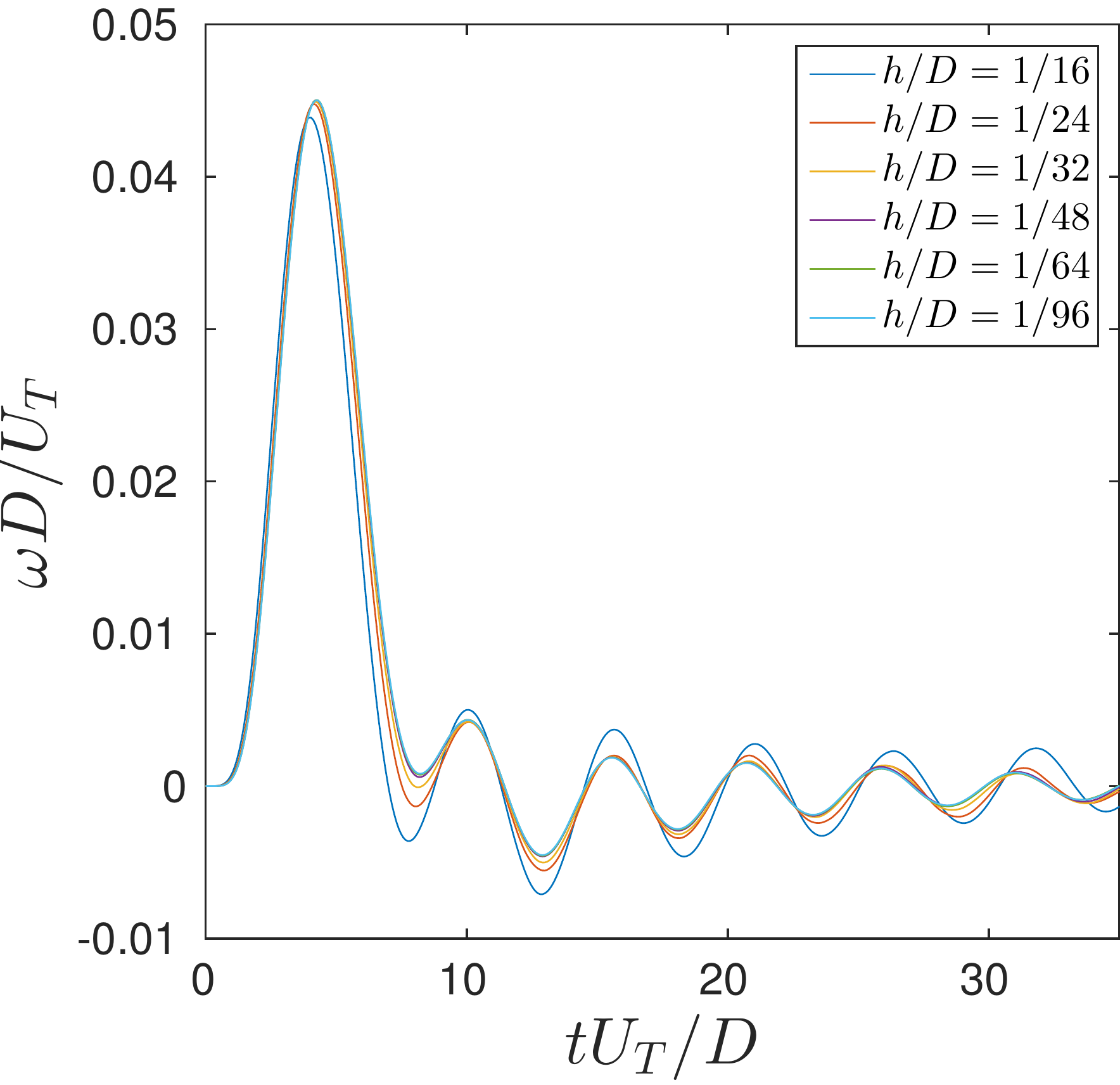}
    \end{minipage}%
    \hfill
    \begin{minipage}[t]{0.49\linewidth}
        \includegraphics[width=\figWidth]{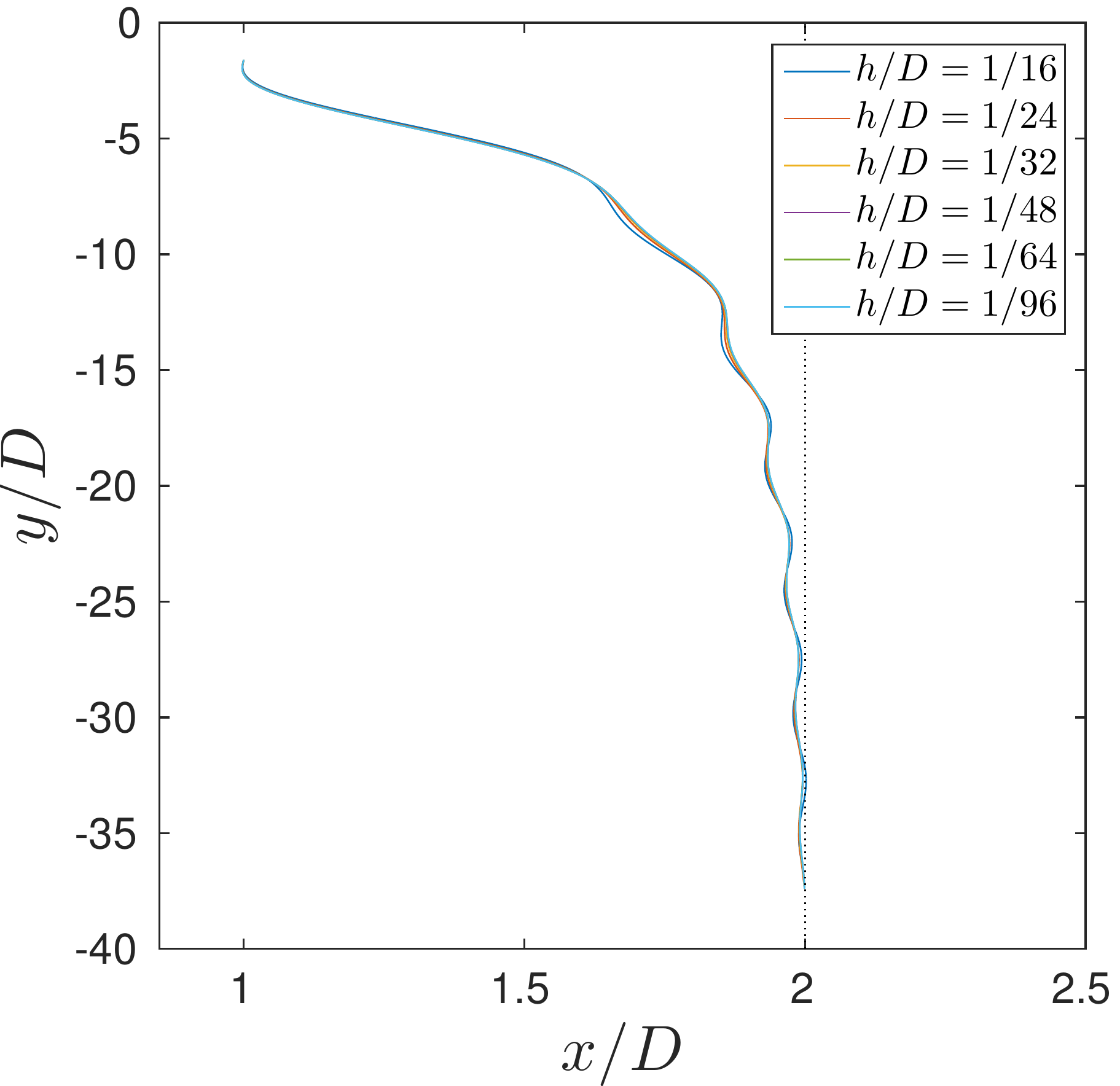}
    \end{minipage}%
    \vspace{-10pt}
    \caption{Left: Normalised angular velocity history for the convergence study at increasing grid resolutions. Right: Position of the disk in the channel for the convergence study at increasing grid resolutions. \label{FIG: convergence omega and position plots}}
\end{figure}
\begin{figure}[hbt]
    \centering
    \begin{minipage}[t]{0.49\linewidth}
        \includegraphics[width=\figWidth]{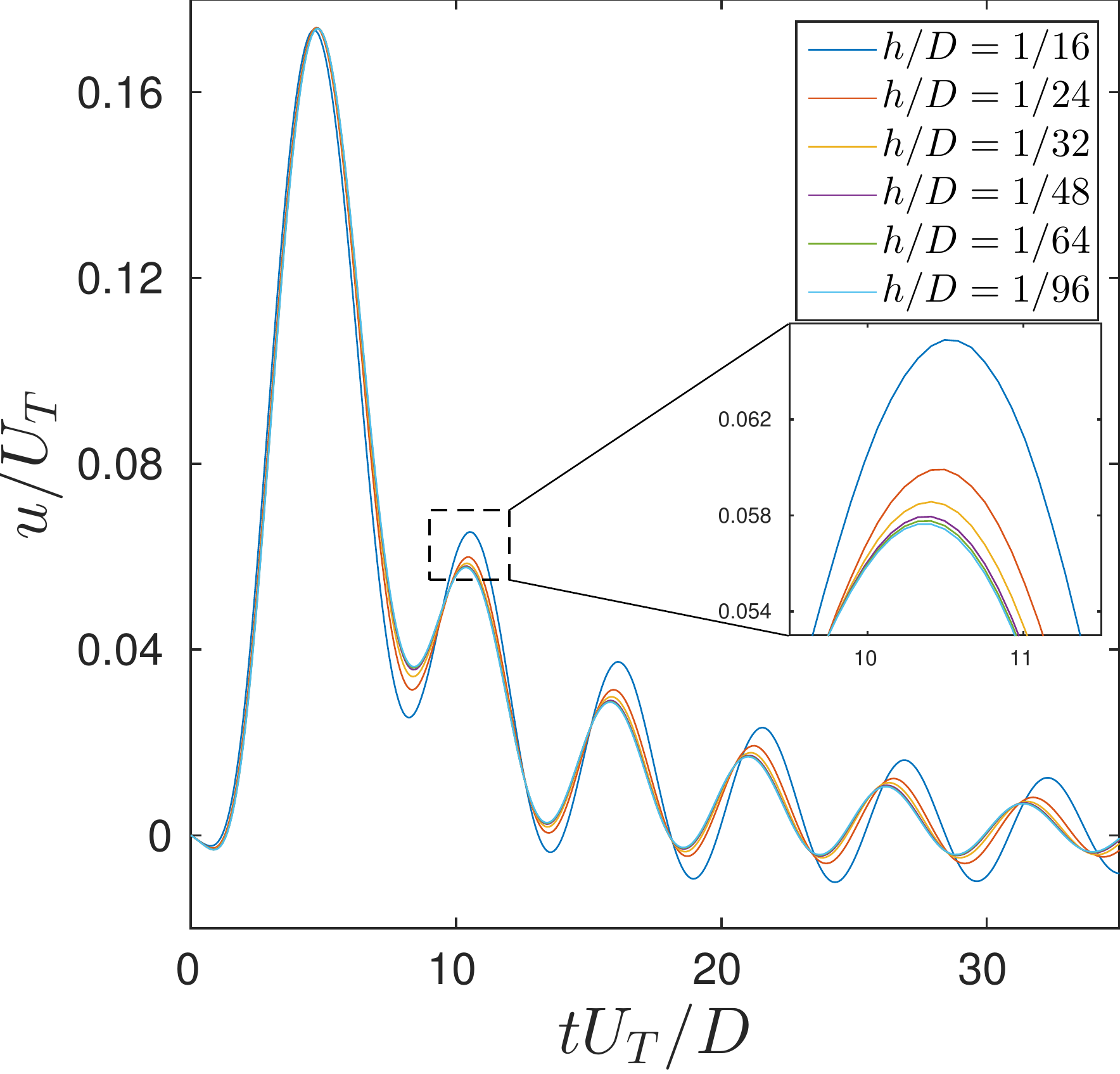}
    \end{minipage}%
    \hfill
    \begin{minipage}[t]{0.49\linewidth}
        \includegraphics[width=\figWidth]{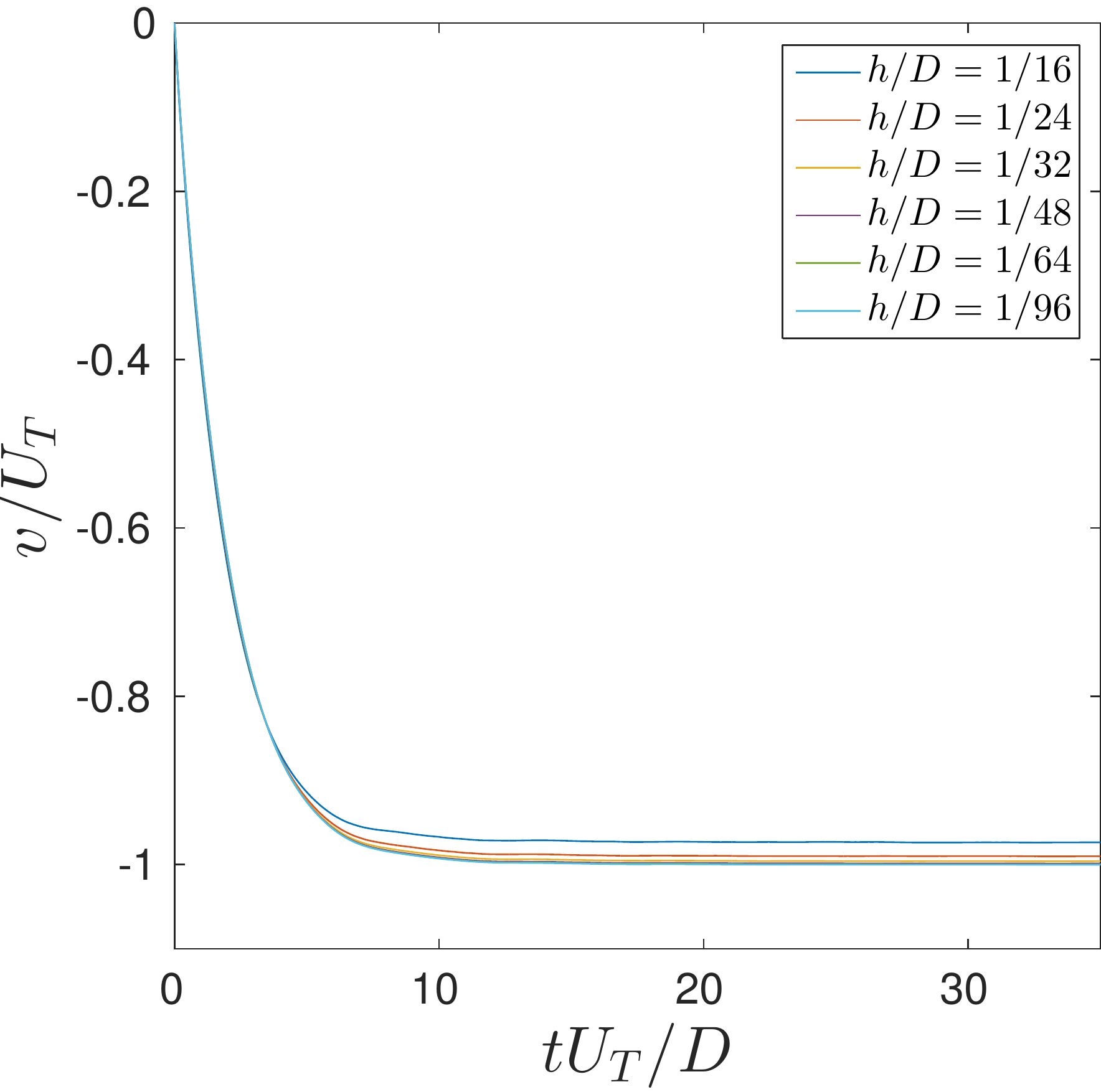}
    \end{minipage}%
    \vspace{-10pt}
    \caption{Left: Normalised horizontal velocity history for the convergence study at increasing grid resolutions. Right: Normalised vertical velocity history for the convergence study at increasing grid resolutions.\label{FIG: convergence u and v}}
\end{figure}

Under the action of gravity the particle rotates in a clockwise sense, as if rolling down the 
wall. Immediately after release it moves a short distance towards the near wall before 
migrating towards its equilibrium position along the channel centreline. 

The wake remains attached but is unsteady. This is reflected in the oscillatory $u^\ast$ and 
$\omega^\ast$ velocity time histories, shown in figures \ref{FIG: convergence u and v} and 
\ref{FIG: convergence omega and position plots}. The angular velocity time history 
exhibits a large initial peak after which it is rapidly damped to low amplitude oscillations 
about $\omega^\ast=0$. After a very small negative peak, the $u^\ast$ time history 
exhibits a large positive peak, much like $\omega^\ast$, with damped successive peaks. However, 
unlike in $\omega^\ast$, the following $u^\ast$ oscillations have a non-zero mean value as the 
disk drifts towards its equilibrium position. In contrast, the vertical velocity, $v^\ast$, shows 
the particle rapidly reaching a steady settling velocity, unaffected by the attached unsteady 
wake.

Fig.~\ref{FIG: walltime error plots} show relative {\em errors} (meaning differences compared to the
reference solution) in $v^\ast$ and $\omega^\ast$, which were 
calculated as $\epsilon_v=(v^\ast - v^\ast_\text{ref})/v^\ast_\text{ref}$ and 
$\epsilon_\omega = (\omega - \omega^\ast_\text{ref})/\omega^\ast_\text{ref}$, 
where data from the \textit{GU1} simulation are used as reference values.
Relative errors in $v^\ast$ and $\omega^\ast$ taken early on in the simulation, at $t=0.2\,\mathrm{s}$, 
and late in the simulation, at $t=1.0\,\mathrm{s}$, show {greater than} second order 
convergence for both components early on but a decrease in convergence rate for $\omega^\ast$ 
as the simulation progresses. The gravitational acceleration is impulsively turned on at $t=0$, and this 
non-smooth forcing could have a detrimental effect on the convergence rates.

\begin{figure}[hbt]
    \centering
    \begin{minipage}[t]{0.49\linewidth}
        \includegraphics[width=\figWidth]{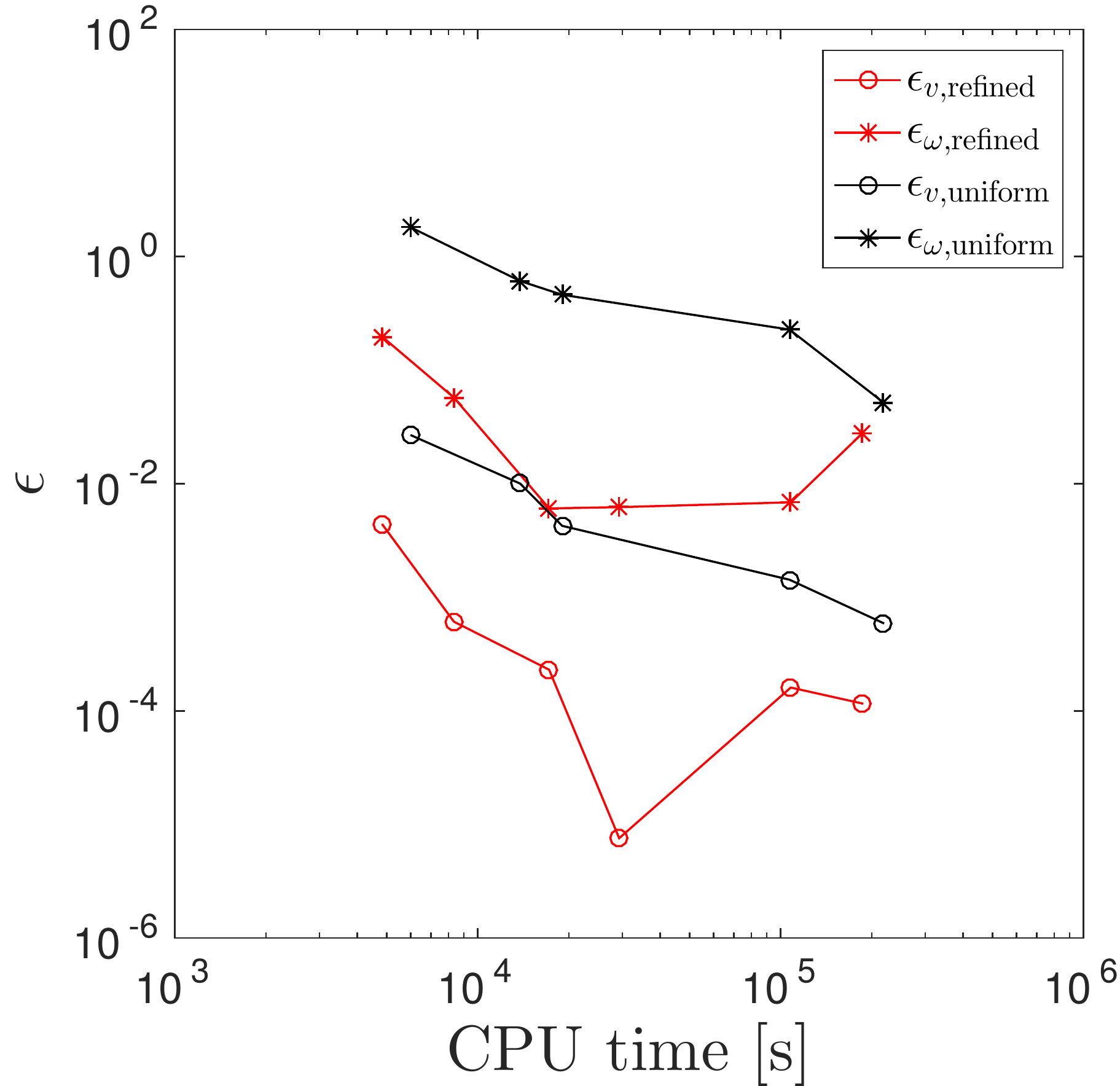}
    \end{minipage}%
    \hfill
    \begin{minipage}[t]{0.49\linewidth}
        \includegraphics[width=\figWidth]{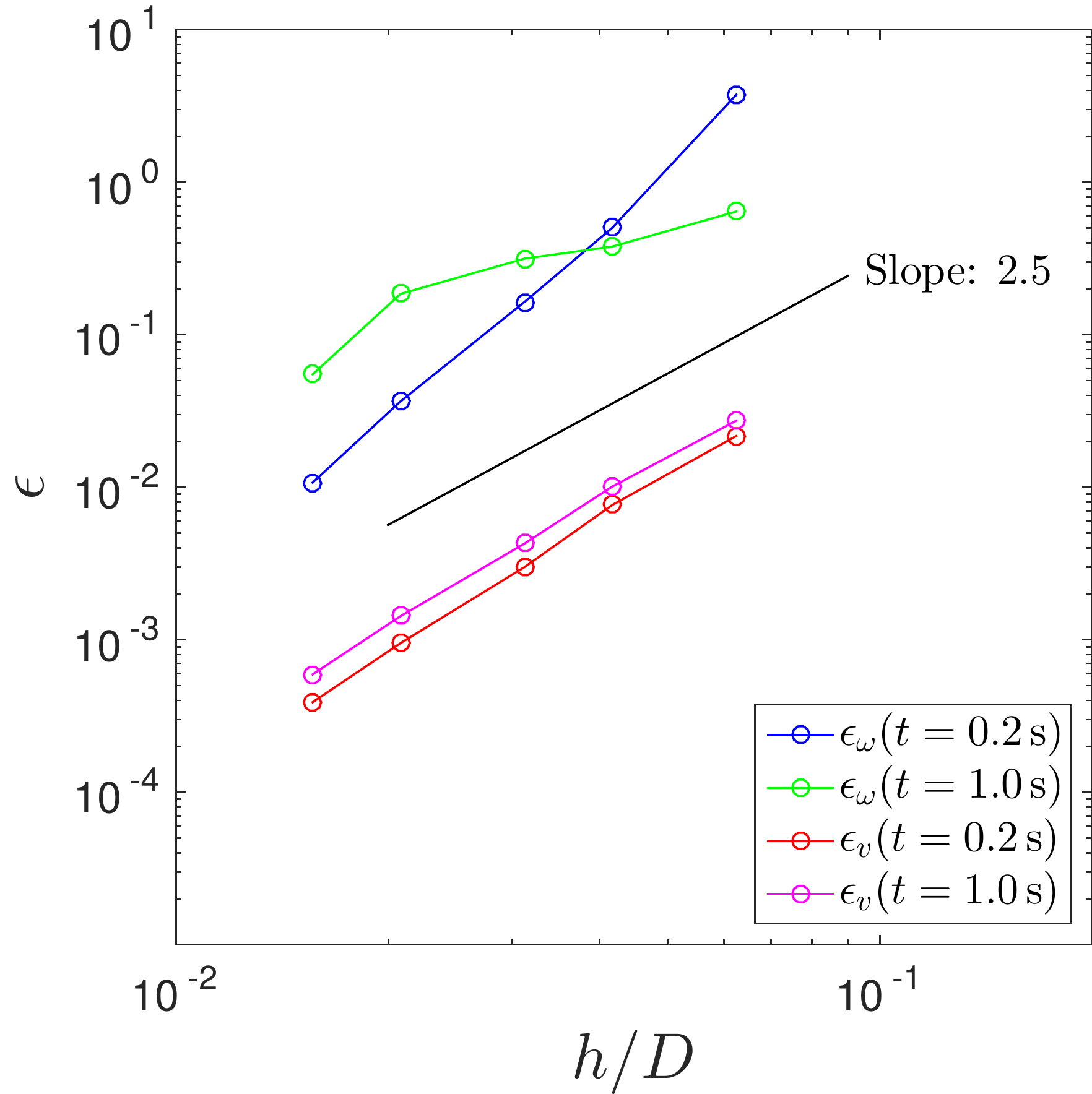}
    \end{minipage}%
    \vspace{-10pt}
    \caption{Left: Comparison of relative error magnitude in $v^\ast$ and $\omega^\ast$ at $t=1\,\mathrm{s}$ against required CPU time using uniform and refinement grids. Right: relative vertical and angular velocity errors at early ($t=0.2\,\mathrm{s}$) and late ($t=1.0\,\mathrm{s}$) stages of the simulation. \label{FIG: walltime error plots}}
\end{figure}

The test was repeated using a series of refined grids. These were constructed using a fine grid 
immediately surrounding the particle surface, smoothly connected to a coarse background 
grid using a transition grid. This construction can be seen in Fig.~\ref{FIG: refinement grid}. 
Detailed descriptions of the grids are provided in table \ref{TAB: grid description}, 
where uniform grids are denoted by the prefix $\textit{GU}$ and refinement grids by 
$\textit{GR}$.

\begin{table}[hbt]\tableFont
    \centering
    \begin{tabular}{@{} l r r r r r r @{}}
        \toprule
        Grid    &   $v/U_T$    &    $u/U_T$     &   $\omega D/U_T\e{-4}$    &   $\epsilon_v\e{-4}$  &   $\epsilon_u$  &   $\epsilon_\omega$ \\
        \midrule
        \textit{GU1}    &   1.0000  &   0.0033  &   6.4761   & ---   &   --- &   --- \\
        \textit{GU2}    &   0.9994  &   0.0034  &   6.1663  &   5.9299  &   0.0288  &   0.0521 \\
        \textit{GU3}    &   0.9994  &   0.0034  &   6.8561  &   14.000  &   0.0969  &   0.2275  \\
        \textit{GU4}    &   0.9957  &   0.0041  &   9.2127  &   43.000  &   0.2775  &   0.4609  \\
        \textit{GU5}    &   0.9900  &   0.0046  &   9.4015  &   100.00  &   0.4513  &   0.6093  \\
        \textit{GU6}    &   0.9734  &   0.0052  &   0.0016  &   266.00  &   0.6329  &   1.7913  \\
        \textit{GR1}    &   0.9995  &   0.0034  &   6.5961  &   1.1599  &   0.0292  &   0.0277  \\
        \textit{GR2}    &   0.9998  &   0.0033  &   6.4250  &   1.6028  &   0.0141  &   0.0069  \\
        \textit{GR3}    &   1.0000  &   0.0031  &   6.4250  &   0.0755  &   0.0186  &   0.0062 \\
        \textit{GR4}    &   1.0002  &   0.0031  &   5.8773  &   2.3025  &   0.0161  &   0.0061  \\
        \textit{GR5}    &   1.0006  &   0.0032  &   6.1766  &   6.0949  &   0.0013  &   0.0573  \\
        \textit{GR6}    &   1.0043  &   0.0026  &   6.9848  &   43.000  &   0.1915  &   0.1956  \\
        \bottomrule
    \end{tabular}
    \caption{Absolute values and relative errors for the convergence study taken at $t=1.0\,\text{s}$.\label{TAB: conv study table 1}}
\end{table}

Table \ref{TAB: conv study table 1} shows absolute values of $u^\ast$, $v^\ast$ and $\omega^\ast$ at 
$t=1.0\,\mathrm{s}$ as well as relative errors, where data from $\textit{GU1}$ were 
taken as reference values. As before, it is evident that $v^\ast$ is fairly insensitive 
to the grid resolution, while $u^\ast$ and $\omega^\ast$ show a large dependence on the 
near surface resolution. With a high resolution surface grid capturing the 
boundary layer, the motion of the disk can be captured quite accurately, even with 
a large surface to background grid resolution ratio. In fact, the $\textit{GR5}$ grid 
with a resolution ratio of $6:1$ reproduced solutions of $\textit{GU1}$ with a 
maximum error of $5\,\%$, with a more than 23 fold reduction in number of grid points. 
Fig.~\ref{FIG: walltime error plots} shows the relative error in $v^\ast$ and $\omega^\ast$ at 
$t=1\,\mathrm{s}$ against the total CPU time of the calculation.

\subsection{Settling disk impacting a wall}

This test simulates the fall of a rigid circular disk in a bounded domain and its impact with the 
bottom boundary. This test has been performed by other researchers using DLM/FDM \citep{Glowinski2001}, an FEM 
fictitious boundary method \citep{Wan2006}, and an immersed boundary lattice Boltzmann method 
\citep{Hu2015}. The computational domain has a width of $W=8\,D$, a 
height of $H=24\,D$ and grid characteristic length scales $\textit{DLS}=D/16$ and $\textit{SLS}=D/96$, 
where $D=0.25\,\mathrm{cm}$ is the disk diameter. The disk is initially placed 
along the centreline of the domain, $8\,D$ from the top boundary.
The disk has density $\rho_d=1.25\,\mathrm{g/cm^3}$ and the kinematic viscosity of 
the fluid is $\nu=0.1\,\mathrm{cm^2/s}$. The results are non-dimensionalised {as in \eqref{EQ: non-dimensionalization},
where the characteristic velocity scale $U_s$ is an estimated terminal velocity,}
\begin{equation}
    U_s = \sqrt{\frac{\pi D}{2}\left(\frac{\rho_d-\rho_f}{\rho_f}\right)g}. \label{EQ: terminal vel estimate}
\end{equation}
The present results (Fig.~\ref{FIG: symmetric cyl rho 1.25}) are in good agreement with the 
previous studies. The disk reaches the terminal settling velocity at $t^\ast=20$, with 
a terminal particle Reynolds number of $\operatorname{Re_T=17.45}$, consistent with the 
literature. 
As the disk approaches the bottom wall the results differ 
slightly. The studies compared against in Fig.~\ref{FIG: symmetric cyl rho 1.25} all 
exhibit a rebound of the disk from the bottom boundary while the present results do not. 

\begin{figure}[hbt]
	\begin{minipage}[t]{0.49\linewidth}
	\includegraphics[width=\figWidth]{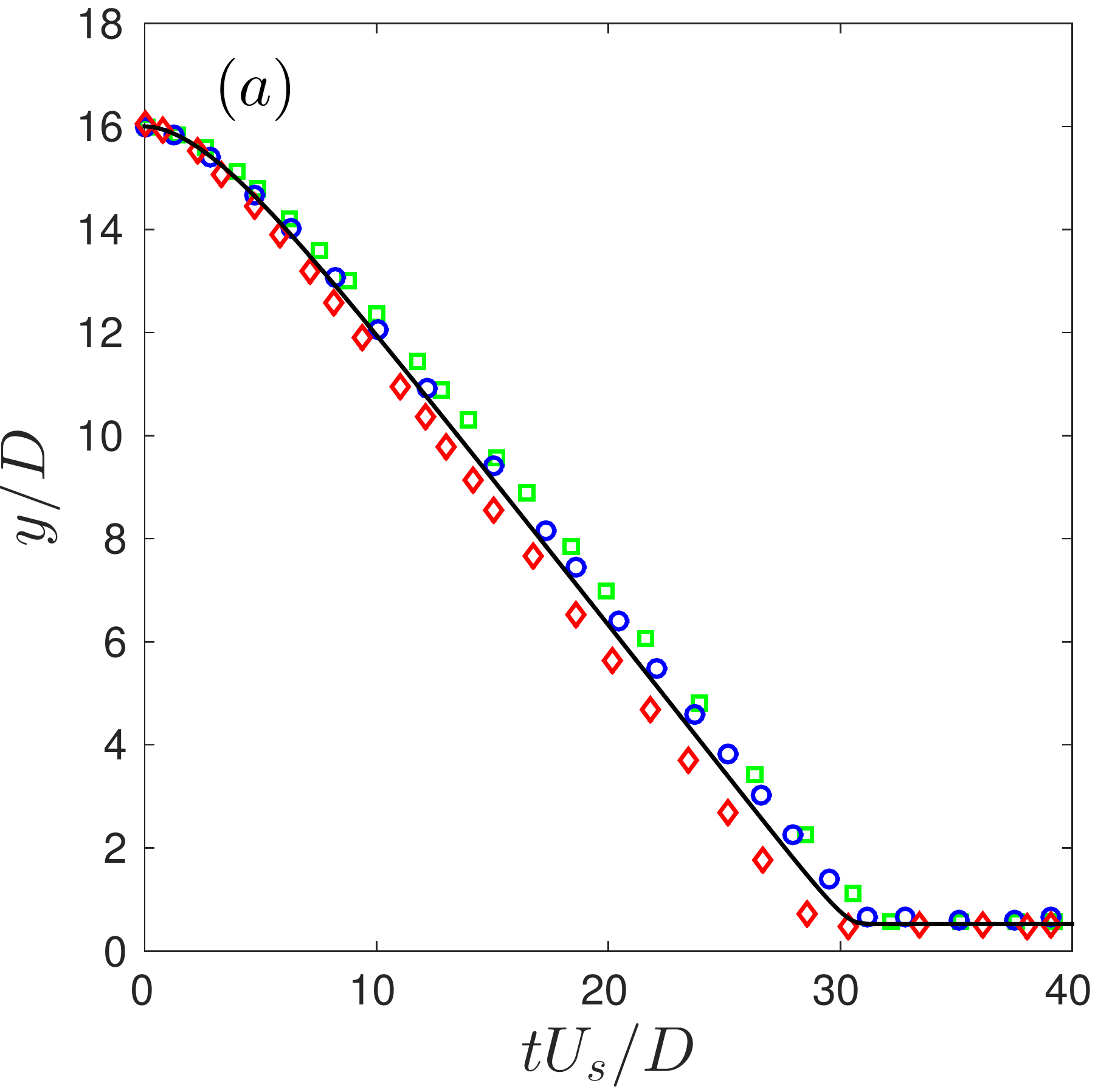}
	\end{minipage}%
	\hfill
	\begin{minipage}[t]{0.49\linewidth}
	\includegraphics[width=\figWidth]{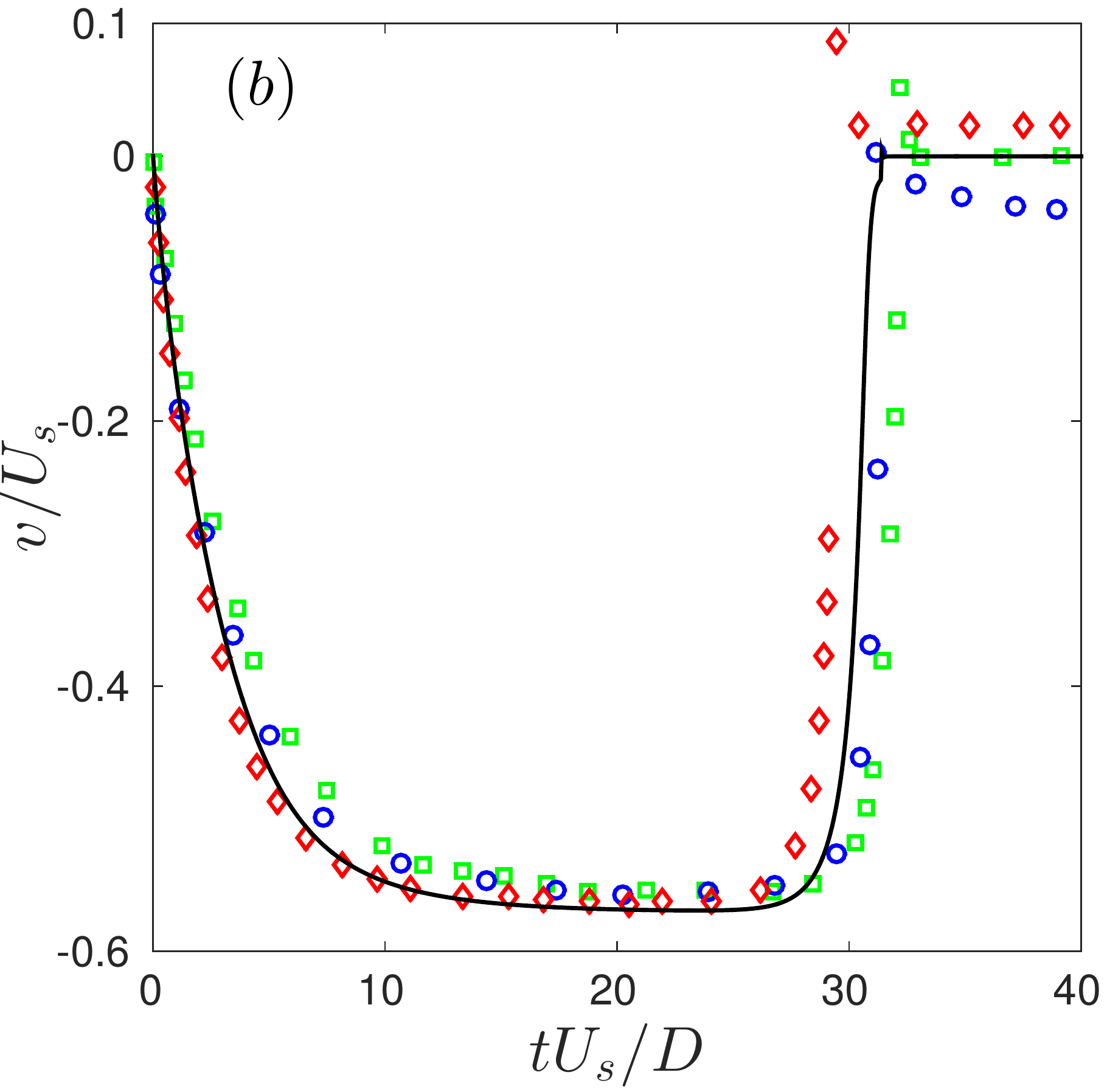}
	\end{minipage}
        \vspace{-10pt}
     	\caption{Histories of the $y^\ast$-coordinate and $v^\ast$ component of the centre of the disk for a low Reynolds number sedimentation of a symmetrically placed disk test case with data 
from Hu~et al.\citep{Hu2015} ($\color{green}\mathsmaller{\square}$), 
Wang et al.\citep{Wan2006} ($\color{blue}{\circ}$), 
Glowinski et al.\citep{Glowinski2001} ($\color{red}\mathsmaller{\lozenge}$) and the present study ($-$). \label{FIG: symmetric cyl rho 1.25}}
\end{figure}

In the present study the grid around the disk and along the bottom of the tank is 
very fine ($\textit{SLS}=D/96$ for both the disk and the bottom of the tank), 
allowing the lubrication forces and flow in the gap to be better resolved.
This slows the particle down more before ``contact'' is made with the 
wall\footnote{{In fact an infinitely smooth disk with flow governed by the Navier-Stokes equations should never actually  
contact the wall but in this settling case only approach the wall exponentially slowly with the gap becoming ever thinner.}}.
Additionally, the present study used a conservation of linear momentum based hard-sphere 
collision model approach to model the collision between the disk and the bottom boundary. 
The previous studies compared against here all used repulsive potential type methods. We can 
estimate a Stokes number for the particle to comment on the ``correctness'' of the 
present results. The Stokes number, $\operatorname{Stk}$, is the ratio between the 
particle and fluid relaxation times, $\tau_p$ and $\tau_f$, respectively. Taking 
$\tau_f=R/U_T$, where $R$ is the disk radius, 
$\tau_p=mv/F_d$, and $F_d$ is the drag force on the disk, then 
$\operatorname{Stk}=mv^2/(RF_d)$.
Once the disk reaches its terminal settling velocity the drag force balances with the force 
due to gravity, so $F_d=\pi R^2g(\rho_d-\rho_f)$ and thus the Stokes number is
$ \operatorname{Stk}={v^2}/(Rg(1-\frac{\rho_f}{\rho_d}))$.
Here, the Stokes number is approximately 2. It has been demonstrated that for 3D cases particles 
settling with $\operatorname{Stk}<10$ there is no rebound after contact is made with the 
bottom boundary \citep{Ardekani2008,Joseph2001}. Assuming this holds true for the 2D equivalent, then 
the above results indicate that the repulsive potential collision model is a poor 
sub-grid model for low speed impacts. The rebound evident in the study of Glowinski~\etal\citep{Glowinski2001} 
indicates that a higher grid resolution is required to adequately resolve the lubrication forces 
than the hydrodynamic interactions during free-fall. While the current approach is adequate here 
it is clear that in many situations resolving the gap is not practical and  prohibitively expensive.
Qiu \etal\cite{Qiu2015}  presented a novel solution to computing incompressible flow in thin gaps using pressure degrees of 
freedom on virtual solid surfaces to provide solid--fluid coupling in the gap region, which when 
extended to no-slip boundaries could be a good alternative for this sort of problem.

\subsection{Settling of two offset disks}

Two offset cylinders settling in a quiescent fluid are simulated to 
demonstrate the drafting, kissing, tumbling behaviour observed experimentally by 
Fortes~\etal\citep{Fortes1987}. This is a difficult problem to simulate 
owing to the non-linear nature of the particle motion and the particle--particle and 
particle--wall interactions. Results are compared to previous studies of 
Patankar~\etal\citep{Patankar2000,Patankar2001a}, Wan and Turek~\citep{Wan2006}, Niu~\etal\citep{Niu2006}, 
Zhang and Prosperetti~\citep{Zhang2003} and Feng and Michaelides~\citep{Feng2004} for a low Reynolds number case and 
Uhlmann~\citep{Uhlmann2005} for a moderate Reynolds number case. 
The low Reynolds number case uses a computational domain of width $10\,D$ and height $40\,D$, with the 
particles of diameter $0.2\,\mathrm{cm}$ placed along the vertical centreline, $4\,D$ and $6\,D$ from the top boundary. The 
high Reynolds number case uses a computational domain of width $8\,D$ and height $24\,D$, with the 
particles of diameter $0.25\,\mathrm{cm}$ placed $4\,D$ and $6\,D$ from the top boundary, and offset by $D/250$ and $-D/250$ from the 
vertical centreline.
For the low Reynolds number case, 
the grid characteristic length scales are $\textit{DLS}=D/19$ and $\textit{SLS}=D/76$ whilst for 
the moderate Reynolds number case $\textit{DLS}=D/24$ and $\textit{SLS}=D/128$. 
In each case, both the top and bottom particles have the same density ratio. For 
the low Reynolds number case, the density ratio is $\rho_r=1.01$ and the fluid has 
kinematic viscosity $\nu=0.1\,\mathrm{cm^2/s}$. For the moderate Reynolds number case, the density ratio is 
$\rho_r=1.5$ and the kinematic viscosity is $\nu=0.01\,\mathrm{cm^2/s}$. In both cases the gravitational 
constant was taken as $\boldsymbol{g}=981\,\mathrm{cm}/\mathrm{s}^2$ and the results are 
non-dimensionalised {as in \eqref{EQ: non-dimensionalization}, where the characteristic 
velocity, $U_s$, is again calculated using \eqref{EQ: terminal vel estimate}.}

{
\newcommand{\labelsize}{\footnotesize}
\newcommand{\drawContour}[7]{%
\begin{scope}[#1]
\draw(0.0,0) node[anchor=south west,xshift=-4pt,yshift=+0pt] {\trimfig{figures/#2}{\figWidtha}};
\draw(.55,6.8) node[draw,fill=white,anchor=west,xshift=2pt,yshift=0pt] {\scriptsize #5};
\begin{scope}[xshift=-4pt,yshift=-3pt]
  \draw (\xcb,\ycb) node[anchor=south west,xshift=0.cm,yshift=.5cm,rotate=-90] {\trimfigcb{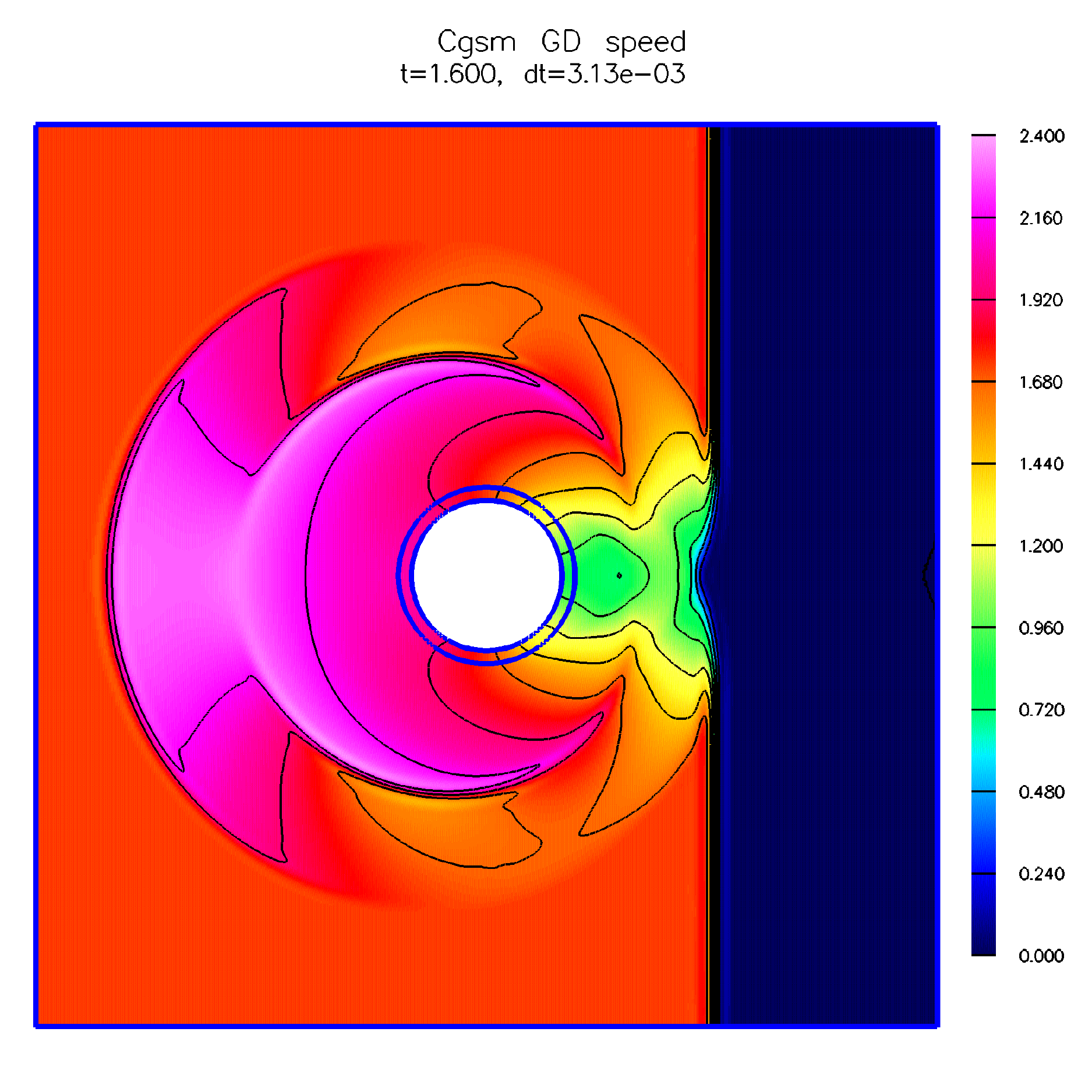}{\cbWidth}{\cbHeight}};
\end{scope}
\end{scope}
}
\newcommand{\cbWidth}{.2cm}
\newcommand{\cbHeight}{2.25cm}
\newcommand{\xcb}{.5cm}
\newcommand{\ycb}{-.2cm}
\setlength{\ycbTop}{\ycb+\cbHeight}
\setlength{\ycbMid}{\ycb+\cbHeight*\real{.5}}
\newcommand{\trimfigcb}[3]{\includegraphics[width=#2, height=#3, clip, trim=17cm 2.35cm 1.65cm 2.35cm]{#1}}
%
\newcommand{\figWidtha}{7.05cm}
\newcommand{\figWidthd}{5cm}
\newcommand{\trimfig}[2]{\trimh{#1}{#2}{.0}{.0}{.0}{.0}}
\newcommand{\depth}{24}
\def\rad{.5}
\newcommand{\plotDisk}{
\fill[fill=red!20,draw=red,line width=1.5pt] 
      plot[samples=100, domain=0.:360] ( {\rad*cos(\x)} , {\rad*sin(\x)} ) -- cycle ;
}
\begin{figure}[htb]
\begin{center}
\begin{tikzpicture}[scale=1]
  \useasboundingbox (-.5,.45) rectangle (16.,7.);  
%
%
\begin{scope}[xshift=0cm,yshift=0cm]
 \drawContour{xshift= 0.00cm,yshift=0.0cm}{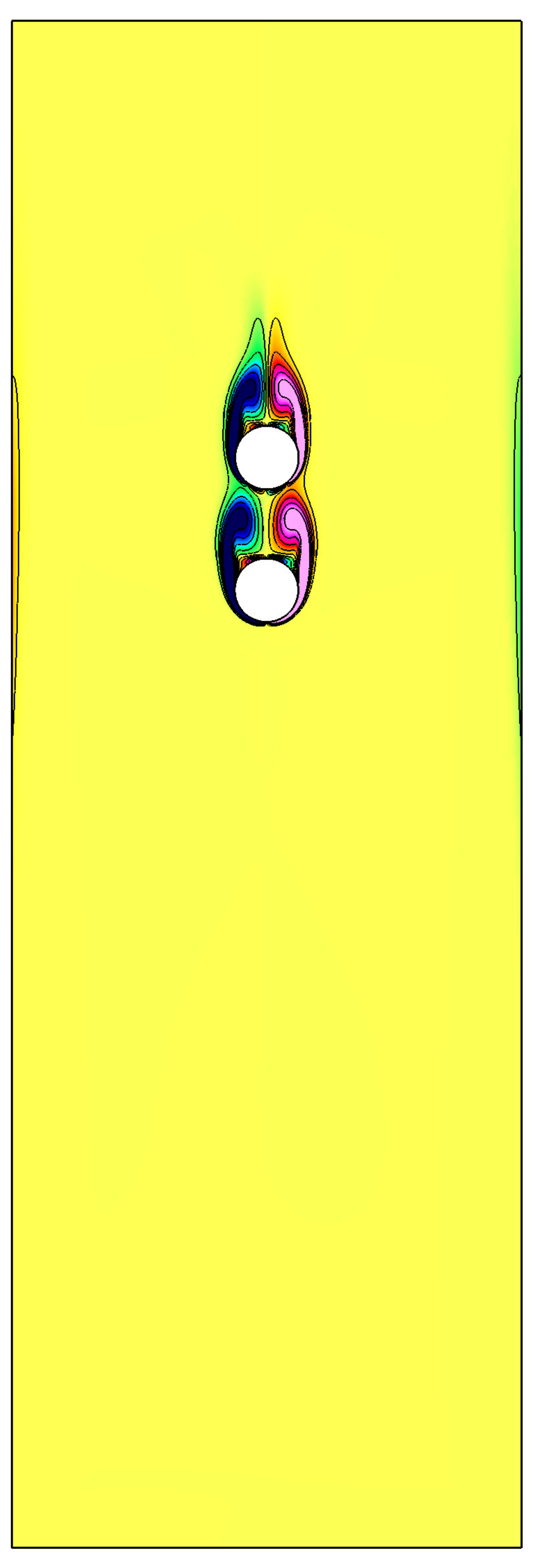}{$\xi$}{$\xi$}{$t=1$}{0.0}{0.23};
 \drawContour{xshift= 2.75cm,yshift=0.0cm}{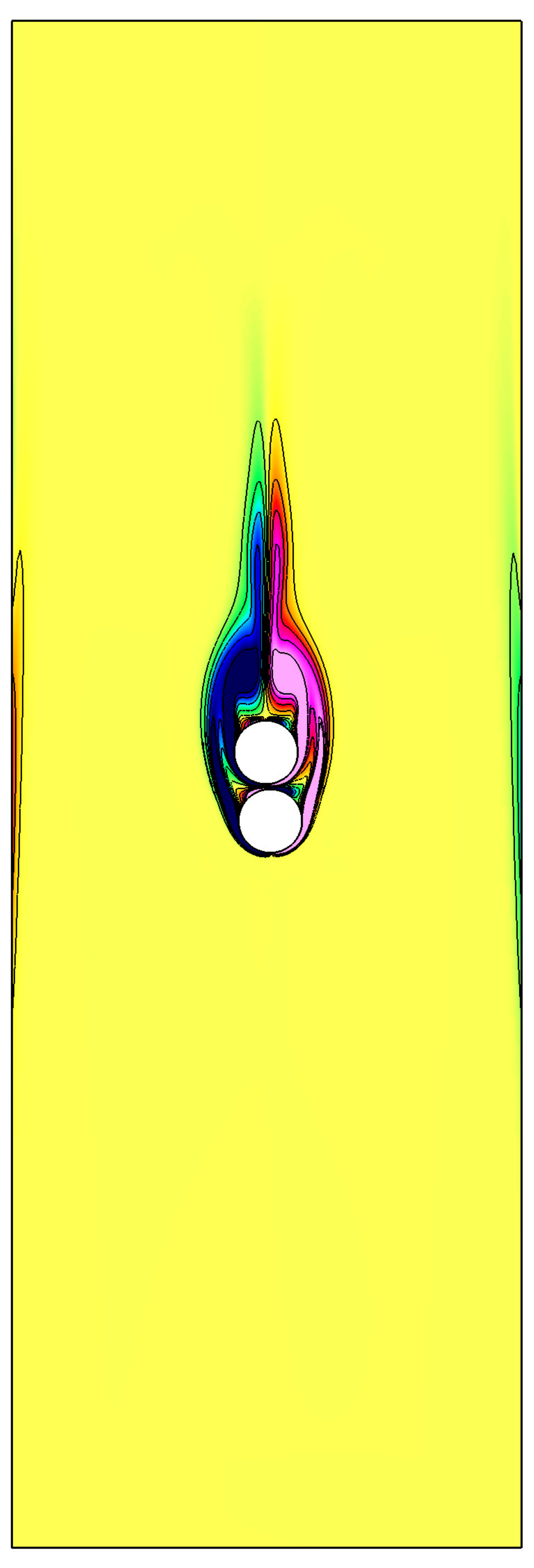}{$\xi$}{$\xi$}{$t=2$}{0.0}{0.54};
 \drawContour{xshift= 5.50cm,yshift=0.0cm}{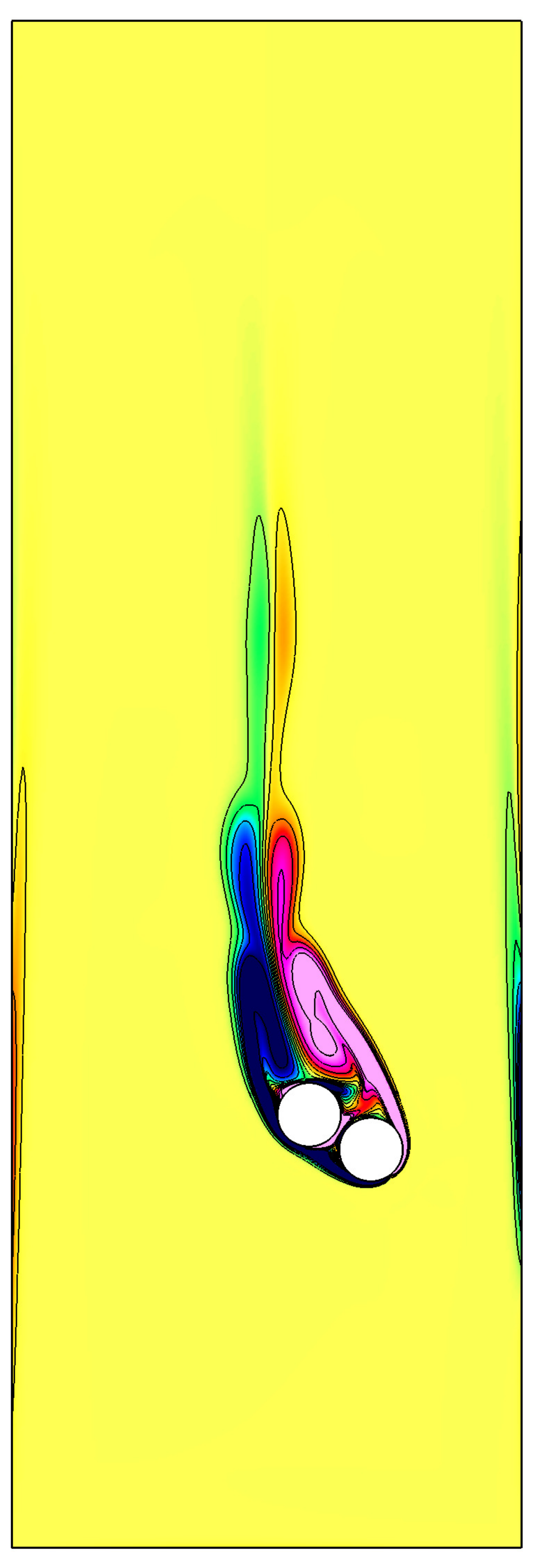}{$\xi$}{$\xi$}{$t=3$}{0.0}{0.98};
 \drawContour{xshift= 8.25cm,yshift=0.0cm}{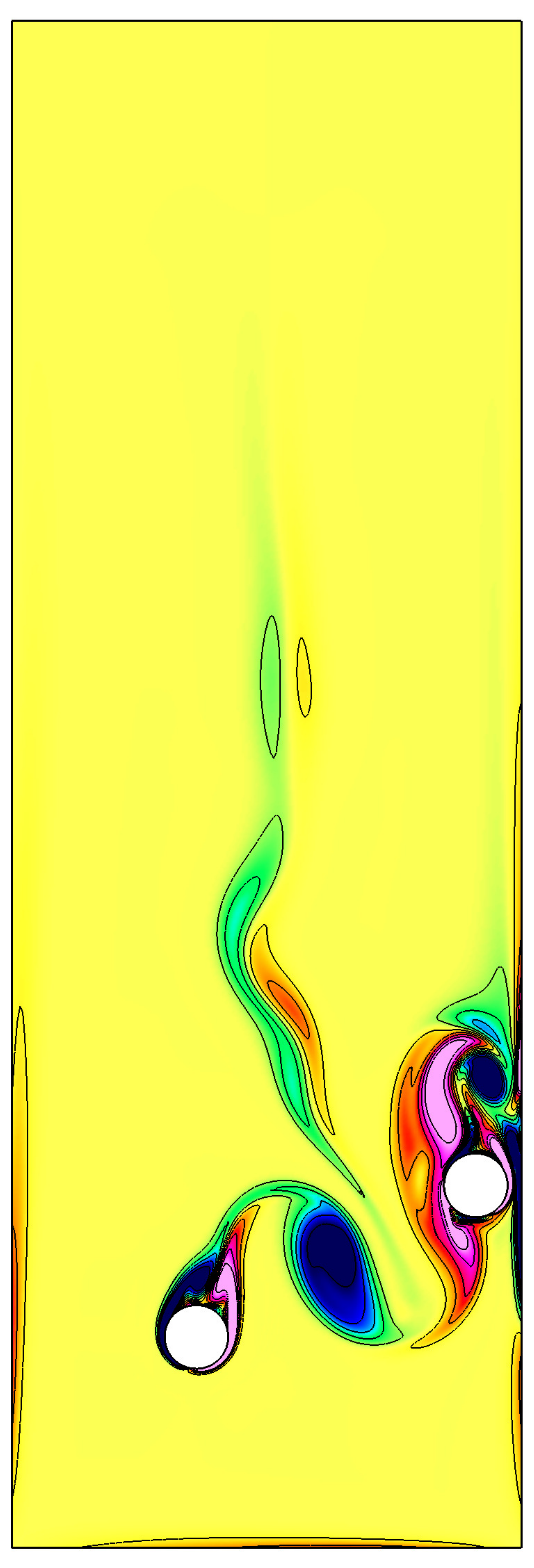}{$\xi$}{$\xi$}{$t=4$}{0.0}{1.04};
 \drawContour{xshift=11.00cm,yshift=0.0cm}{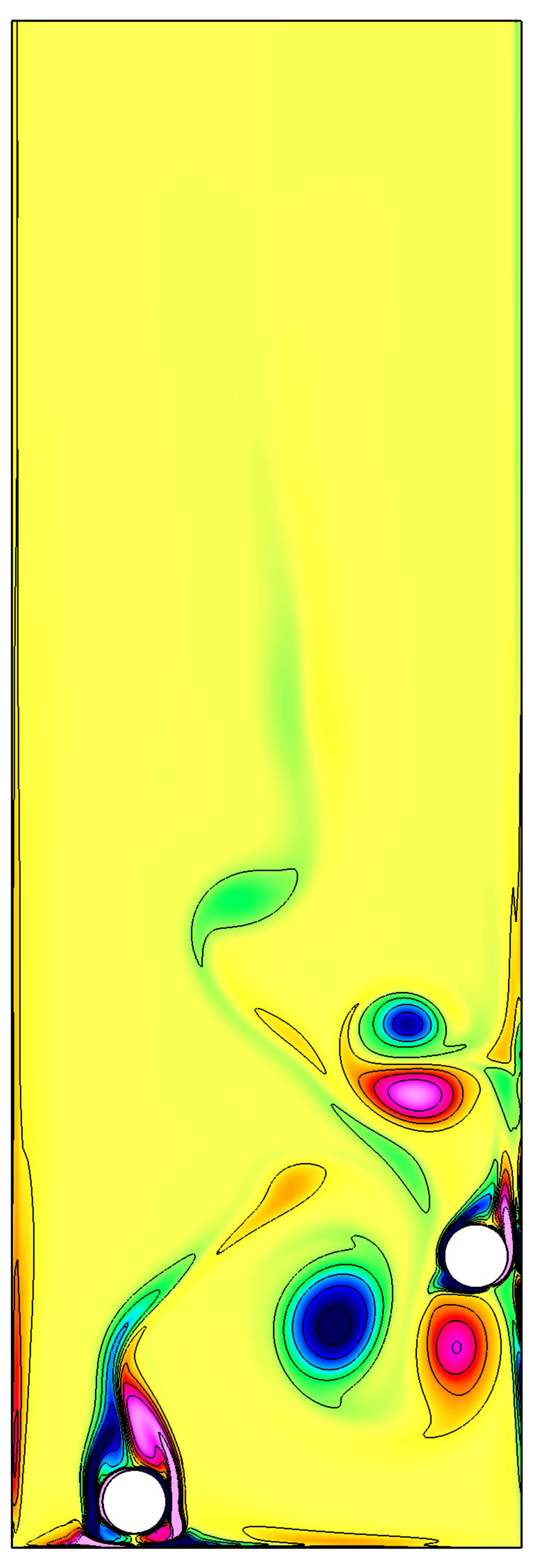}{$\xi$}{$\xi$}{$t=5$}{0.0}{1.04};
\end{scope}
\end{tikzpicture}
\end{center}
    \caption{Settling of two disks in a quiescent fluid. Contours of the vorticity at five different time, with a
    vorticity scale between $-3.6\leq\xi D/U_s\leq3.6$.\label{FIG: Uhlmann vort contour plots}}
\end{figure}
}

\begin{figure}
	\begin{minipage}[t]{0.49\linewidth}
		\includegraphics[width=\figWidth]{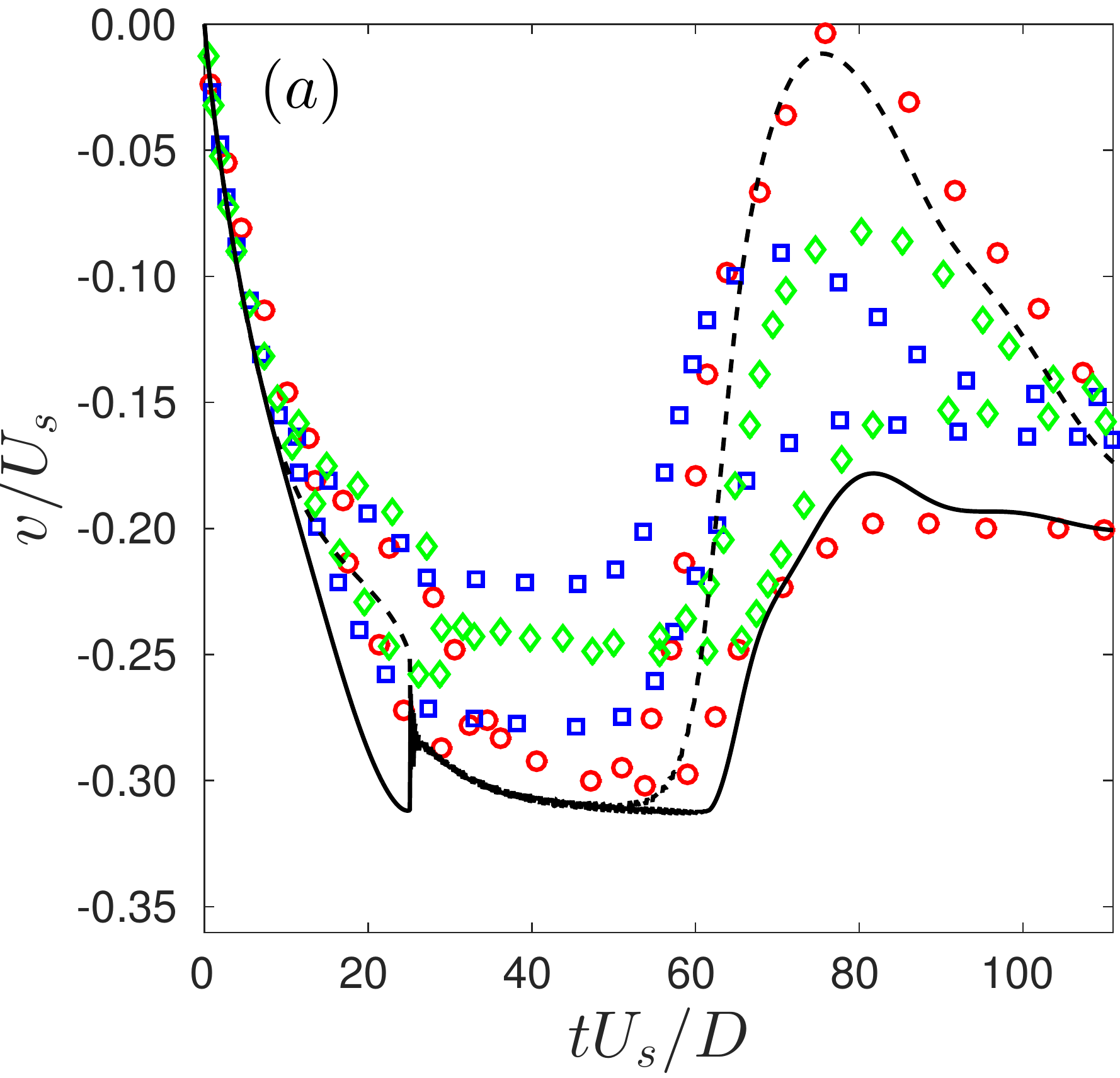}
	\end{minipage}%
	\hfill
	\begin{minipage}[t]{0.49\linewidth}
		\includegraphics[width=\figWidth]{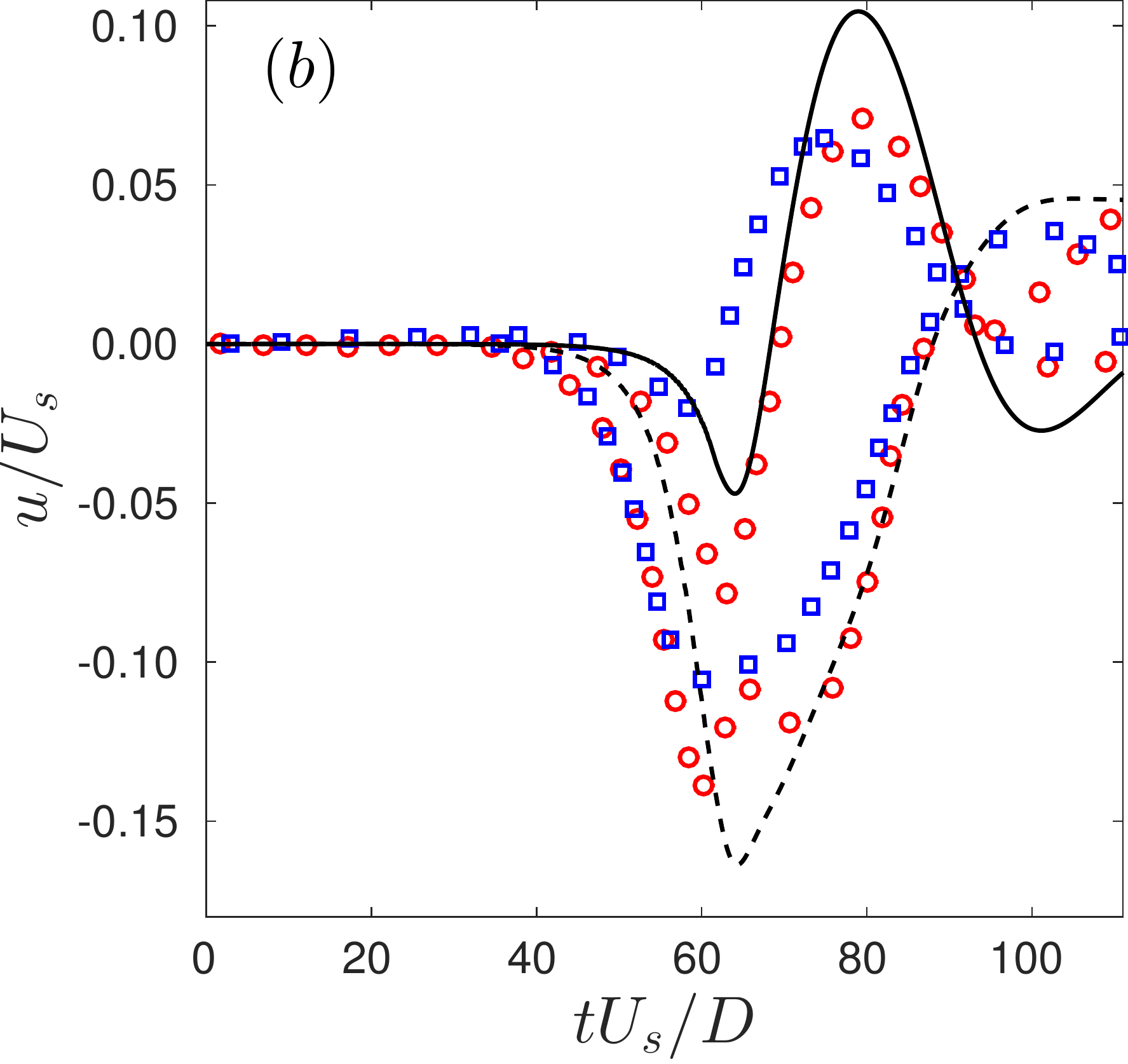}
	\end{minipage}%
        \vspace{-10pt}
	\caption{Histories of the $(a)$ $v^\ast$ and $(b)$ $u^\ast$ velocity components of the centre of the disks for the low Reynolds number drafting, kissing, and tumbling test case, with $\rho_d=1.01$, $\nu=0.1$  where the solid line denotes the (initially) top disk and the dashed line the bottom disk, with data from: 
$(a)$ Patankar~\citep{Patankar2001a} ($\color{green}\mathsmaller{\lozenge}$), 
Patankar~et al.~\citep{Patankar2000} ($\color{blue}\mathsmaller{\square}$) and 
Feng~et al.~\citep{Feng2004} ($\color{red}\circ$); 
$(b)$ Patankar~et al.~\citep{Patankar2000} ($\color{blue}\mathsmaller{\square}$) and 
 Feng~et al.~\citep{Feng2004} ($\color{red}\circ$) overlayed. 
\label{FIG: other vel}}
\end{figure}

\begin{figure}
	\begin{minipage}[t]{0.49\linewidth}
		\includegraphics[width=\figWidth]{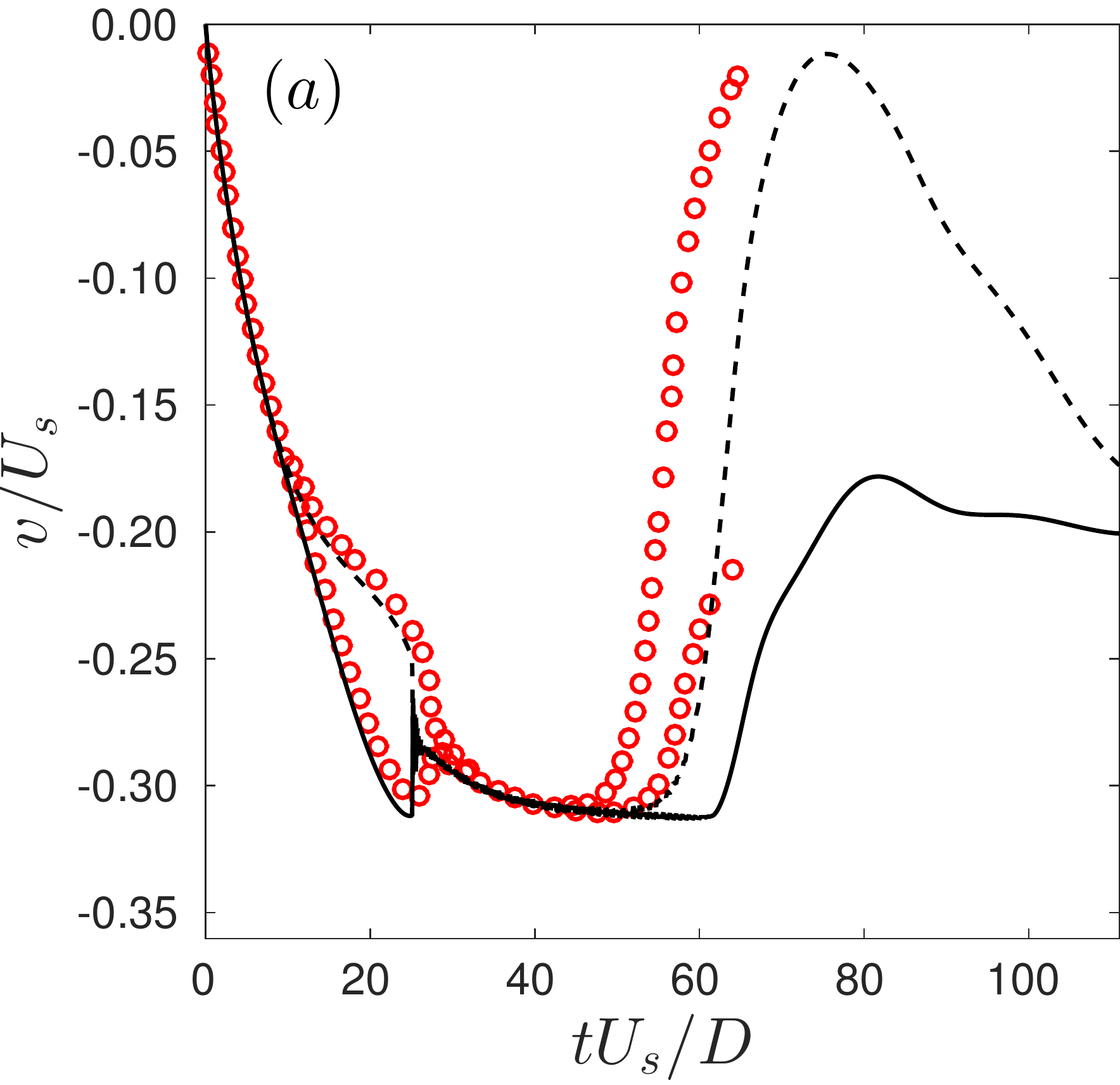}
	\end{minipage}%
	\hfill
	\begin{minipage}[t]{0.49\linewidth}
		\includegraphics[width=\figWidth]{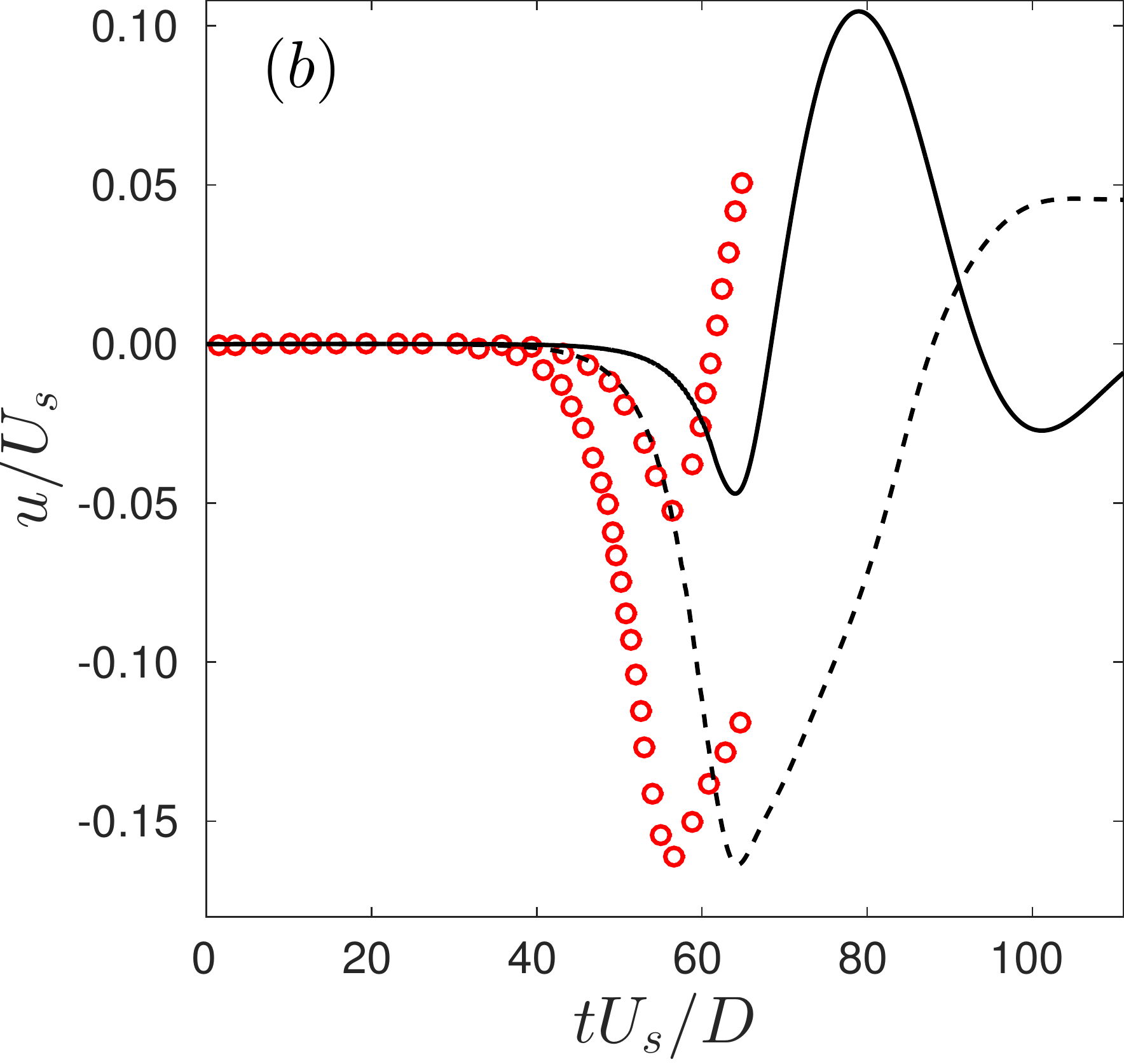}
	\end{minipage}%
        \vspace{-10pt}
    \caption{Time histories of the $(a)$ $v^\ast$ and $(b)$ $u^\ast$ velocity components of the centre of the disks for the low Reynolds number drafting, kissing, and tumbling test case with $\rho_d=1.01$, $\nu=0.1$, where the solid line denotes the (initially) top disk and the dashed line the bottom disk, and 
   data from Zhang~et al.~\citep{Zhang2003} overlayed ($\color{red}\circ$). 
\label{FIG: Patankar vel}}
\end{figure}

Fig.~\ref{FIG: Uhlmann vort contour plots} shows the positions of the particles as they 
sediment, interacting with each other and the domain boundary, along with 
instantaneous vorticity contours. The observed dynamical interactions are in good agreement 
with those observed in the quasi two-dimensional experiments in~\citep{Fortes1987}. 
Initially, the two particles begin moving from rest under 
the influence of gravity with the same acceleration. As the wake forms behind the lower 
particle the top particle becomes shielded in the resultant low pressure region. This allows 
the top particle to draft behind the lower particle, similar to cyclists in a peloton. This is 
the ``drafting'' stage. Eventually, the top particle makes near contact with the lower particle 
(they ``kiss'') and effectively form an elongated body with axis parallel to the 
fall. This configuration is inherently unstable and the elongated 
body rotates to align its long 
axis perpendicular to the fall. This is the ``tumbling'' stage described in~\citep{Fortes1987}. 
The particles separate and the lower particle is overtaken by the top particle, which 
continues to sediment with a slightly negative $u^\ast$ velocity. The other particle impacts the 
wall, after which it, too, sediments with a slightly negative $u^\ast$ velocity.

The results for the low Reynolds number case compare well qualitatively with those of 
\citep{Patankar2000,Patankar2001a,Feng2004} but not quantitatively (see Fig.~\ref{FIG: other vel}). Though 
quantitative agreement is not necessarily apparent amongst the results of these studies 
themselves, what is apparent is that their settling velocities are all lower than those found 
in the present study. Although different methods were used, all three of these simulations 
used low grid resolutions around the particles, particularly~\citep{Feng2004}. 
The study in~\citep{Zhang2003} used a higher resolution grid, with 20 computational 
nodes per particle diameter. 
Very good quantitative agreement is found between that study and the present one, as is 
evident from figure \ref{FIG: Patankar vel}, though there 
is a discrepancy in the duration of the ``kissing'' contact and the onset of ``tumbling''. 
The onset of ``tumbling'' is caused by the build up of numerical error, so this is expected 
to be solver specific. In the absence of numerical error, or bias introduced by the grid, the 
disks would not leave the ``kissing'' stage~\citep{Uhlmann2005}.

{
\newcommand{\figWidtha}{4.75cm}
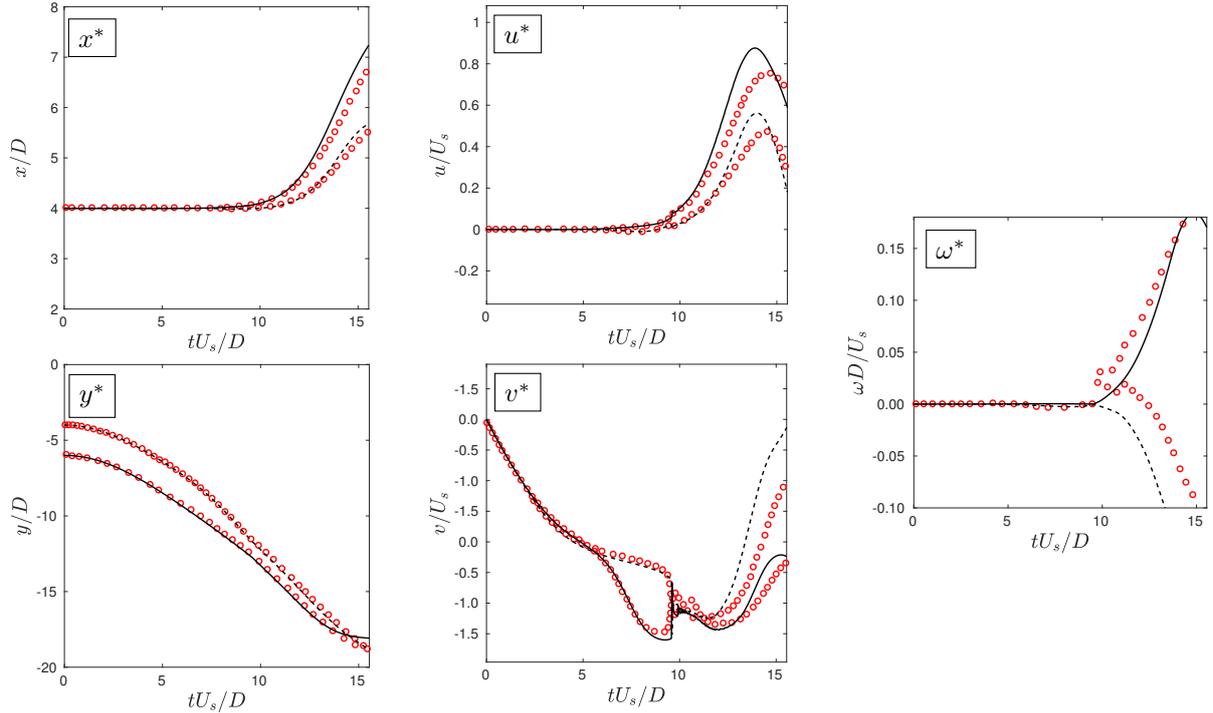
\begin{figure}[htb]
\begin{center}
\begin{tikzpicture}[scale=1]
  \useasboundingbox (0,.45) rectangle (16.,9.);  
  \begin{scope}[xshift=2.5cm,yshift=2cm]
    \draw (0.0,0) node {\includegraphics[width=\figWidtha]{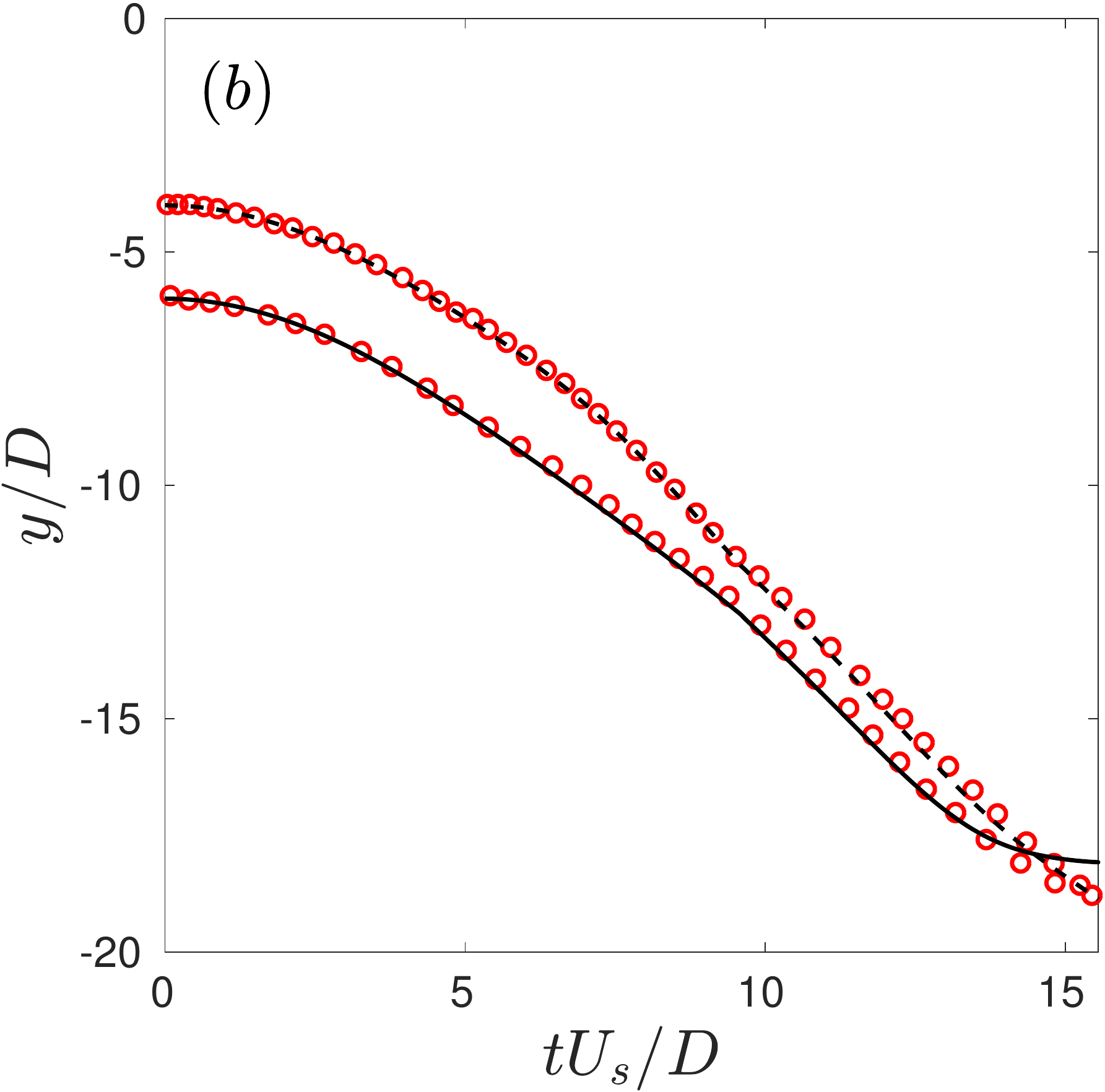}};
    \draw (0,4.75) node {\includegraphics[width=\figWidtha]{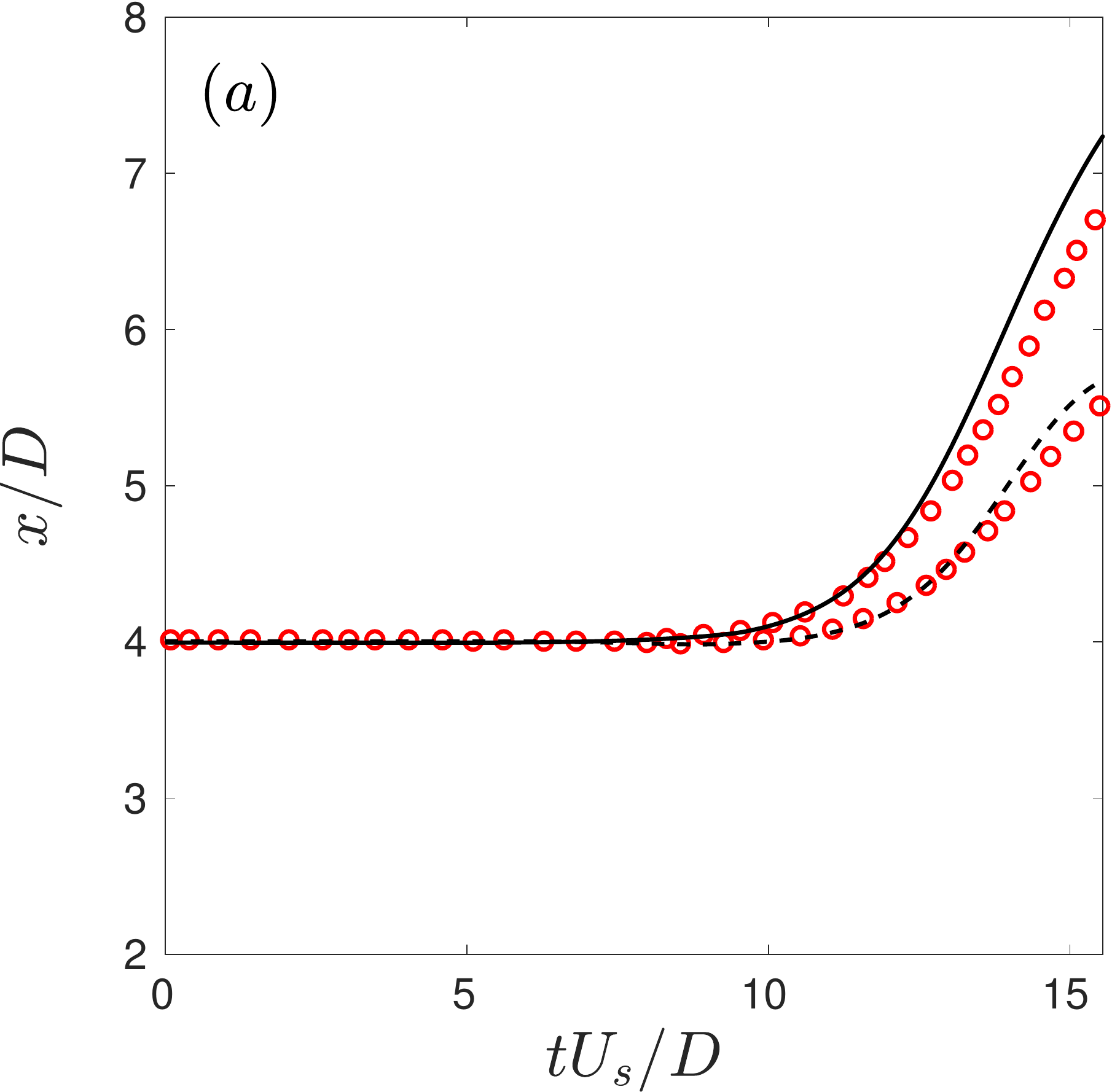}};

    \draw (5.5,0.0) node {\includegraphics[width=\figWidtha]{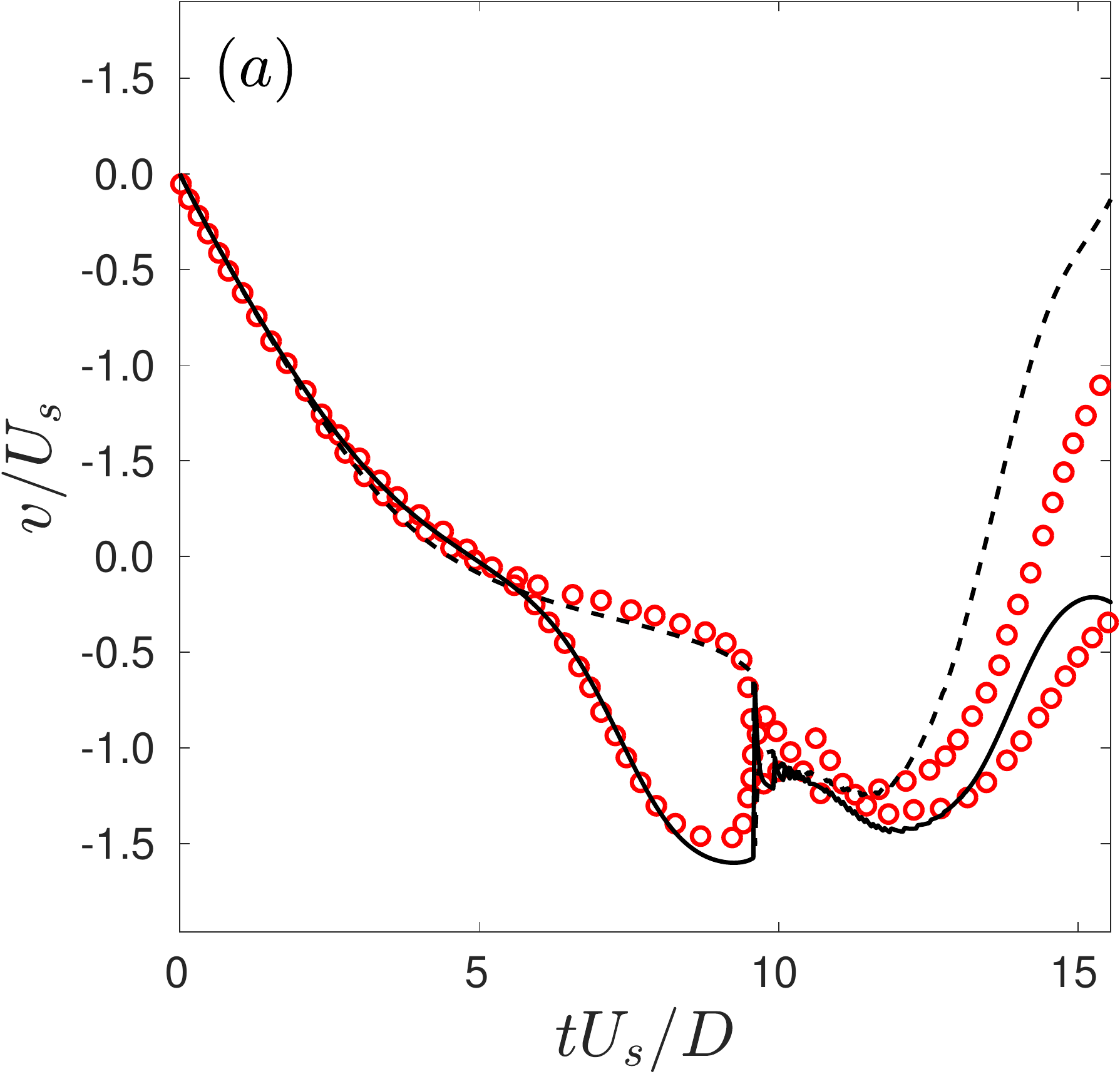}};
    \draw (5.5,4.75) node {\includegraphics[width=\figWidtha]{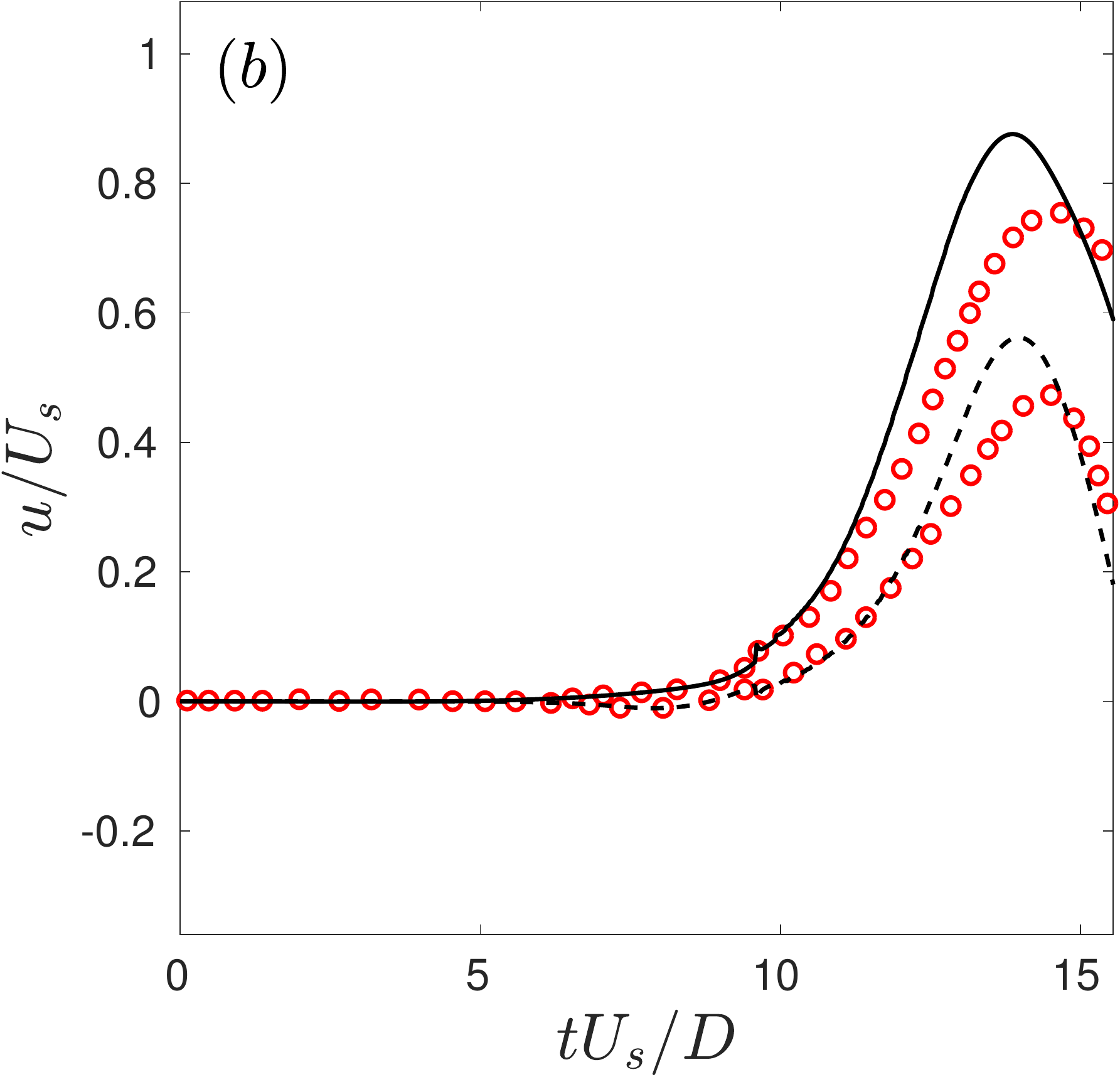}};

    \draw(11,2) node {\includegraphics[width=\figWidtha]{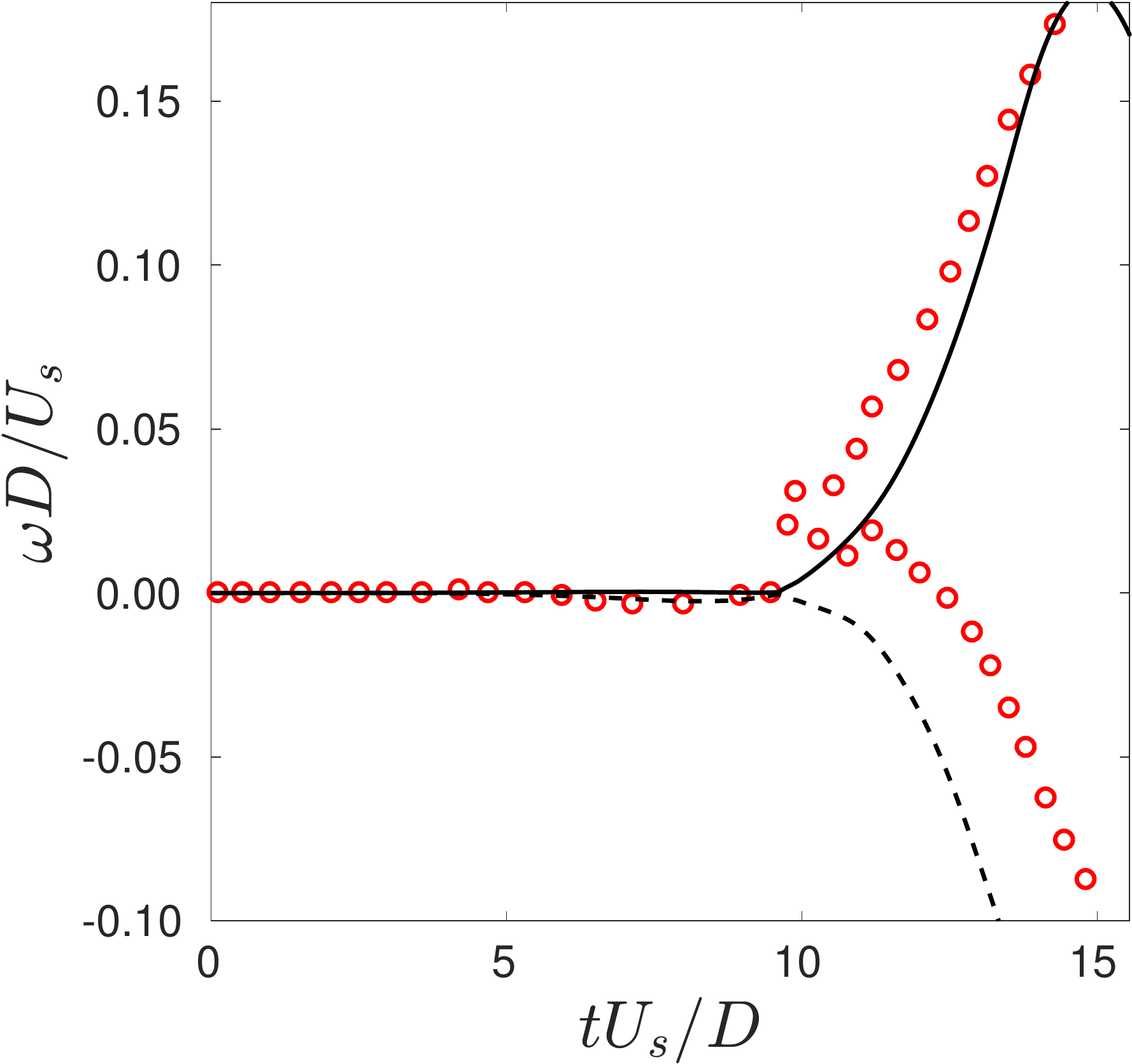}};

    \draw (-1.3,6.6) node[draw,fill=white] {\small $x^\ast$};
    \draw (-1.3,1.85) node[draw,fill=white] {\small $y^\ast$};

    \draw (4.3,6.65) node[draw,fill=white] {\small $u^\ast$};
    \draw (4.3,1.9) node[draw,fill=white] {\small $v^\ast$};

    \draw (10.,3.8) node[draw,fill=white] {\small $\omega^\ast$};
  \end{scope}
\end{tikzpicture}
\end{center}
  \caption{Time histories of $x^\ast$, $y^\ast$, $v^\ast$, $u^\ast$ and $\omega^*$ for the moderate Reynolds number drafting, kissing, and tumbling test case with  $\rho_d=1.5$, $\nu=0.01$, where the solid line denotes the (initially) top disk and the dashed line the bottom, and data from Uhlmann~\citep{Uhlmann2005} overlayed ($\color{red}\circ$).}
  \label{fig:settlingDisksTimeHistory}
\end{figure}
}


Results for the moderate Reynolds number case 
are shown in Fig.~\ref{fig:settlingDisksTimeHistory}.
These are 
compared to results from~\citep{Uhlmann2005}, who used an immersed boundary method on 
high resolution grids. Both qualitatively and quantitatively the results are in excellent 
agreement for the particle positions and $u,v$ velocity components, with the only differences found during the initial contact and subsequent 
``kissing'' stage, due to the different collision models. 
All of the aforementioned studies used a repulsive force based model, while 
a conservation of linear momentum model is used here. 
Fig.~\ref{fig:settlingDisksTimeHistory} shows 
good qualitative but poor quantitative agreement for the angular velocity component. This is 
a very sensitive metric~\citep{Uhlmann2005} and it is likely that the differences are due, 
in large part, to the different collision mechanisms used. The novel approach 
of Kempe and Fr\"ohlich~\citep{kempe2012improved}, which uses a sub-grid lubrication force correction and conserves 
angular momentum, would probably be a more appropriate collision mechanism for this case.

\subsection{Two particle wake interaction \label{SEC: pure wake interaction}}

\begin{figure}[hbt]\scriptsize
    \centering
    \def\svgwidth{.6\linewidth}
    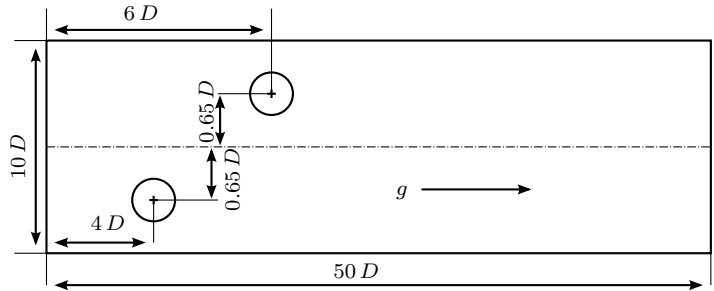
    \vspace{-15pt}
    \caption{Problem geometry for the two particle wake interaction test case. \label{FIG: wake interaction geo}}
\end{figure}

This is a case presented by Uhlmann~\citep{Uhlmann2005} to test the fluid--structure interaction, with 
particular emphasis on examining the effect of wake interactions between the particles on 
the angular velocity. Two particles of differing densities settle in an otherwise quiescent 
fluid. The heavier particle passes the lighter particle, subjecting it to perturbations from its 
wake. The particles do not collide and therefore no collision model is required, making it an 
attractive benchmark case. 

{
\newcommand{\twoWakeFigWidth}{8cm}
\newcommand{\cbWidth}{.2cm}
\newcommand{\cbHeight}{8cm}
\newcommand{\trimfigcb}[3]{\includegraphics[width=#2, height=#3, clip, trim=17cm 2.35cm 1.65cm 2.35cm]{#1}}
\begin{figure}[htb]
\begin{center}
\begin{tikzpicture}[scale=1]
  \useasboundingbox (0,.3) rectangle (16.,7.5);  
  \begin{scope}[xshift=4.cm,yshift=1cm]
    \draw (0.0,5.81) node {\includegraphics[width=\twoWakeFigWidth]{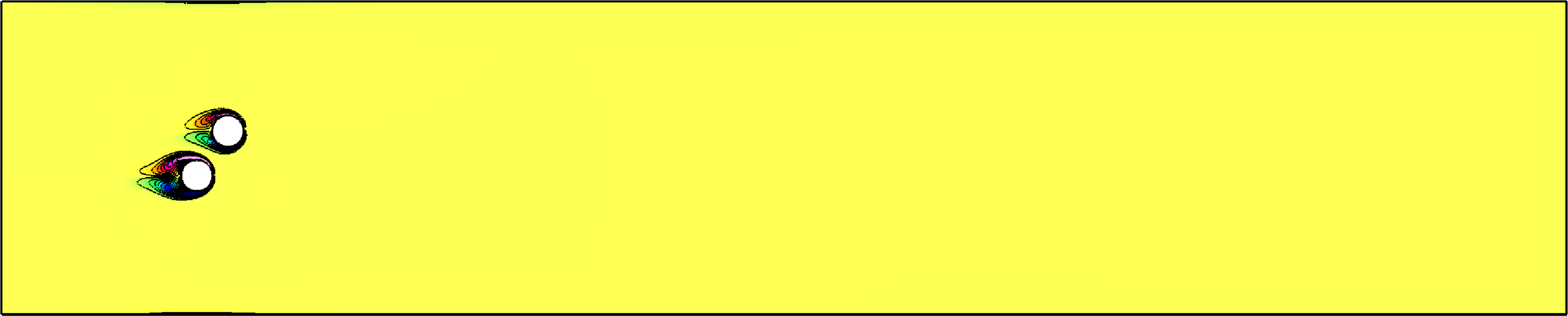}};
    \draw (0.0,3.92) node {\includegraphics[width=\twoWakeFigWidth]{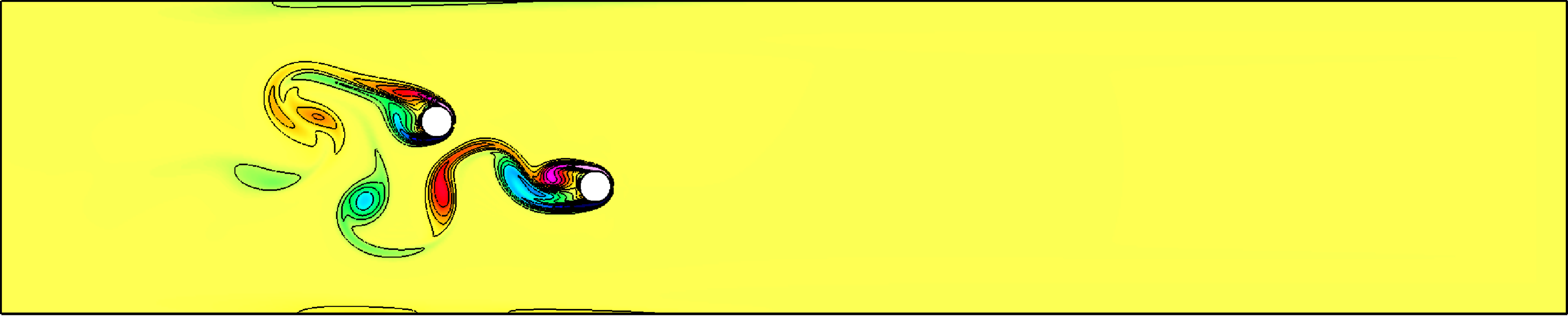}};
    \draw (0.0,2.05) node {\includegraphics[width=\twoWakeFigWidth]{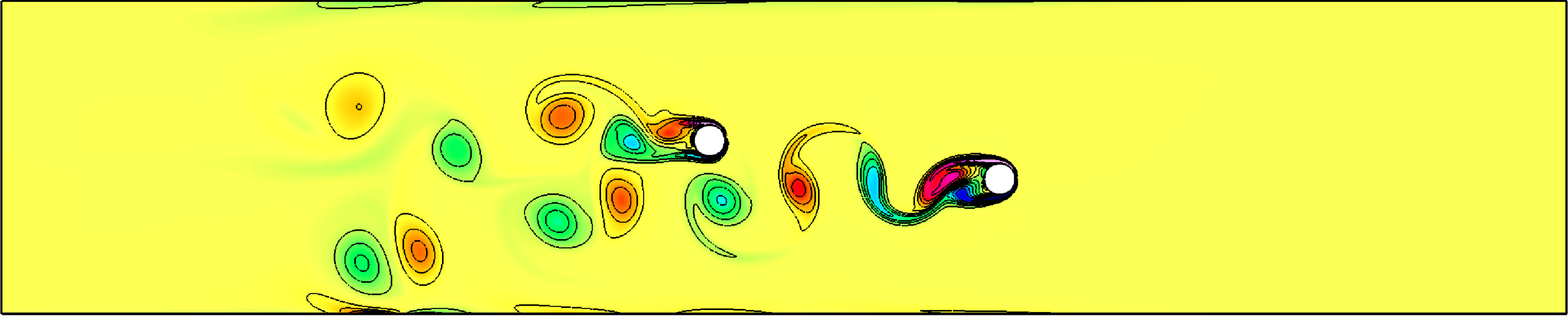}};
    \draw (0.0,0.19 ) node {\includegraphics[width=\twoWakeFigWidth]{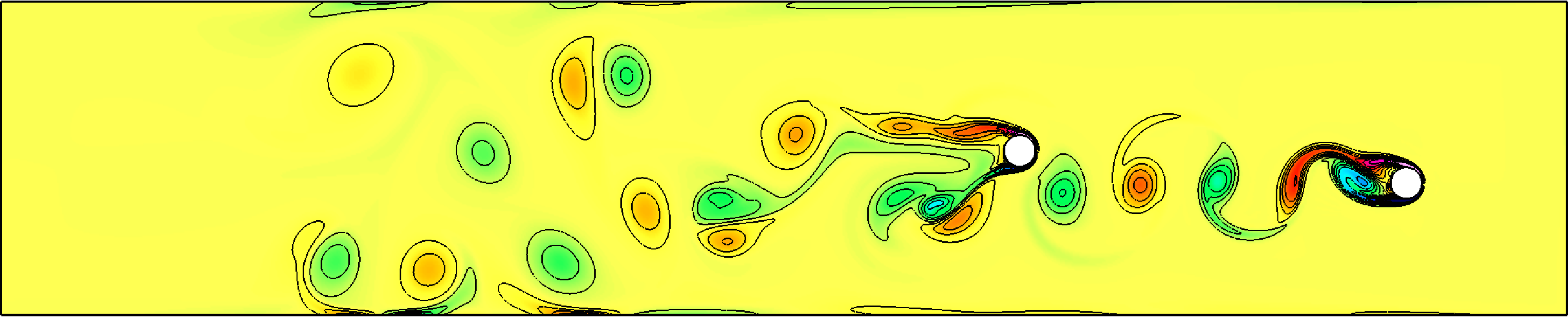}};
    \draw (8.25,3.0) node {\includegraphics[width=\twoWakeFigWidth]{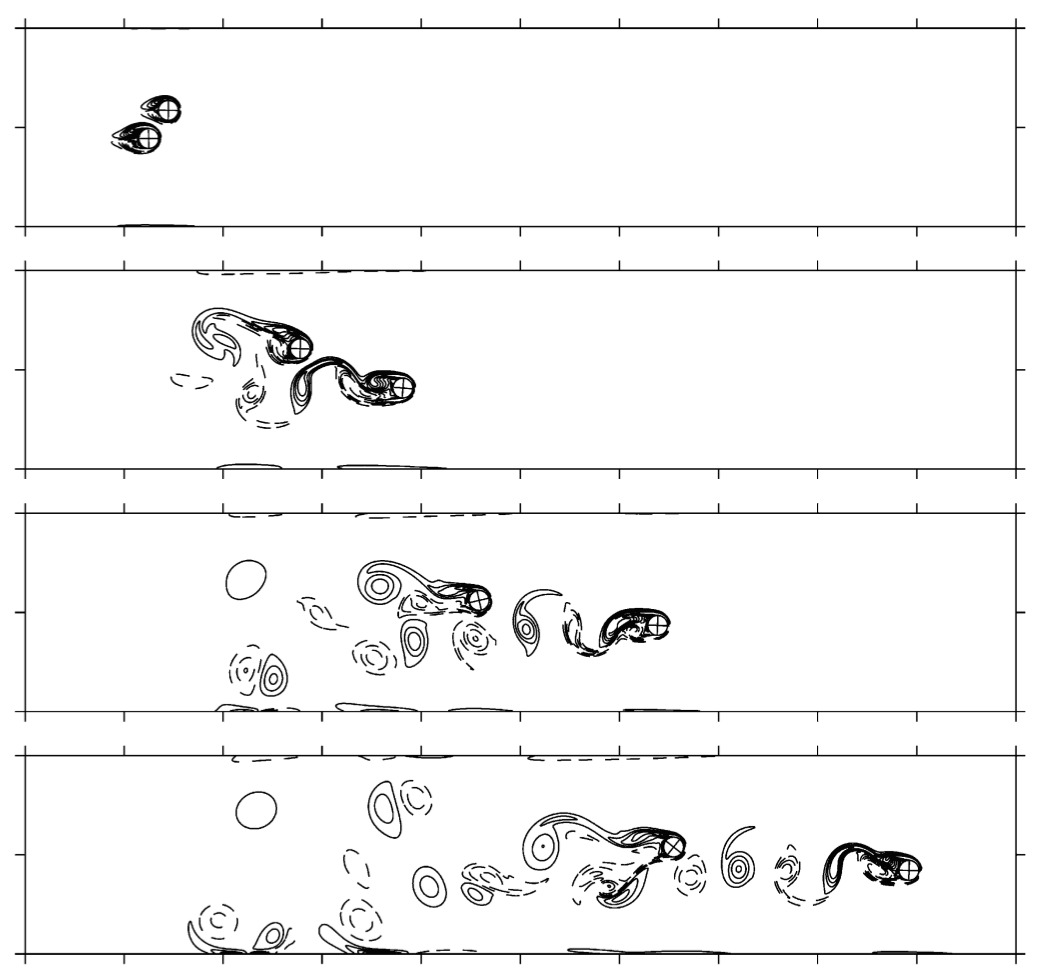}};
  \end{scope}
  \begin{scope}[xshift=8cm,yshift=0pt]
     \draw(0,2.00) node[draw,fill=white,anchor=east,xshift=-2pt,yshift=-9pt] {\scriptsize$t=8.0$};
     \draw(0,3.86) node[draw,fill=white,anchor=east,xshift=-2pt,yshift=-9pt] {\scriptsize$t=5.6$};
     \draw(0,5.75) node[draw,fill=white,anchor=east,xshift=-2pt,yshift=-9pt] {\scriptsize$t=3.2$};
     \draw(0,7.65) node[draw,fill=white,anchor=east,xshift=-2pt,yshift=-9pt] {\scriptsize$t=0.8$};
  \end{scope}
  \begin{scope}[xshift=-4pt,yshift=-3pt]
    \draw (0,0) node[anchor=south west,xshift=0.cm,yshift=.5cm,rotate=-90] {\trimfigcb{figures/colourBarLines}{\cbWidth}{\cbHeight}};
     \draw (0,0) node[anchor=north,xshift=10pt,yshift=6pt] {\scriptsize$-30$};
     \draw (4,0) node[anchor=north,yshift=6pt] {\scriptsize $\xi$};
     \draw (8,0) node[anchor=north,xshift=-2pt,yshift=6pt] {\scriptsize $3.0$};
  \end{scope}


%
%

\end{tikzpicture}
\end{center}
	\caption{Contours of vorticity at times $t=0.8$, $t=3.2$, $t=5.6$, $t=8.0$  
       for the pure wake interaction test case compared to plots from Uhlmann~\citep{Uhlmann2005} 
       taken at the same times and with the same vorticity extrema. Left: present study 
        using a quasi uniform grid with $\textit{DLS=D/40}$ and $\textit{SLS=D/40}$. Right: results from~\citep{Uhlmann2005} 
        computed on a uniform grid of resolution $D/40$.}
	\label{FIG: wake interaction vorticity plots}
\end{figure}
}
\bogus{ 
{
\newcommand{\twoWakeFigWidth}{8cm}
\begin{figure}
	\centering
	\begin{minipage}[c]{0.475\linewidth}
        \vspace{.75em}
        \includegraphics[width=\twoWakeFigWidth]{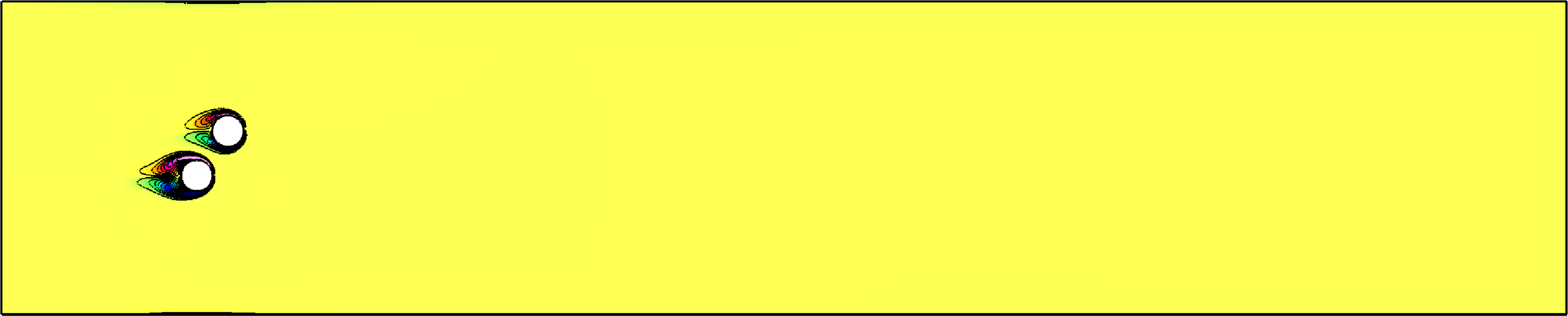}\vspace{.7em}
        \includegraphics[width=\twoWakeFigWidth]{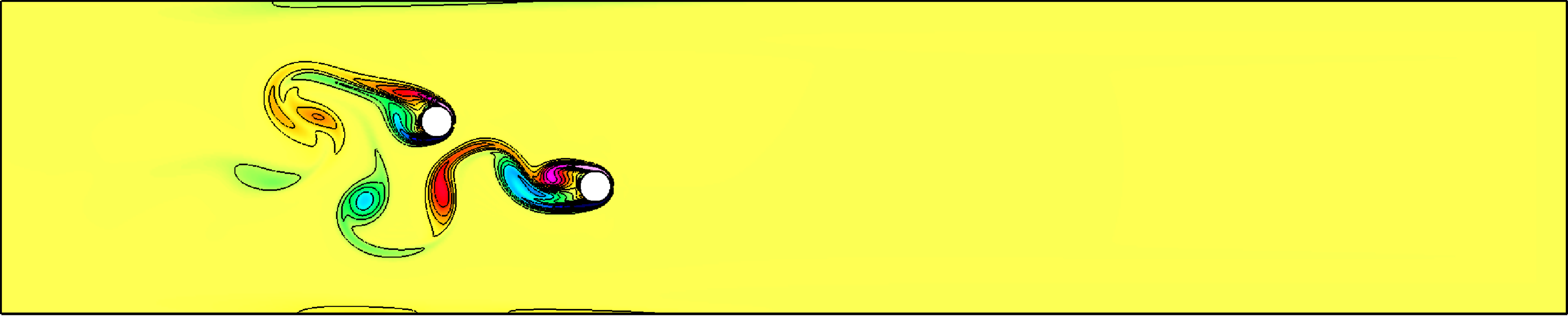}\vspace{.7em}
        \includegraphics[width=\twoWakeFigWidth]{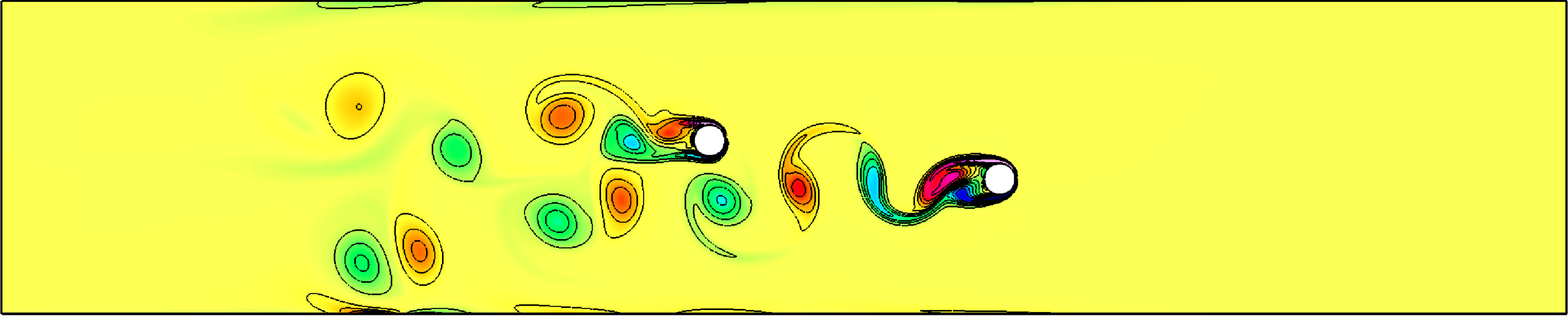}\vspace{.7em}
        \includegraphics[width=\twoWakeFigWidth]{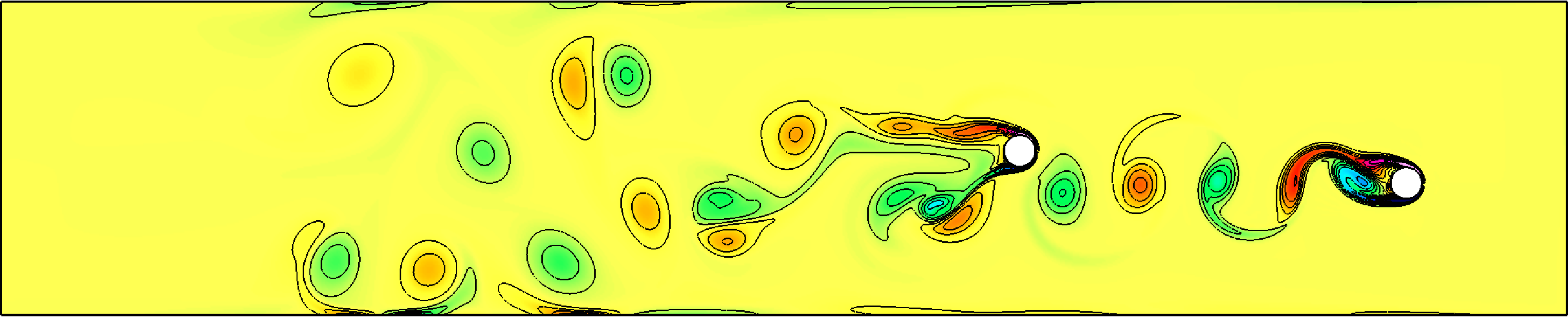}\vspace{.7em}
	\end{minipage}%
	\hfill
	\begin{minipage}[c]{0.49\linewidth}
		\includegraphics[width=\twoWakeFigWidth]{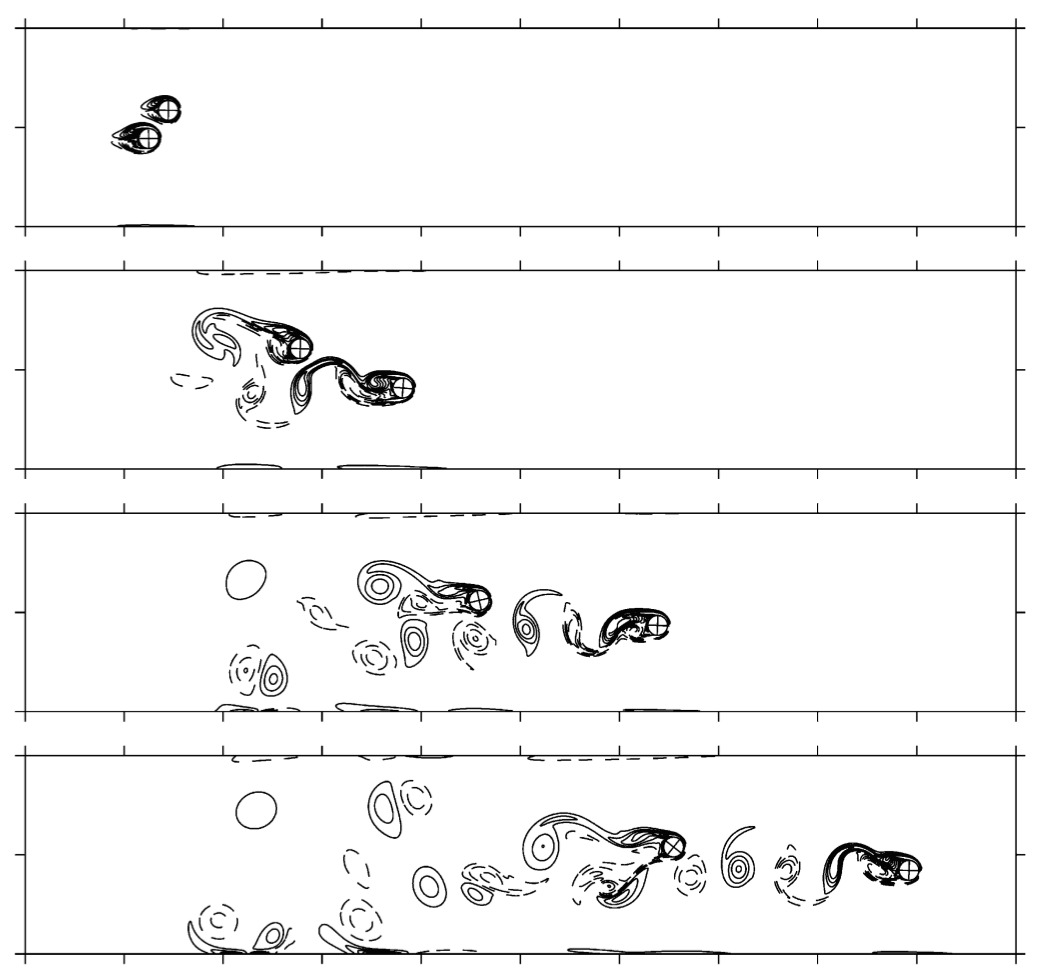}
	\end{minipage}%
	\caption{Instantaneous contours of vorticity, where the rainbow colour map shows negative vorticity in blue, the zero level in yellow and positive vorticity in red, between $-30\leq\xi\leq3$ at $t=0.8$, $
		t=3.2$, $t=5.6$, $t=8.0$ (from top to bottom) for the pure wake interaction test case compared to plots from 
		\citet{Uhlmann2005} taken at the same times and with the same vorticity extrema. Left: present study 
        using a quasi uniform grid with $\textit{DLS=D/40}$ and $\textit{SLS=D/40}$. Right: \citep{Uhlmann2005} 
        computed on a uniform grid of resolution $D/40$. 
	\label{FIG: wake interaction vorticity plots}}
\end{figure}
}
} 

The computational domain, shown in Fig.~\ref{FIG: wake interaction geo}, has a width of $10\,D$, a height of $50\,D$ and the 
grid characteristic length scales are $\textit{DLS}=D/40$ and $\textit{SLS}=D/100$, where the 
particle diameter is $D=0.2\,\mathrm{m}$. A heavier particle of density ratio $\rho_{r,1}=1.5$ is 
initially positioned at $\boldsymbol{x}=(-0.65\,D,4\,D)$ from the channel centreline and 
top boundary respectively, while the lighter particle of density ratio 
$\rho_{r,2}=1.25$ is positioned at $\boldsymbol{x}=(0.65\,D,6\,D)$. Both particles are 
initially at rest and the fluid of kinematic viscosity $\nu=0.0008\,\mathrm{m^2/s}$ is quiescent at
$t=0\,\mathrm{s}$. The gravitational constant is set to $\boldsymbol{g}=9.81\,\mathrm{m/s^2}$. The 
results are non-dimensionalised as {in \eqref{EQ: non-dimensionalization}, where the characteristic 
    velocity $U_s$ is calculated using \eqref{EQ: terminal vel estimate} and the density ratio of the 
heavier disk, \viz $\rho_{r,1}=1.5$.}

{
\newcommand{\figWidtha}{4.75cm}
\begin{figure}[htb]
\begin{center}
\begin{tikzpicture}[scale=1]
  \useasboundingbox (0,.25) rectangle (16.,9.);  
  \begin{scope}[xshift=2.5cm,yshift=2cm]
    \draw (0.0,0) node {\includegraphics[width=\figWidtha]{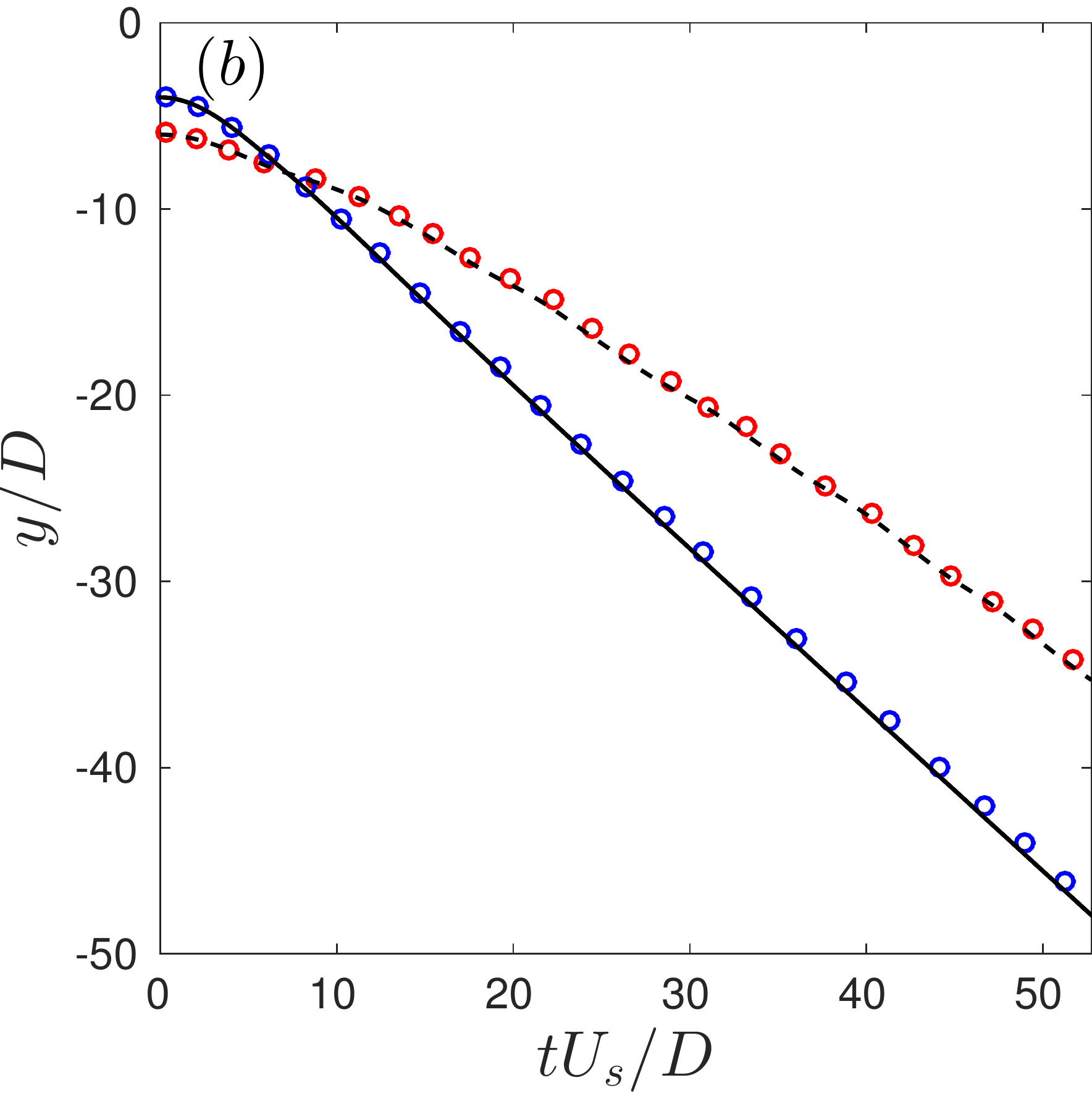}};
    \draw (0,4.75) node {\includegraphics[width=\figWidtha]{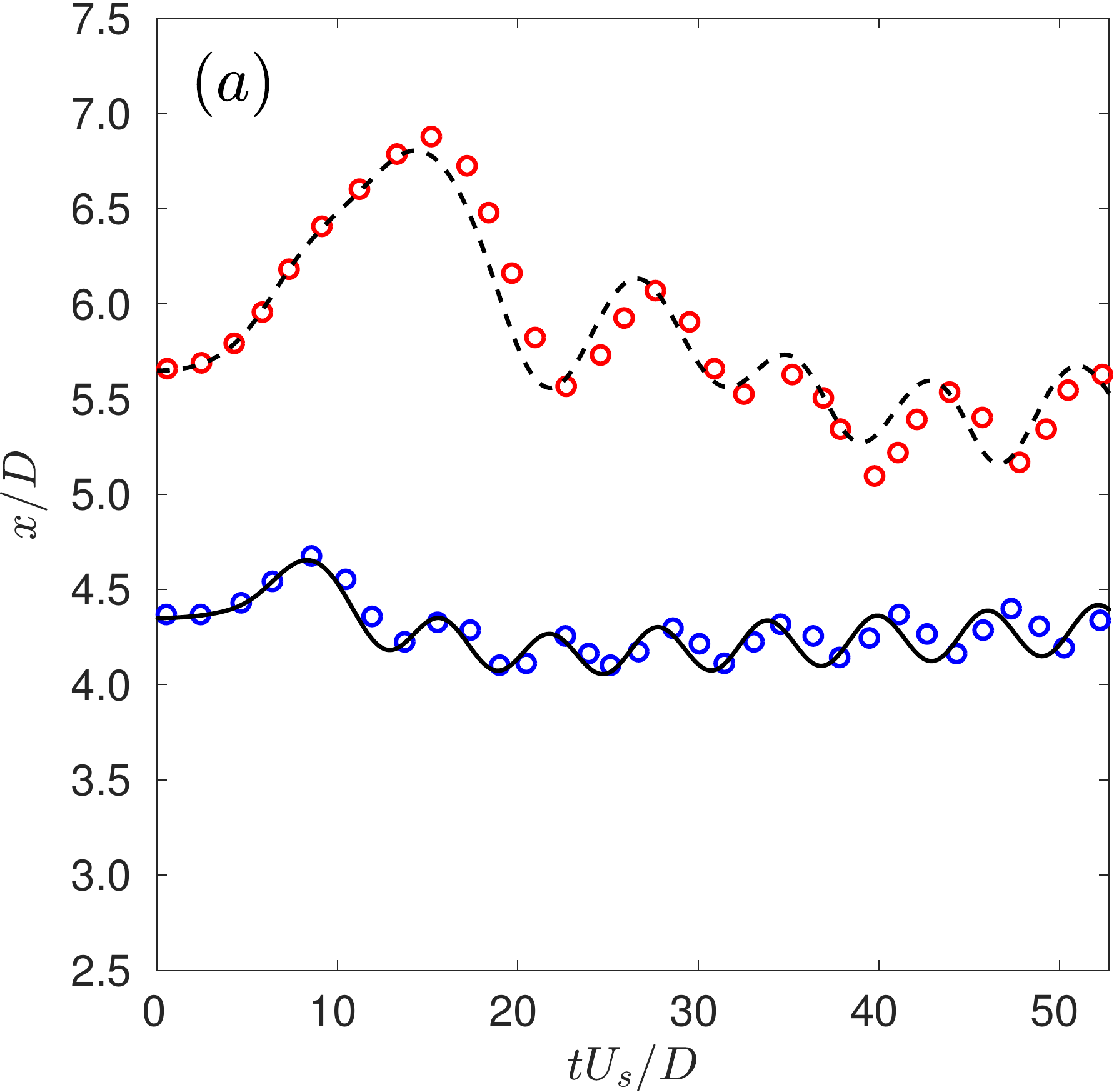}};

    \draw (5.5,0.0) node {\includegraphics[width=\figWidtha]{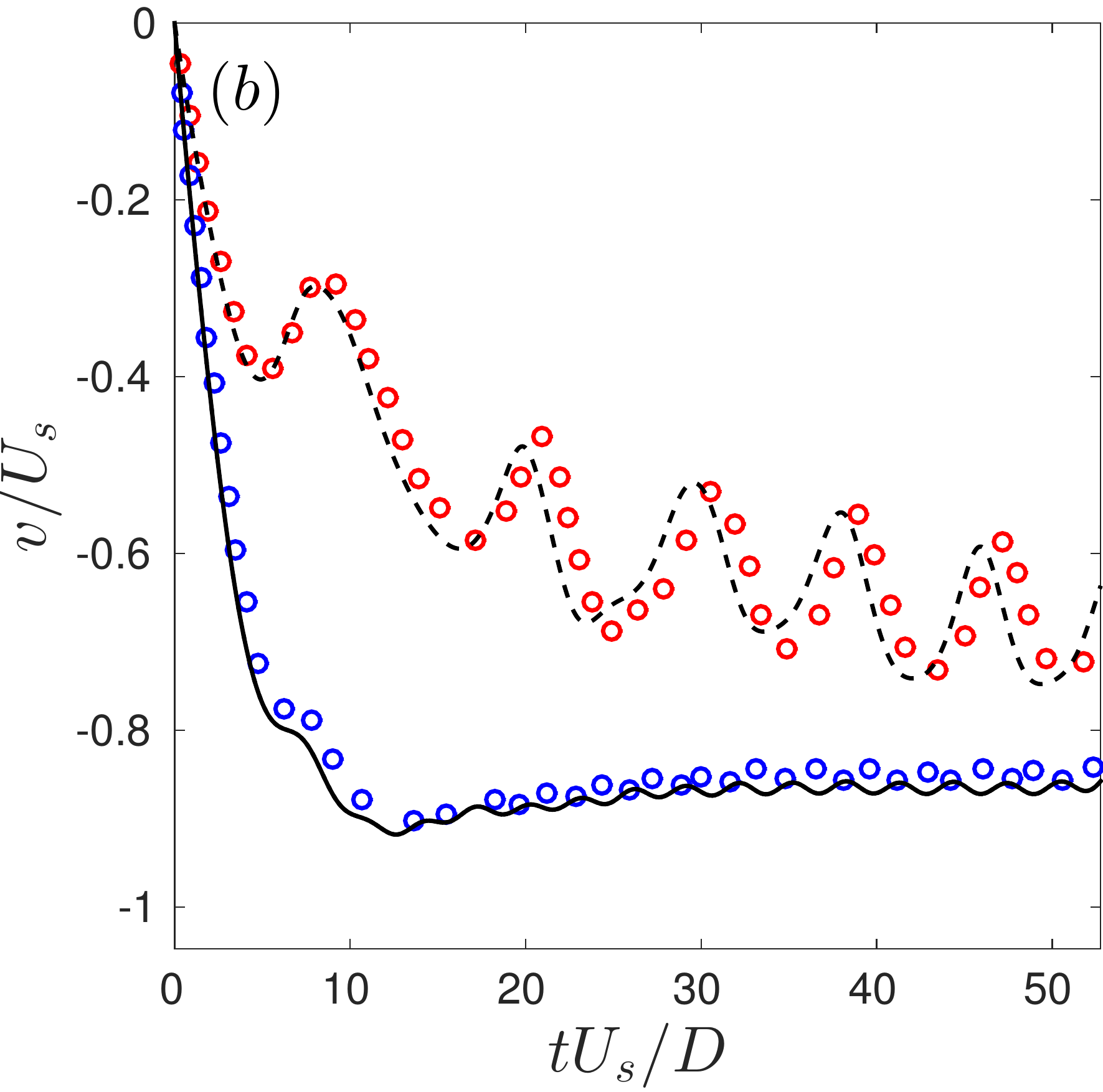}};
    \draw (5.5,4.75) node {\includegraphics[width=\figWidtha]{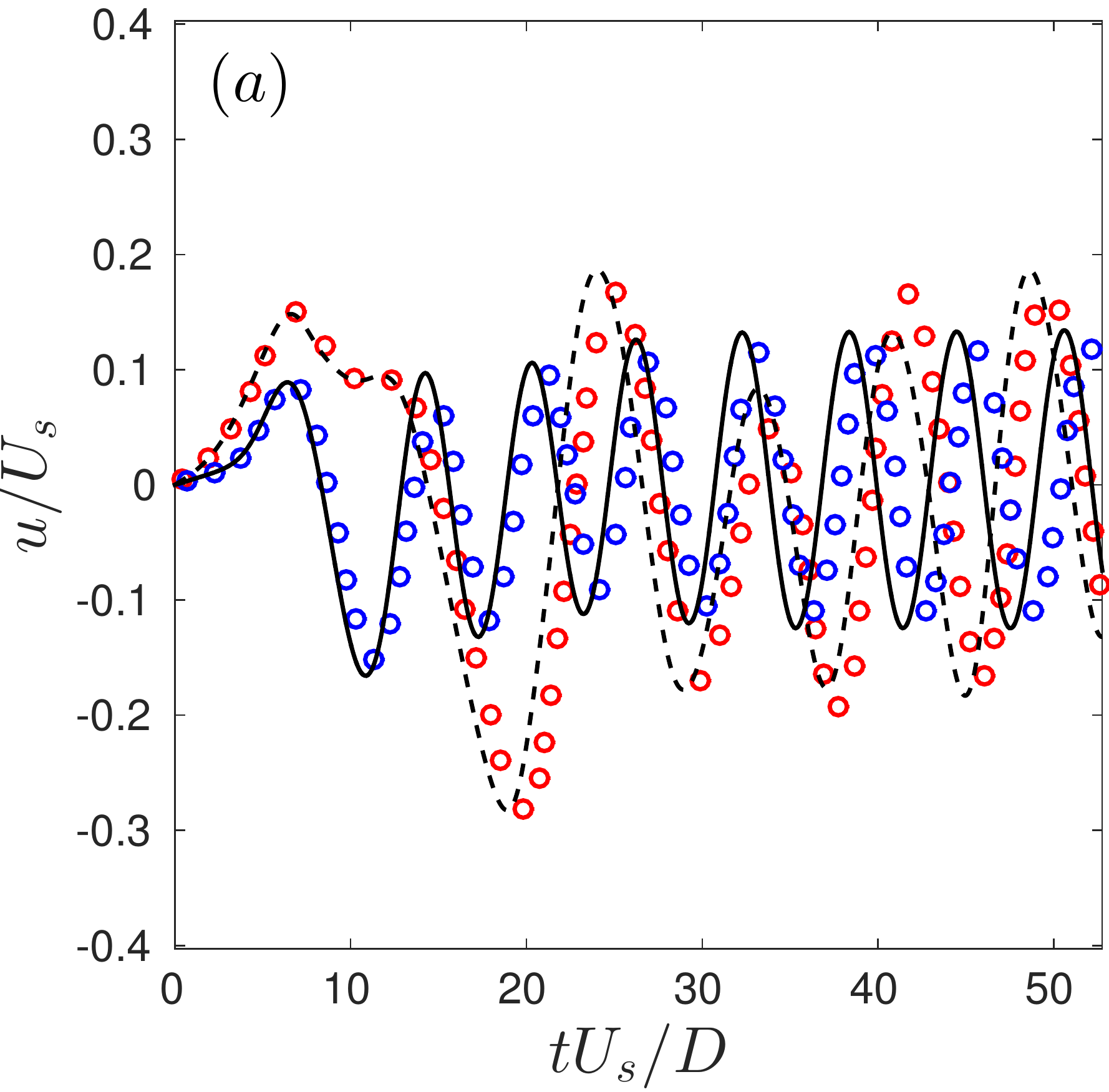}};

    \draw(11,2) node {\includegraphics[width=\figWidtha]{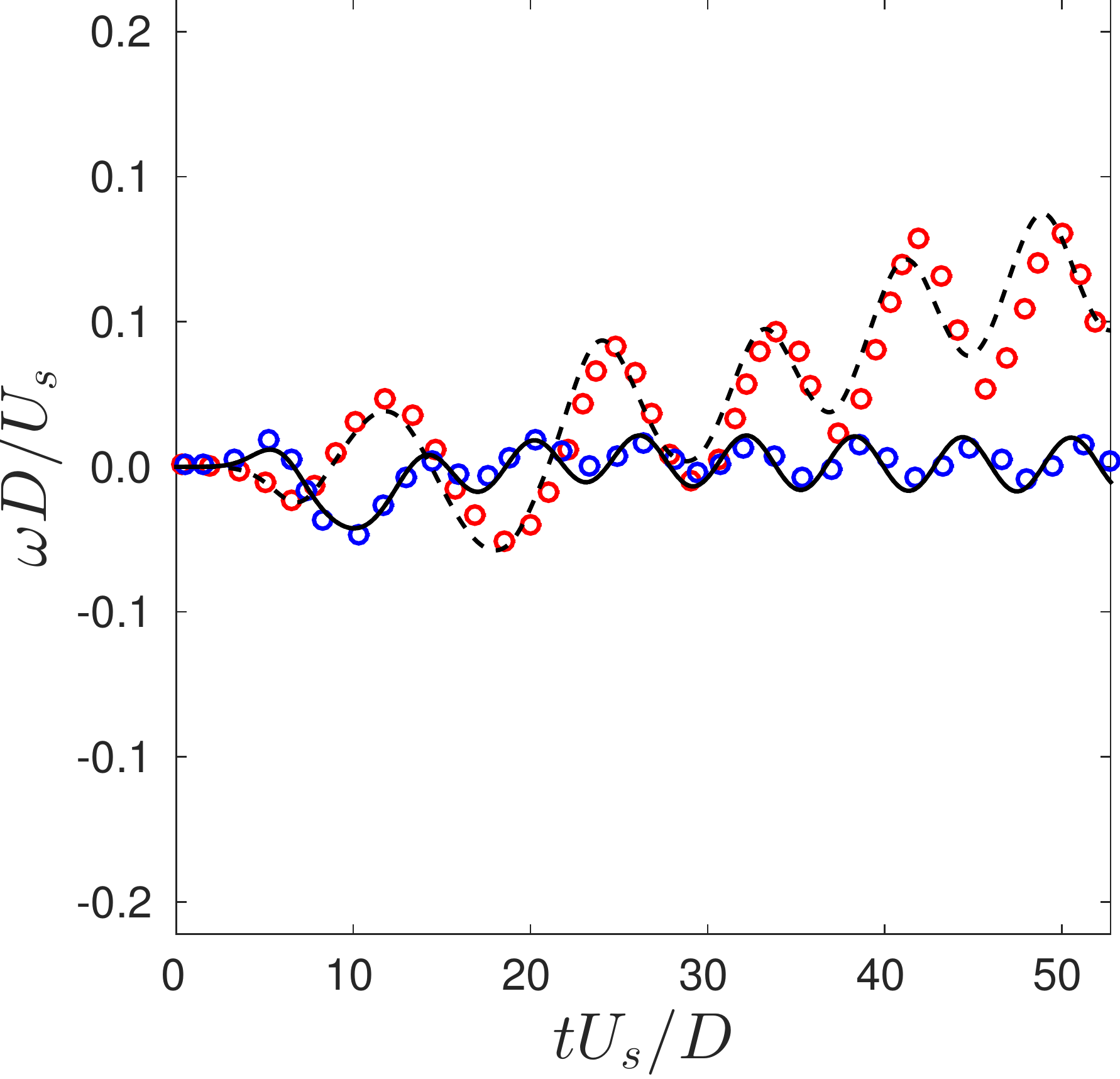}};

    \draw (-1.3,6.6) node[draw,fill=white] {\small $x^\ast$};
    \draw (-1.2,2.25) node[draw,fill=white] {\small $y^\ast$};

    \draw (4.3,6.65) node[draw,fill=white] {\small $u^\ast$};
    \draw (4.35,1.9) node[draw,fill=white] {\small $v^\ast$};

    \draw (10.,3.8) node[draw,fill=white] {\small $\omega^\ast$};
  \end{scope}
\end{tikzpicture}
\end{center}
  \caption{Time histories of $x^\ast$, $y^\ast$, $v^\ast$, $u^\ast$ and $\omega^*$ for the pure wake interaction
  test case with $\rho_{r,1}=1.5$, $\rho_{r,2}=1.25$, $\nu=0.0008\,\mathrm{m^2/s}$ and data from Uhlman~\citep{Uhlmann2005} overlayed.}
  \label{fig:wakeInteractionTimeHistory}
\end{figure}
}


The maximum particle Reynolds numbers of the heavy and light particle are 280 and 230, 
respectively~\citep{Uhlmann2005}. Uhlmann~\citep{Uhlmann2005} used a uniform grid resolution 
of $\textit{DLS}=D/40$, allowing for fewer than three grid points across the estimated 
boundary layer depth for both the light and heavy particles. Given the findings of the 
convergence study in section \ref{SEC: convergence study} it is not likely that the 
solutions at this grid resolution are grid independent. A brief convergence study was 
performed, shown in table \ref{TAB: wake interaction convergence}, and solutions were found to converge at $\textit{SLS}=D/100$, allowing for 6 
points across the boundary layers for both the light and heavy particle. 

\begin{table}\tableFont
    \centering
        \begin{tabular}{@{} r r r r r r r @{}}
        \toprule
        $\textit{SLS}$ & $u/U_s$ & $v/U_s$ & $\omega D/U_s$ & $\epsilon_u$  & $\epsilon_v$  & $\epsilon_\omega$ \\
        \midrule
        D/200    &   0.14550 &   -0.36470    &   -0.01123    &   --- &   --- &   --- \\
        D/150     &   0.14554 &   -0.36483    &   -0.01225    &   0.00028 &   0.00033 &   0.00181 \\
        D/100     &   0.14567 &   -0.36518    &   -0.01132    &   0.00121 &   0.00130 &   0.00831 \\
        D/50     &   0.14680 &   -0.36705    &   -0.01205    &   0.00896 &   0.00643 &   0.07353 \\
        D/40     &   0.01479 &   -0.36847    &   -0.01293    &   0.01644 &   0.01033 &   0.15205 \\
        \bottomrule
    \end{tabular}
\caption{Absolute values and relative errors for disk 2 taken at $t=1\,\mathrm{s}$ of the wake interaction grid independence study.\label{TAB: wake interaction convergence}}
\end{table}

Fig.~\ref{FIG: wake interaction vorticity plots} shows successive snapshots of the 
instantaneous vorticity field with snapshots from~\citep{Uhlmann2005} below. The evolving 
flow field and particle positions match well. 
Fig.~\ref{fig:wakeInteractionTimeHistory} shows the time histories of the particles
position, velocity and angular velocity.

The converged solutions of the present study match well with those in~\citep{Uhlmann2005}, 
although a phase shift is apparent in the oscillatory components and the settling velocity 
is slightly higher in the present study. 

The largest differences are found in the horizontal 
velocity components, particularly for the light particle. These differences are probably due 
in large part to the differences in the angular velocity components, which will 
affect vortex shedding and lift on the particles. The amplitudes of the 
angular velocity component oscillations for the heavier particle match well but a slight 
phase shift is apparent. This is reflected in the horizontal velocity components for the heavier particle 
by matching amplitudes but markedly different periods of oscillation. The angular velocity 
components of the lighter particle differ in both amplitude and period of oscillation, leading 
to more pronounced differences in the horizontal velocity components of the two studies.

\subsection{Settling sphere}
As a final validation case we compare experimental \citep{tenCate2002} and numerical \citep{Yang2015} results on the motion of a single 
sphere in a closely confined container to 
numerical results produced by the current method.
Four cases were run, with 
Reynolds numbers ranging from 1.5 to a moderate 31.9. The material parameters used in each case are 
detailed in table \ref{TAB: settling sphere}, with $\boldsymbol{g}=9.81\,\mathrm{m/s^2}$ throughout.
For each case the grid characteristic length scales 
are $\textit{DLS}=D/10$ and $\textit{SLS}=D/38$, allowing for approximately six points across the 
estimated boundary layer depth for the highest Reynolds number case.

\begin{table}[hbt]
\tableFont
    \centering
    \begin{tabular}{@{} c r r r r @{}}
        \toprule
        \multirow{2}{*}{Case number}     &   $\rho_f$    &   $\mu_f$     &   $\operatorname{Re}$     &   $\operatorname{Stk}$\\
                                         &  [$\mathrm{kg}/\mathrm{m}^3$] &  [$\mathrm{N}\,\mathrm{s}/\mathrm{m}^2$] & [--]    &[--]\\
                                     1   &  970 &   0.373 &   1.5 &   0.19\\
        2   &   965 &   0.212 &   4.1 &   0.53    \\
        3   &   962 &   0.113 &   11.6    &   1.50    \\
        4   &   960 &   0.058  &   31.9    &   4.13    \\
        \bottomrule
    \end{tabular}
    \caption{Parameters used for the four settling sphere cases. \label{TAB: settling sphere}}
\end{table}

The sphere undergoes three distinct periods of motion after its release from rest: an initial acceleration 
followed by a period of steady fall at a terminal settling velocity and finally a deceleration as 
it approaches the wall. As the Reynolds number is increased the three stages become progressively shorter. 
Similar to Yang and Stern~\citep{Yang2015} the wall collisions were not considered here and the simulations stopped before 
the sphere made contact with the wall. 

The present results are shown in Fig.~\ref{FIG: sphere h and v} and match satisfactorily 
with the experimental results of~ten Cate~\etal\citep{tenCate2002} although 
some slight differences remain: the terminal settling velocity in case 1 is found to be approximately $4.6\,\%$ 
lower here and in case 2 the sphere begins the wall induced deceleration sooner than in~\citep{tenCate2002}. 
This earlier deceleration in case 2 is also present in the results of~\citep{Yang2015}, as is the lower 
terminal settling velocity of case 1. The benefit of using overset grids is again evident; the 
grid used above consists of $5.18\e{5}$ grid points while the results are as good as those produced 
on a uniform grid over three times the size.

\begin{figure}
    \centering
    \begin{minipage}[t]{0.49\linewidth}
        \includegraphics[width=\figWidth]{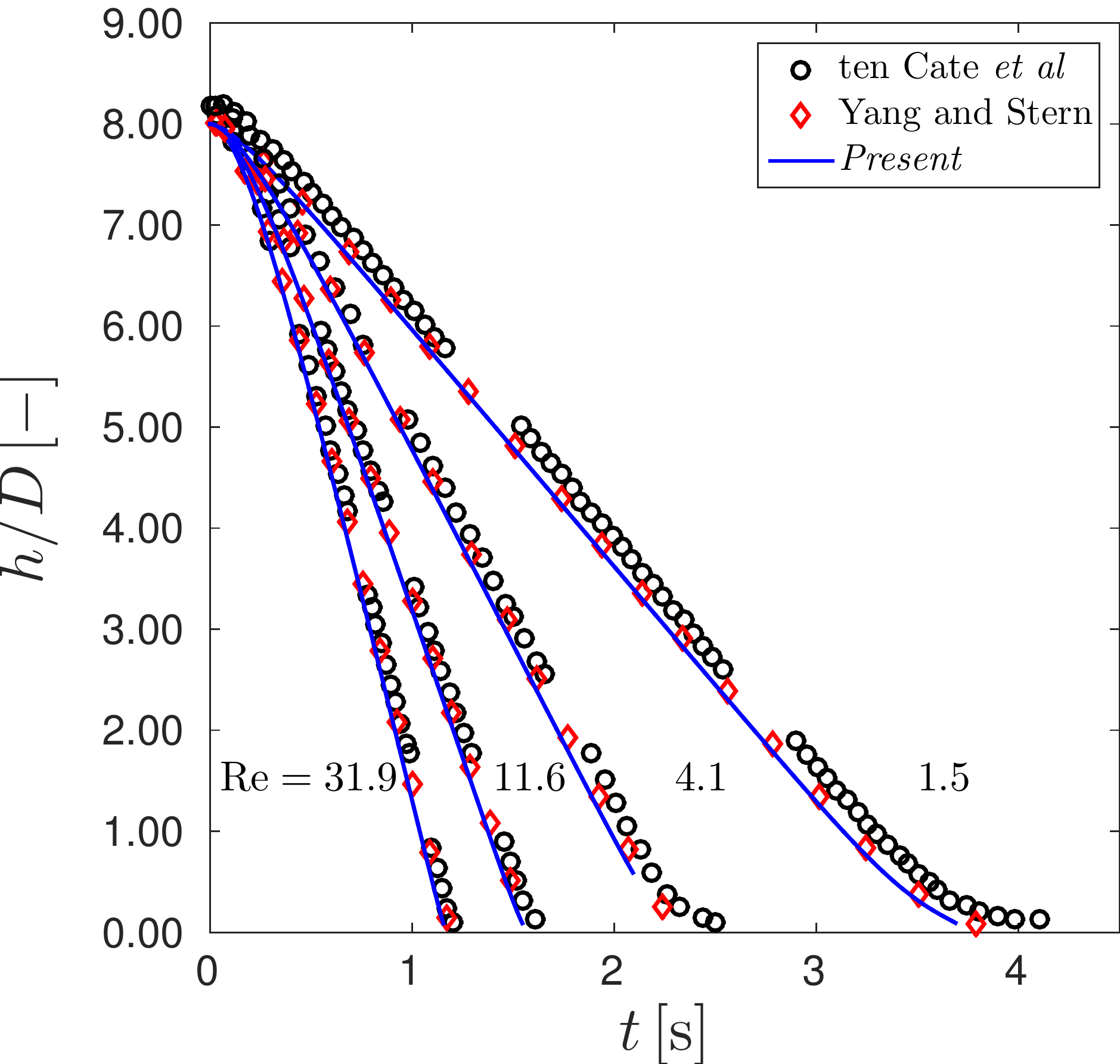}
    \end{minipage}%
    \hfill
    \begin{minipage}[t]{0.49\linewidth}
        \includegraphics[width=\figWidth]{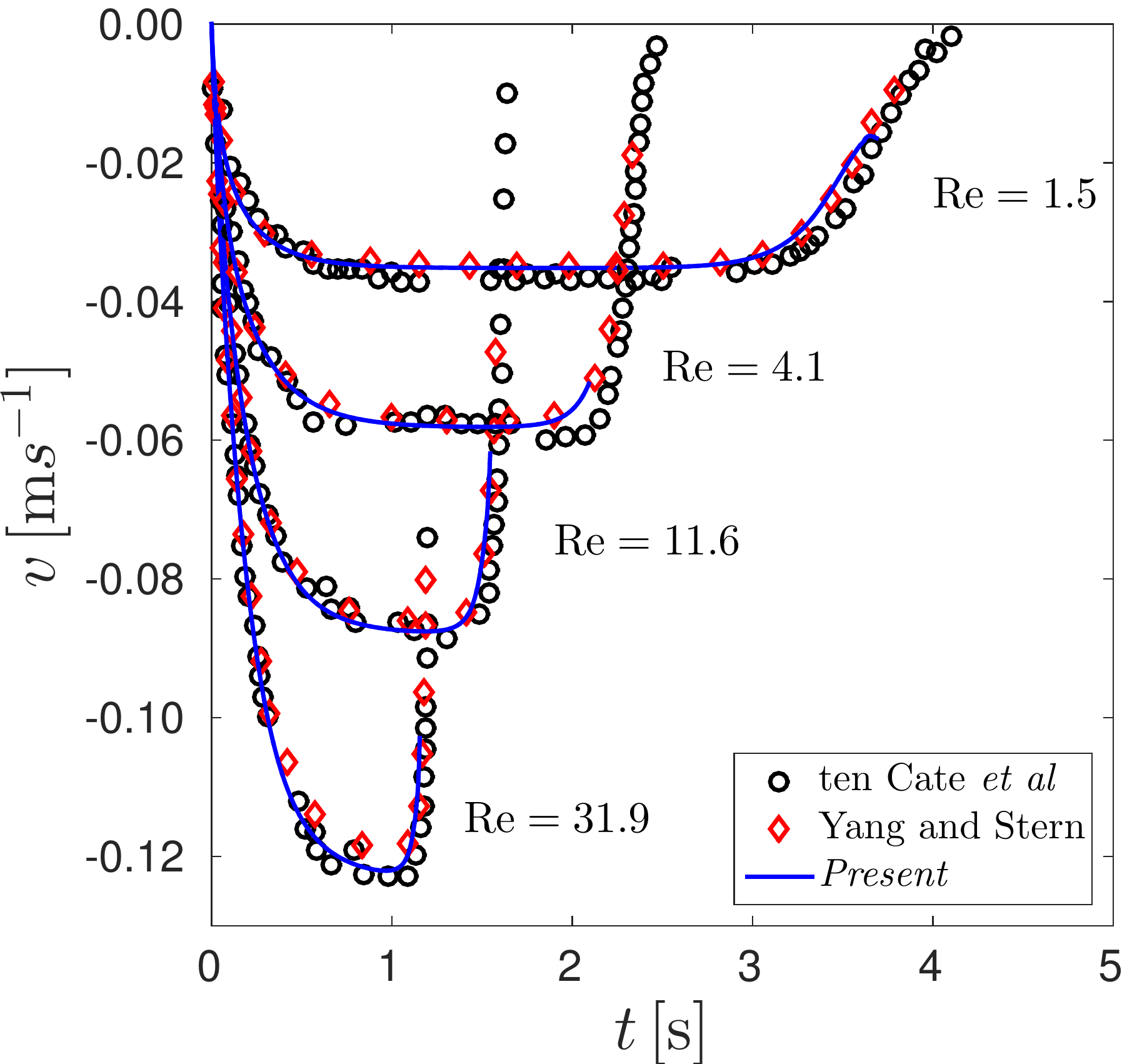}
    \end{minipage}%
    \vspace{-10pt}
    \caption{Left: non-dimensionalised vertical position of the sphere compared to numerical results of 
Yang~et al.~\citep{Yang2015} ($\color{red}\mathsmaller{\lozenge}$) and 
experimental results of ten Cate~et al.~\citep{tenCate2002} ($\mathsmaller\circ$). 
Right: dimensional vertical velocity of the sphere compared to numerical results of~\citep{Yang2015} ($\color{red}\mathsmaller{\lozenge}$) and experimental results of~\citep{tenCate2002} ($\mathsmaller\circ$). \label{FIG: sphere h and v}}
\end{figure}

\section{Conclusion}

We evaluated the overset grid method for DNS of viscous, incompressible 
fluid flow with rigid, moving bodies.
Several FSI benchmark test cases were carried out for 
verification and validation purposes. A systematic convergence test was carried out using six 
uniformly refined grids and six with local refinement near the particle surface. Local refinement 
was found to produce results deviating no more than five percent from the reference solutions, with a 
more than 23 fold decrease in grid point count and a subsequent 13 fold decrease in CPU time. 

Finally, results for a sphere settling in a small tank 
at various Reynolds numbers compared well with both experimental results of ten Cate~\etal\citep{tenCate2002} and 
recent numerical results of Yang~\etal\citep{Yang2015}, using only one third the number of grid points as 
the latter study.

The popular test cases presented in this work are all, to varying degrees, inertially dominated and 
exhibit viscous boundary layers that must be fully resolved to accurately simulate the 
behaviour of the rigid bodies in the flow. The second-order accurate boundary fitted method demonstrated 
here was found to produce reasonably converged results with approximately six grid points across the 
estimated boundary layer depth. By using a coarse---but fine enough to resolve wake structures---Cartesian 
background grid and refined, boundary-fitted grids, grid point counts were greatly reduced, even 
in two-dimensional problems. 

The overset grid method has shown promising capabilities for fully-resolved DNS of small numbers of 
rigid particles. With a more sophisticated collision mechanism, 
\eg the multi-scale approach of 
Kempe~\etal\citep{Kempe2012b} or the DEM approach of~Wachs~\citep{Wachs2009}, fully-resolved DNS of larger numbers of 
arbitrarily shaped particles could be performed. Without modification to the underlying discretisation 
technique other types of flow, for example arbitrarily moving bodies in non-Newtonian flows, could be 
examined.

\section*{References}
 \bibliographystyle{elsarticle-harv}
\bibliography{bibliography/denseSuspensions,bibliography/henshawPapers,bibliography/henshaw,bibliography/fsi}

\begin{thebibliography}{76}
\expandafter\ifx\csname natexlab\endcsname\relax\def\natexlab#1{#1}\fi
\expandafter\ifx\csname url\endcsname\relax
  \def\url#1{\texttt{#1}}\fi
\expandafter\ifx\csname urlprefix\endcsname\relax\def\urlprefix{URL }\fi

\bibitem[{Almgren et~al.(1998)Almgren, Bell, Colella, Howell, and
  Welcome}]{Almgren1998}
Almgren, A.~S., Bell, J.~B., Colella, P., Howell, L.~H., Welcome, M.~L., 1998.
  A conservative adaptive projection method for the variable density
  incompressible {Navier--Stokes} equations. J.\@ Comput.\@ Phys.\@ 142, 1--46.

\bibitem[{Ardekani and Rangel(2008)}]{Ardekani2008}
Ardekani, A.~M., Rangel, R.~H., 2008. Numerical investigation of
  particle--particle and particle--wall collisions in a viscous fluid. J.\@
  Fluid Mech.\@ 596, 437--466.

\bibitem[{Bagchi(2007)}]{Bagchi2007}
Bagchi, P., 2007. Mesoscale simulation of blood flow in small vessels.
  Biophys.\@ J.\@ 92, 1858--1877.

\bibitem[{Balay et~al.(2013)Balay, Brown, Buschelman, Eijkhout, Gropp, Kaushik,
  Knepley, McInnes, Smith, and Zhang}]{Balay2013}
Balay, S., Brown, J., Buschelman, K., Eijkhout, V., Gropp, W.~D., Kaushik, D.,
  Knepley, M.~G., McInnes, L.~C., Smith, B.~F., Zhang, H., 2013. {PETS}c users
  manual. Tech. rep., Argonne National Laboratory.

\bibitem[{Banks et~al.(2016)Banks, Henshaw, Kapila, and
  Schwendeman}]{flunsi2016}
Banks, J.~W., Henshaw, W.~D., Kapila, A., Schwendeman, D.~W., 2016. An
  added-mass partitioned algorithm for fluid-structure interactions of
  compressible fluids and nonlinear solids. J. Comput. Phys. 305,
  1037--1064\citeCount{0}.

\bibitem[{Banks et~al.(2012)Banks, Henshaw, and Schwendeman}]{fsi2012}
Banks, J.~W., Henshaw, W.~D., Schwendeman, D.~W., 2012. Deforming composite
  grids for solving fluid structure problems. J. Comput. Phys. 231~(9),
  3518--3547\citeCount{5}.

\bibitem[{Banks et~al.(2014{\natexlab{a}})Banks, Henshaw, and
  Schwendeman}]{Banks2014}
Banks, J.~W., Henshaw, W.~D., Schwendeman, D.~W., 2014{\natexlab{a}}. An
  analysis of a new stable partitioned algorithm for {FSI} problems. {P}art
  {I}: incompressible flow and elastic solids. J.\@ Comput.\@ Phys.\@ 269,
  108--137.

\bibitem[{Banks et~al.(2014{\natexlab{b}})Banks, Henshaw, and
  Schwendeman}]{fis2014}
Banks, J.~W., Henshaw, W.~D., Schwendeman, D.~W., 2014{\natexlab{b}}. An
  analysis of a new stable partitioned algorithm for {FSI} problems. {Part II}:
  Incompressible flow and structural shells. J. Comput. Phys. 268,
  399--416\citeCount{2}.

\bibitem[{Banks et~al.(2013)Banks, Henshaw, and Sj{\"o}green}]{lrb2013}
Banks, J.~W., Henshaw, W.~D., Sj{\"o}green, B., 2013. A stable {FSI} algorithm
  for light rigid bodies in compressible flow. J. Comput. Phys. 245,
  399--430\citeCount{1}.

\bibitem[{Batchelor(1967)}]{Batchelor1967}
Batchelor, G.~K., 1967. An introduction to fluid dynamics. Cambridge University
  Press.

\bibitem[{Behr and Tezduyar(1994)}]{Behr1994}
Behr, M., Tezduyar, T.~E., 1994. Finite element strategies for large-scale flow
  simulations. Comput.\@ Methods Appl.\@ Mech.\@ Engrg.\@ 112, 3--24.

\bibitem[{Borazjani et~al.(2008)Borazjani, Ge, and
  Sotiropoulos}]{Borazjani2008}
Borazjani, I., Ge, L., Sotiropoulos, F., 2008. Curvilinear immersed boundary
  method for simulating fluid structure interaction with complex {3D} rigid
  bodies. J. Comput. Phys. 227~(16), 7587--7620.

\bibitem[{Brady(1988)}]{Brady1988}
Brady, J.~F., 1988. Stokesian dynamics. Annu.\@ Rev.\@ Fluid Mech.\@ 20,
  111--157.

\bibitem[{Broering et~al.(2012)Broering, Lian, and
  Henshaw}]{BroeringLianHenshaw2012}
Broering, T., Lian, Y., Henshaw, W., 2012. Numerical investigation of energy
  extraction in a tandem flapping wing configuration. AIAA Journal 50~(11),
  2295--2307\citeCount{2}.

\bibitem[{Chan(2009)}]{Chan2009}
Chan, W.~M., 2009. Overset grid technology development at {NASA Ames Research
  Center}. Comput. Fl. 38~(3), 496--503.

\bibitem[{Chandar and Damodaran(2010)}]{ChandarDamodaran2010}
Chandar, D. D.~J., Damodaran, M., 2010. Numerical study of the free flight
  characteristics of a flapping wing in low {R}eynolds numbers. AIAA J.
  Aircraft 47~(1), 141--150.

\bibitem[{Chesshire and Henshaw(1990)}]{CGNS}
Chesshire, G.~S., Henshaw, W.~D., 1990. Composite overlapping meshes for the
  solution of partial differential equations. J. Comput. Phys. 90~(1),
  1--64\citeCount{349}.

\bibitem[{Chesshire and Henshaw(1994)}]{CGCN94}
Chesshire, G.~S., Henshaw, W.~D., July 1994. A scheme for conservative
  interpolation on overlapping grids. SIAM J. Sci. Comput. 15~(4),
  819--845\citeCount{60}.

\bibitem[{Dougherty and Kuan(1989)}]{Dougherty}
Dougherty, F.~C., Kuan, J.-H., 1989. Transonic store separation using a
  three-dimensional {Chimera} grid scheme. paper 89-0637, AIAA.

\bibitem[{English et~al.(2013)English, Qiu, Yu, and Fedkiw}]{English2013}
English, R.~E., Qiu, L., Yu, Y., Fedkiw, R., 2013. An adaptive discretization
  of incompressible flow using a multitude of moving {Cartesian} grids. J.\@
  Comput.\@ Phys.\@ 254, 107--154.

\bibitem[{Fadlun et~al.(2000)Fadlun, Verzicco, Orlandi, and
  Mohd-Yusof}]{Fadlun2000}
Fadlun, E.~A., Verzicco, R., Orlandi, P., Mohd-Yusof, J., 2000. Combined
  immersed-boundary finite-difference methods for three-dimensional complex
  flow simulations. J.\@ Comput.\@ Phys.\@ 161, 35--60.

\bibitem[{Falgout and Yang(2002)}]{Falgout2002}
Falgout, R.~D., Yang, U.~M., 2002. hypre: a library of high performance
  preconditioners. Springer Berlin Heidelberg.

\bibitem[{Feng and Michaelides(2004)}]{Feng2004}
Feng, Z.-G., Michaelides, E.~E., 2004. The immersed boundary-lattice
  {Boltzmann} method for solving fluid--particles interaction problems. J.\@
  Comput.\@ Phys.\@ 195, 602--628.

\bibitem[{Ferziger and Peri\'{c}(2002)}]{Ferziger2002}
Ferziger, J.~H., Peri\'{c}, M., 2002. Computational methods for fluid dynamics,
  3rd Edition. Springer.

\bibitem[{Fortes et~al.(1987)Fortes, Joseph, and Lundgren}]{Fortes1987}
Fortes, A.~F., Joseph, D.~D., Lundgren, T.~S., 1987. Nonlinear mechanics of
  fluidization of beds of spherical particles. J.\@ Fluid Mech.\@ 177,
  467--483.

\bibitem[{Glowinski et~al.(2001)Glowinski, Pan, Hesla, Joseph, and
  P\'eriaux}]{Glowinski2001}
Glowinski, R., Pan, T.~W., Hesla, T.~I., Joseph, D.~D., P\'eriaux, J., 2001. A
  fictitious domain approach to the direct numerical simulation of
  incompressible viscous flow past moving rigid bodies: {Application} to
  particulate flow. J.\@ Comput.\@ Phys.\@ 169, 363--426.

\bibitem[{Haeri and Shrimpton(2012)}]{Haeri2012}
Haeri, S., Shrimpton, J.~S., 2012. On the application of immersed boundary,
  ficticious domain and body-conformal mesh methods to many particle multiphase
  flows. Int.\@ J.\@ Multiphas.\@ Flow 40, 38--55.

\bibitem[{Henshaw(1994)}]{Henshaw1994}
Henshaw, W.~D., 1994. A fourth-order accurate method for the incompressible
  {Navier--Stokes} equations on overlapping grids. J.\@ Comput.\@ Phys.\@ 113,
  13--25.

\bibitem[{Henshaw(1998)}]{OGEN}
Henshaw, W.~D., 1998. Ogen: An overlapping grid generator for {O}verture.
  Research Report UCRL-MA-132237, Lawrence Livermore National Laboratory.

\bibitem[{Henshaw(2005)}]{automg}
Henshaw, W.~D., 2005. On multigrid for overlapping grids. SIAM J. Sci. Comput.
  26~(5), 1547--1572\citeCount{16}.

\bibitem[{Henshaw and Petersson(2001)}]{}
Henshaw, W.~D., Petersson, N.~A., 2001. A split-step scheme for the
  incompressible {Navier--Stokes} equations. Tech. rep., Centre for {Applied
  Scientific Computing}, {Lawrence Livermore National Laboratory}, Livermore,
  CA, 94551.

\bibitem[{Henshaw and Petersson(2003)}]{splitStep2003}
Henshaw, W.~D., Petersson, N.~A., 2003. A split-step scheme for the
  incompressible {Navier-Stokes} equations. In: Hafez, M.~M. (Ed.), Numerical
  Simulation of Incompressible Flows. World Scientific, pp.
  108--125\citeCount{30}.

\bibitem[{Henshaw and Schwendeman(2006)}]{mog2006}
Henshaw, W.~D., Schwendeman, D.~W., 2006. Moving overlapping grids with
  adaptive mesh refinement for high-speed reactive and non-reactive flow. J.
  Comput. Phys. 216~(2), 744--779\citeCount{54}.

\bibitem[{Hu(1996)}]{Hu1996}
Hu, H.~H., 1996. Direct simulation of flows of solid--liquid mixtures. Int.\@
  J.\@ Multiphas.\@ Flow 22~(2), 335--352.

\bibitem[{Hu et~al.(2001)Hu, Patankar, and Zhu}]{Hu2001}
Hu, H.~H., Patankar, N.~A., Zhu, M.~Y., 2001. Direct numerical simulations of
  fluid--solid systems using the arbitrary {Lagrangian}--{Eulerian} technique.
  J.\@ Comput.\@ Phys.\@ 169, 427--462.

\bibitem[{Hu et~al.(2015)Hu, Li, Shu, and Niu}]{Hu2015}
Hu, Y., Li, D., Shu, S., Niu, X., 2015. Modified momentum exchange method for
  fluid--particle interactions in the lattice {Boltzmann} method. Phys.\@
  Rev.\@ E 91.

\bibitem[{Hughes et~al.(1981)Hughes, Liu, and Zimmermann}]{Hughes1981}
Hughes, T. J.~R., Liu, W.~K., Zimmermann, T.~K., 1981. Lagrangian--{Eulerian}
  finite element formulation for incompressible viscous flows. Comput.\@
  Methods Appl.\@ Mech.\@ Engrg.\@ 29, 329--349.

\bibitem[{Johnson and Tezduyar(1996)}]{Johnson1996}
Johnson, A.~A., Tezduyar, T.~E., 1996. Simulation of multiple spheres falling
  in a liquid-filled tube. Comput.\@ Methods Appl.\@ Mech.\@ Engrg.\@ 134,
  351--373.

\bibitem[{Joseph et~al.(2001)Joseph, Zenit, Hunt, and Rosenwinkel}]{Joseph2001}
Joseph, G.~G., Zenit, R., Hunt, M.~L., Rosenwinkel, A.~M., 2001. Particle--wall
  collisions in a viscous fluid. J.\@ Fluid Mech.\@ 433, 329--346.

\bibitem[{Kempe and Fr\"ohlich(2012{\natexlab{a}})}]{Kempe2012b}
Kempe, T., Fr\"ohlich, J., 2012{\natexlab{a}}. Colision modelling for the
  interface-resolved simulation of spherical particles in viscous fluids. J.\@
  Fluid Mech.\@ 709, 445--489.

\bibitem[{Kempe and Fr\"ohlich(2012{\natexlab{b}})}]{Kempe2012a}
Kempe, T., Fr\"ohlich, J., 2012{\natexlab{b}}. An improved immersed boundary
  method with direct forcing for the simulation of particle laden flows. J.\@
  Comput.\@ Phys.\@ 231, 3663--3684.

\bibitem[{Kempe and Fr{\"o}hlich(2012)}]{kempe2012improved}
Kempe, T., Fr{\"o}hlich, J., 2012. An improved immersed boundary method with
  direct forcing for the simulation of particle laden flows. J. Comput. Phys.
  231~(9), 3663--3684.

\bibitem[{Kim and Choi(2006)}]{Kim2006}
Kim, D., Choi, H., 2006. Immersed boundary method for flow around an
  arbitrarily moving body. J.\@ Comput.\@ Phys.\@ 212, 662--680.

\bibitem[{Kushch et~al.(2002)Kushch, Sangani, Spelt, and Koch}]{Kushch2002}
Kushch, V.~I., Sangani, A.~S., Spelt, P. D.~M., Koch, D.~L., 2002.
  Finite-{Weber}-number motion of bubbles through a nearly inviscid liquid.
  J.\@ Fluid Mech.\@ 460, 241--280.

\bibitem[{K{\"u}ttler and Wall(2008)}]{KuttlerWall2008}
K{\"u}ttler, U., Wall, W.~A., 2008. Fixed-point fluid--structure interaction
  solvers with dynamic relaxation. Computational Mechanics 43~(1), 61--72.

\bibitem[{Lani et~al.(2012)Lani, Sj\"ogreen, Yee, and
  Henshaw}]{LaniSjogreenYeeHenshaw2012}
Lani, A., Sj\"ogreen, B., Yee, H.~C., Henshaw, W.~D., 2012. Variable high-order
  multiblock overlapping grid methods for mixed steady and unsteady multiscale
  viscous flows, part {II}: hypersonic nonequilibrium flows. Commun. Comput.
  Phys. 13~(2), 583--602\citeCount{4}.

\bibitem[{Li et~al.(2016)Li, Henshaw, Banks, Schwendeman, and
  Main}]{beamins2016}
Li, L., Henshaw, W.~D., Banks, J.~W., Schwendeman, D.~W., Main, G.~A., 2016. A
  stable partitioned {FSI} algorithm for incompressible flow and deforming
  beams. J. Comput. Phys. 312, 272--306.

\bibitem[{Luo et~al.(2012)Luo, Dai, {Ferreira de Sousa}, and Yin}]{Luo2012}
Luo, H., Dai, H., {Ferreira de Sousa}, P. J. S.~A., Yin, B., 2012. On the
  numerical oscillation of the direct-forcing immersed-boundary method for
  moving boundaries. Comput.\@ Fluids 56, 61--76.

\bibitem[{Meakin(1993)}]{Meakin93}
Meakin, R., 1993. Moving body overset grid methods for complete aircraft
  tiltrotor simulations. paper 93-3350, AIAA.

\bibitem[{Mittal and Iccarino(2005)}]{Mittal2005}
Mittal, R., Iccarino, G., 2005. Immersed boundary methods. Annu.\@ Rev.\@ Fluid
  Mech.\@ 37, 239--261.

\bibitem[{Mohd-Yusof(1997)}]{Mohd1997}
Mohd-Yusof, J., 1997. Combined immersed-boundary/{B}-spline methods for
  simulations of flow in complex geometries. Center for Turbulence Research
  Annual Research Briefs.

\bibitem[{Nelson and Guillot(2006)}]{Nelson2006}
Nelson, E.~B., Guillot, D., 2006. Well cementing, 2nd Edition. Schlumberger.

\bibitem[{Nicolle(2010)}]{Nicolle2010}
Nicolle, A., 2010. Flow through and around groups of bodies. Ph.D. thesis,
  University College London.

\bibitem[{Nicolle and Eames(2011)}]{Nicolle2011}
Nicolle, A., Eames, I., 2011. Numerical study of flow through and around a
  circular array of cylinders. J.\@ Fluid Mech.\@ 679, 1--31.

\bibitem[{Niu et~al.(2006)Niu, Shu, Chew, and Peng}]{Niu2006}
Niu, X.~D., Shu, C., Chew, Y.~T., Peng, Y., 2006. A momentum exchange-based
  immersed boundary-lattice {Boltzmann} method for simulating incompressible
  viscous flows. Phys.\@ Lett.\@ A 354, 173--182.

\bibitem[{Patankar(2001)}]{Patankar2001a}
Patankar, N.~A., 2001. A formulation for fast computations of rigid particulate
  flows. Center for Turbulence Research.

\bibitem[{Patankar et~al.(2000)Patankar, Singh, Joseph, Glowinski, and
  Pan}]{Patankar2000}
Patankar, N.~A., Singh, P., Joseph, D.~D., Glowinski, R., Pan, T.-W., 2000. A
  new formulation of the distributed {Lagrange} multiplier/fictitious domain
  method for particulate flows. Int.\@ J.\@ Multiphas.\@ Flow 26, 1509--1524.

\bibitem[{Patankar and Spalding(1972)}]{Patankar1972}
Patankar, S.~V., Spalding, D.~B., 1972. A calculation procedure for heat, mass
  and momentum transfer in three-dimensional parabolic flows. Int.\@ J.\@ Heat
  Mass Transfer 15, 1787--1806.

\bibitem[{Peskin(1972)}]{Peskin1972}
Peskin, C.~S., 1972. Flow patterns around heart valves: a numerical method.
  J.\@ Comput.\@ Phys.\@ 10, 252--271.

\bibitem[{Petersson(2001)}]{Petersson2001}
Petersson, N.~A., 2001. Stability of pressure boundary conditions for {Stokes}
  and {Navier--Stokes} equations. J.\@ Comput.\@ Phys.\@ 172, 40--70.

\bibitem[{Qiu et~al.(2015)Qiu, Yu, and Fedkiw}]{Qiu2015}
Qiu, L., Yu, Y., Fedkiw, R., 2015. On thin gaps between rigid bodies two-way
  coupled to incompressible flow. J.\@ Comput.\@ Phys.\@ 292, 1--29.

\bibitem[{Sangani and Didwania(1993)}]{Sangani1993}
Sangani, A.~S., Didwania, A.~K., 1993. Dynamic simulations of flows of bubbly
  liquids at large {Reynolds} numbers. J.\@ Fluid Mech.\@ 250, 307--337.

\bibitem[{Seo and Mittal(2011)}]{Seo2011}
Seo, J.~H., Mittal, R., 2011. A sharp-interface immersed boundary method with
  improved mass conservation and reduced spurious pressure oscillations. J.\@
  Comput.\@ Phys.\@ 230, 7347--7363.

\bibitem[{Tang et~al.(2003)Tang, Jones, and
  Sotiropoulos}]{TangJonesSotiropoulos2003}
Tang, H.~S., Jones, S.~C., Sotiropoulos, F., 2003. An overset-grid method for
  {3D} unsteady incompressible flows. J. Comput. Phys. 191, 567--600.

\bibitem[{ten Cate et~al.(2002)ten Cate, Nieuwstad, Derksen, and {Van den
  Akker}}]{tenCate2002}
ten Cate, A., Nieuwstad, C.~H., Derksen, J.~J., {Van den Akker}, H. E.~A.,
  2002. Particle imaging velocimetry experiments and lattice-{Boltzmann}
  simulations on a single sphere settling under gravity. Phys.\@ Fluids 11.

\bibitem[{Tezduyar et~al.(1992)Tezduyar, Behr, and Liou}]{Tezduyar1992}
Tezduyar, T.~E., Behr, M., Liou, J., 1992. A new strategy for finite element
  computations involving moving boundaries and interfaces -- the
  deforming-spatial-domain/space-time procodure: I. {T}he concept and the
  preliminary numerical tests. Comput.\@ Methods Appl.\@ Mech.\@ Engrg.\@ 94,
  339--351.

\bibitem[{Uhlmann(2005)}]{Uhlmann2005}
Uhlmann, M., 2005. An immersed boundary method with direct forcing for the
  simulation of particulate flows. J.\@ Comput.\@ Phys.\@ 209, 448--476.

\bibitem[{Uhlmann and Doychev(2014)}]{Uhlmann2014}
Uhlmann, M., Doychev, T., 2014. Sedimentation of a dilute suspension of rigid
  spheres at intermediate {Galileo} numbers: the effect of clustering upon the
  particle motion. J.\@ Fluid Mech.\@ 752, 310--348.

\bibitem[{Vowinckel et~al.(2014)Vowinckel, Kempe, and
  Fr\"ohlich}]{Vowinckel2014}
Vowinckel, B., Kempe, T., Fr\"ohlich, J., 2014. Fluid--particle interaction in
  turbulent open channel flow with fully-resolved mobile beds. Adv.\@ Water
  Resour.\@ 72, 32--44.

\bibitem[{Wachs(2009)}]{Wachs2009}
Wachs, A., 2009. A {DEM-DLM/FD} method for direct numerical simulation of
  particulate flows: {Sedimentation} of polygonal isometric particles in a
  {Newtonian} fluid with collisions. Comput.\@ Fluids 38, 1608--1628.

\bibitem[{Wachs et~al.(2015)Wachs, Hammouti, Vinay, and Rahmani}]{Wachs2015}
Wachs, A., Hammouti, A., Vinay, G., Rahmani, M., 2015. Accuracy of finite
  volume/staggered grid distributed {Lagrange} multiplier/fictitious domain
  simulations of particulate flows. Comput.\@ Fluids 115, 154--172.

\bibitem[{Wan and Turek(2006)}]{Wan2006}
Wan, D., Turek, S., 2006. Direct numerical simulation of particulate flow via
  multigrid {FEM} techniques and the fictitious boundary method. Int.\@ J.\@
  Numer.\@ Meth.\@ Fl.\@ 51, 531--566.

\bibitem[{Yang and Balaras(2006)}]{Yang2006}
Yang, J., Balaras, E., 2006. An embedded-boundary formulation for large-eddy
  simulation of turbulent flows interacting with moving boundaries. J.\@
  Comput.\@ Phys.\@ 215, 12--40.

\bibitem[{Yang and Stern(2015)}]{Yang2015}
Yang, J., Stern, F., 2015. A non-iterative direct forcing immersed boundary
  method for strongly-coupled fluid-solid interactions. J.\@ Comput.\@ Phys.\@
  295~(779--804).

\bibitem[{Zahle et~al.(2007)Zahle, Johansen, S{\o}rensen, and
  Graham}]{Zahle2007}
Zahle, F., Johansen, J., S{\o}rensen, N.~N., Graham, J. M.~R., 2007. Wind
  turbine rotor-tower interaction using an incompressible overset grid method.
  paper 2007-425, AIAA.

\bibitem[{Zhang and Prosperetti(2003)}]{Zhang2003}
Zhang, Z., Prosperetti, A., 2003. A method for particle simulation. ASME 70,
  64--74.

\end{thebibliography}

\end{document}